\RequirePackage[l2tabu, orthodox]{nag}
\RequirePackage{snapshot}

\documentclass[9pt,twocolumn]{extarticle}

\makeatletter
\if@twocolumn
  \usepackage[dvips,letterpaper,top=0.5in, bottom=0.5in, left=0.75in, right=0.5in,includefoot,heightrounded]{geometry}
\else
  \usepackage[dvips,letterpaper,margin=1in,includefoot,heightrounded]{geometry}
\fi

\usepackage{srcltx}

\usepackage[russian,portuges,english]{babel}

\iflanguage{portuges}
    {\newcommand{\keywordname}{Palavras-chaves}}
    {\newcommand{\keywordname}{Keywords}}

\usepackage{amsmath}
\usepackage{amssymb,amsfonts}

\usepackage{abstract}

\usepackage{graphicx}
\usepackage[usenames,dvipsnames,svgnames,x11names]{xcolor}
\usepackage{subfigure}

\usepackage{booktabs}

\usepackage{setspace}
\usepackage{flushend}

\usepackage{cite}

\usepackage{hyperref}
\usepackage[normalem]{ulem}

\usepackage{enumerate}

\usepackage{multirow}
\usepackage[noend]{algpseudocode}

\usepackage{listings}

\usepackage[short,12hr]{datetime}
       \usepackage{fouriernc}
\makeatletter

\newenvironment{rsmallmatrix}{\null\,\vcenter\bgroup
  \Let@\restore@math@cr\default@tag
  \baselineskip6\ex@ \lineskip1.5\ex@ \lineskiplimit\lineskip
  \ialign\bgroup\hfil$\m@th\scriptstyle##$&&\thickspace\hfil
  $\m@th\scriptstyle##$\crcr
}{%
  \crcr\egroup\egroup\,%
}

\newcommand{\printtitle}{%
\makeatletter
\if@twocolumn

\twocolumn[%
  \maketitle
  \begin{onecolabstract}
    \myabstract
  \end{onecolabstract}
  \begin{center}
    \small
    \textbf{\keywordname}
    \\\medskip
    \mykeywords
  \end{center}
  \bigskip
]
\saythanks
\else
  \maketitle
  \begin{onecolabstract}
    \myabstract
  \end{onecolabstract}
  \begin{center}
    \small
    \textbf{\keywordname}
    \\\medskip
    \mykeywords
  \end{center}
  \bigskip
  \onehalfspacing
\fi
\makeatother
}

\author{%
A.~P.~Rad\"unz
\thanks{Programa de P\'os-Gradua\c c\~ao em Estat\'istica, Universidade Federal de Pernambuco, Recife, Brazil.
E-mail: \url{apr1@de.ufpe.br}}
\and
T.~L.~T.~da~Silveira
\thanks{Instituto de Informática, Universidade Federal do Rio Grande do Sul, Porto Alegre, Brazil.
E-mail: \url{thiago@inf.ufsm.br}}
\and
F.~M.~Bayer
\thanks{Departamento de Estat\'istica and LACESM, Universidade Federal de Santa Maria, Santa Maria, Brazil.
E-mail: \url{bayer@ufsm.br}}
\and
R.~J.~Cintra
\thanks{%
Signal Processing Group, CCEN,
UFPE, Brazil.
E-mail: \url{rjdsc@de.ufpe.br}}
}

\title{%
Data-independent Low-complexity KLT Approximations for Image and Video Coding}

\newcommand{\myabstract}{%
The Karhunen-Lo\`eve transform (KLT) is often used for data decorrelation and dimensionality reduction.
The KLT is able to optimally retain the signal energy in only few transform components,
being mathematically suitable for image and video compression.
However, in practice, because of its high computational cost and dependence on the input signal,
its application in real-time scenarios is precluded.
This work proposes low-computational cost approximations for the KLT.
We focus on the blocklengths $N \in \{4, 8, 16, 32 \}$
because they are widely employed in image and video coding standards such as JPEG and high efficiency video coding (HEVC).
Extensive computational experiments
demonstrate the suitability of the proposed low-complexity transforms
for
image and video compression.
}

\newcommand{\mykeywords}{%
Approximate transform,
image compression,
Karhunen-Lo\`eve transform,
low-complexity transforms,
signed KLT.
}

\date{}

\date{\today\ @ \currenttime}

\date{}

\begin{document}

\printtitle

\section{Introduction}
\label{sec:intro}

The Karhunen-Lo\`eve transform (KLT)~\cite{britanak2010discrete}
is a commonly used tool
for data decorrelation and dimensionality reduction~\cite{jolliffe1986principal,chipman2005interpretable}.
It consists of a linear transformation that maps correlated variables into uncorrelated variables,
sometimes referred to as principal components~\cite{johnson2002applied}.
Usually, only the first coefficients of the transformed data
are
sufficient to represent the signal.
The KLT capability for energy compaction is
paramount
for data compression, since most information can be preserved even
reducing the dimensionality of the data~\cite{du2007hyperspectral}.
In fact, considering  first-order Markov processes, the KLT is an optimal linear transform capable of minimizing the mean square error in data
compression and concentrating energy in few coefficients of the output signal~\cite{britanak2010discrete}.
Although it is a well-established optimal transform in  terms  of  energy  compaction  and
decorrelation~\cite{ochoa2019discrete}, the KLT is not widely applied because its computation depends
on the covariance matrix of the input data.
Indeed, such data-dependent requirements can hinder the development of fast algorithms for an efficient implementation of the transform.

However,
if the input data is a first-order Markov process with
known correlation coefficient $\rho$, then it was shown in~\cite{ray1970further} that we can derive an analytical solution for the elements of the KLT matrix.
Nevertheless, even with the transform matrix known,
computational complexity of its implementation can be infeasible for practical data compression scenarios.
In this context, several fast approximations for the KLT have been proposed~\cite{jain1976fast, lan1993fast,lan1994improved, pirooz1998new, yanyun2004updating, cagnazzo2006low, sole2009joint, blanes2012divide, hao2003reversible,wongsawat2006integer} aiming at reducing
the computational costs.
Although such methods
generate fast approximations for the KLT,
their scope is relatively limited
because
the data-dependence is still present;
in some cases either depending on the covariance matrix of
the input data~\cite{lan1993fast,lan1994improved,pirooz1998new,yanyun2004updating,cagnazzo2006low,blanes2012divide,wongsawat2006integer}
or
on
the correlation coefficient in case of first-order Markovian signals~\cite{jain1976fast}.

When considering first-order Markovian random signals, \cite{ahmed1974discrete} and \cite{clarke1981relation,clarke1984relation} have shown that the discrete cosine transform (DCT-II) and the discrete sine transform (DST-I) are asymptotic approximations for the KLT,
with the correlation coefficient of the input signal tending to unity and to zero, respectively~\cite{ahmed1974discrete}.
Both the DCT and the DST are independent of the input signal, allowing the development of computationally efficient fast algorithms.
The DCT is widely adopted in image and video compression standards such as JPEG~\cite{wallace1992jpeg}
and high efficiency video coding (HEVC)~\cite{pourazad2012hevc},
just to name a few.
However, the use of this transform can still be prohibitive in contexts under severe restrictions on
processing power or energy autonomy~\cite{cintra2014low,cintra2018low,bouguezel2008low, sheltami2016data}.
In fact, DCT realizations that require multiplications implemented
in floating-point arithmetic-based hardware~\cite{potluri2014improved}
demand significant circuitry complexity and energy consumption~\cite{britanak2010discrete}.
In this sense, several multiplication-free approximations for the DCT have been proposed~\cite{cintra2014low, cintra2011dct, bouguezel2008low, bouguezel2012binary,Jridi2015, lengwehasatit2004scalable, haweel2016fast,da2017multiplierless,huang2019deterministic, almurib2017approximate,tablada2017dct,oliveira2019low,canterle2020multiparametric}, including the signed DCT (SDCT)~\cite{haweel2001new}.
The SDCT is
derived by applying the signum function to the elements of the DCT matrix,
thus resulting in a matrix of
trivial multiplicands $\{-1, +1\}$.
Therefore, the transform computation
requires only additions.
Such reduction in the arithmetic cost implies in a lower computational cost,
favoring applications in real-time and
in low-consumption devices~\cite{britanak2010discrete}.

The present work employs
the signum function as a means to obtain computationally efficient alternatives to
the KLT for first-order Markov processes.
We follow an entirely different approach
when compared with the fast KLT approximations
already known in the literature.
Here, we focus
on the proposition of deterministically defined multiplierless low-complexity approximations for the KLT that does not depend on the input signal
and
is capable of
coping with a wide range of correlation coefficients.
Our analyses are devoted to the blocklengths $N\in\{4, 8, 16, 32 \}$
because of their relevance in image and video standards as JPEG~\cite{wallace1992jpeg} and HEVC~\cite{pourazad2012hevc}.
In order to find the best-performing low-cost approximations,
we propose a constrained optimization approach
according
to suitable figures of merit
for the KLT analysis.
The considered approximation method
is specifically tailored
to furnish low-complexity transformations
appropriate for dedicated highly-efficient circuitry design.
The resulting KLT approximations are sought to be numerically evaluated according to
coding performance \cite{jayant1984digital,nikara2001unified}, and similarity/proximity metrics~\cite{britanak2010discrete,cintra2011dct} with respect to the exact KLT.
The obtained transforms are then embedded into (i)~a JPEG-like image compression scheme, and (ii)~an HEVC reference software for video coding assessment.

To the best of our knowledge, literature lacks
KLT approximations that combine the following properties:
\begin{enumerate}
\item
deterministic definition;

\item
suitability for fast algorithm design;

\item
data-independence; and

\item
capability of processing data at a wide range correlation.
\end{enumerate}
We aim, therefore, at a proposition of a new class of KLT approximations that addresses these gaps.
The main goal of our paper is to propose low-complexity approximate
transforms for the KLT considering different values of the correlation coefficients $\rho$, so low and mid-correlated
signals could be properly treated as well.
Since, to the best of our knowledge, the literature lacks efficient KLT-based
methods considering lowly correlated data, there is no competing method for a fair comparison.

This paper is structured as follows.
In Section \ref{S:KLT}, we revise the mathematical formulation of the KLT for first-order Markovian signals and define the general framework for signed KLT (SKLT).
Section \ref{S:busca} describes
the computational approach for obtaining
new transforms and
presents
these transforms for different lengths
attaining
optimality according to the proposed figures of merit based on classical metrics.
Section \ref{S:busca} also presents fast algorithms for the proposed transforms.
In  Section~\ref{S:applic}, we assess the proposed $4$-, $8$-, $16$-, and $32$-point SKLT in image and video coding.
Section \ref{S:conclu}
concludes the paper.

\section{Signed KLT}
\label{S:KLT}

\subsection{Karhunen-Lo\`eve Transform for the First-order Markov Process}

The KLT is a linear transformation represented by an orthogonal matrix $\mathbf{K}_N^{(\rho)}$ which decorrelates an input signal
$\mathbf{x} =
\begin{bmatrix}
x_0 & x_1 & \ldots & x_{N-1}
\end{bmatrix}^\top$ resulting in uncorrelated signal
$\mathbf{y}=
\begin{bmatrix}
y_0 & y_1 & \ldots & y_{N-1}
\end{bmatrix}^\top$.
The $(i,j)$th elements of the transform matrix $\mathbf{K}_N^{(\rho)}$, for an arbitrary value of $\rho \in [0,1]$, are given by~\cite{britanak2010discrete}%
\begin{align}\label{eq:u}
\begin{split}
k_{ij} = \sqrt{\frac{2}{N+ \lambda_j}} \sin \left[\omega_j \left( i - \frac{N-1}{2}\right)+ \frac{(j+1)\pi}{2} \right], \\
i,j = 0,1,\ldots,N-1,
\end{split}
\end{align}
where the eigenvalues of the transformed signal $\mathbf{y}$ covariance matrix are obtained by
\begin{equation}
\lambda_j = \frac{1-\rho^2}{1+\rho^2 -2\rho \cos \omega_j}, \quad j=0,1,\ldots,N-1,
\end{equation}
and $\omega_1, \omega_2, \ldots, \omega_{N}$ are the $N$ solutions of the non-linear equation%
\begin{equation}
\tan N \omega = \frac{-(1-\rho^2)\sin \omega}{(1+\rho^2)\cos \omega - 2\rho}.
\end{equation}

It is a well-known fact that adjacent pixels from natural images are highly correlated~\cite{rao2000transform}, being $\rho = 0.95$ a widely adopted assumption~\cite{britanak2010discrete}. When the correlation of the input signal tends the unity, $\rho \rightarrow 1$,
the KLT converges to the DCT~\cite{ahmed1974discrete}.

For instance, if $N = 8$ and $\rho = 0.95$, then the KLT matrix is given by
\begin{equation}
\mathbf{K}_8^{(0.95)}  =
\left[
\begin{rsmallmatrix}
0.338	&	0.351	&	0.360	&	0.364	&	0.364	&	0.360	&	0.351	&	0.338	\\
0.481	&	0.420	&	0.286	&	0.101	&	-0.101	&	-0.286	&	-0.420	&	-0.481	\\
0.467	&	0.207	&	-0.179	&	-0.456	&	-0.456	&	-0.179	&	0.207	&	0.467	\\
0.423	&	-0.085	&	-0.487	&	-0.278	&	0.278	&	0.487	&	0.085	&	-0.423	\\
0.360	&	-0.347	&	-0.356	&	0.351	&	0.351	&	-0.356	&	-0.347	&	0.360	\\
0.283	&	-0.488	&	0.094	&	0.415	&	-0.415	&	-0.094	&	0.488	&	-0.283	\\
0.195	&	-0.462	&	0.460	&	-0.190	&	-0.190	&	0.460	&	-0.462	&	0.195	\\
0.100	&	-0.279	&	0.416	&	-0.490	&	0.490	&	-0.416	&	0.279	&	-0.100\\
\end{rsmallmatrix}
\right].
\end{equation}

\subsection{KLT Approximations}\label{S:SKLT}
Our approach is based on the technique used in~\cite{haweel2001new} for proposing the classical signed DCT (SDCT).
The
proposed transform,
as well as the SDCT, is
motivated by
the reduction of the
total number of arithmetic operations required for
the computation of the transform
at the cost of some accuracy loss~\cite{haweel2001new}.
The technique considers the signum function to generate a matrix approximation
for the KLT.
Thus, we propose the following approximate transformation matrix:
\begin{equation} \label{eq:Tsign}
\widehat{\mathbf{T}}_N^{(\rho)}
\triangleq
\frac{1}{\sqrt{N}}
\operatorname{sign}
\left(
\mathbf{K}_N^{(\rho)}
\right)
,
\end{equation}
where
\begin{equation}\label{eq:sign}
\operatorname{sign}(x) =
\begin{cases}
1, &  \text{if $x>0$,}\\
0, &  \text{if $x=0$,} \\
-1, & \text{if $x<0$},
\end{cases}
\end{equation}
and $\mathbf{K}_N^{(\rho)}$ is the KLT matrix of
order
$N$ with a predefined correlation coefficient $\rho$,
which entries are given by~\eqref{eq:u}.
When applied to a matrix, the signum function operates element-wise.

In other words, we map a given KLT matrix to a low-complexity matrix close to it. Note that, if $\rho = 0$, then the Equation \eqref{eq:Tsign} degenerates into the null matrix. Also if $\rho = 1$, then $\mathbf{K}_N^{(1)}$ is the DCT and the resulting approximation is the signed DCT~\cite{haweel2001new}. Therefore, in practice, our analysis is constrained to $0 < \rho < 1$. Note that because of the non-linearity and discrete nature of the signum function, different KLT matrices might be mapped to the same approximate matrix.
For instance, considering $\rho_1 = 0.7$ and $\rho_2 = 0.9$, we obtain that
\begin{equation}
\widehat{\mathbf{T}}_8^{(\rho_1)} = \widehat{\mathbf{T}}_8^{(\rho_2)} = \frac{1}{\sqrt{8}}
\left[
\begin{rsmallmatrix}
1	&	1	&	1	&	1	&	1	&	1	&	1	&	1	\\
1	&	1	&	1	&	1	&	-1	&	-1	&	-1	&	-1	\\
1	&	1	&	-1	&	-1	&	-1	&	-1	&	1	&	1	\\
1	&	-1	&	-1	&	-1	&	1	&	1	&	1	&	-1	\\
1	&	-1	&	-1	&	1	&	1	&	-1	&	-1	&	1	\\
1	&	-1	&	1	&	1	&	-1	&	-1	&	1	&	-1	\\
1	&	-1	&	1	&	-1	&	-1	&	1	&	-1	&	1	\\
1	&	-1	&	1	&	-1	&	1	&	-1	&	1	&	-1	\\
\end{rsmallmatrix}
\right]
.
\end{equation}

Exhaustively computing all KLT matrices in the range $\rho \in [10^{-3}, 1-10^{-3}]$ in steps of $10^{-3}$ returns $999$ matrices.
However, the number of approximations is much lower: $1$, $2$, $9$, and $37$ different approximations for $N = 4$, $8$, $16$, and $32$, respectively. In view of the above, a methodology for selecting best-performing approximations is necessary, which is the topic of the next section.

Although the KLT is an orthogonal matrix, the proposed transforms are not constrained to be,
therefore, given the proposed transform $\widehat{\mathbf{T}}_N^{(\rho)}$, the transformed signal is given by
\begin{equation}
\mathbf{y}=\widehat{\mathbf{T}}_N^{(\rho)} \cdot \mathbf{x} ,
\end{equation}
and the inverse transformation can be written
as
\begin{equation}
{\mathbf{y}=(\widehat{\mathbf{T}}_N^{(\rho)}})^{-1} \cdot \mathbf{x} .
\end{equation}

\section{Optimal SKLT} \label{S:busca}
In this section, we describe an optimization problem, aiming at the identification of best-performing SKLT matrices, according to the figures of merit detailed next.

\subsection{Figures of Merit for Approximate Transforms} \label{S:fig}

Approximate transform methods~\cite{haweel2001new,bouguezel2008low,
	cintra2011dct,cintra2014low,cintra2018low} are usually assessed in terms of
(i)~coding
metrics such as the coding gain~\cite{jayant1984digital} and transform efficiency~\cite{nikara2001unified}, which measure the power of decorrelation and energy compression;
and
(ii)~proximity metrics with respect
to the exact transform, such as the mean-square error~\cite{britanak2010discrete} and total error energy~\cite{cintra2011dct}, which measure similarities or dissimilarities between
approximate and exact transforms.
In the following, let $\widehat{\mathbf{T}}_N$ be a candidate matrix to be assessed.
\subsubsection{Unified Coding Gain}
The unified coding gain of a transform $\widehat{\mathbf{T}}_N$ is given by~\cite{katto1992short}%
\begin{equation}
{\rm Cg} ({\widehat{\mathbf{T}}_N}) =
10 \cdot
\log_{10}
\Biggl\{
\prod_{k=1}^N
\frac{1}{\sqrt[N]{A_k \cdot B_k}}
\Biggr\}
,
\end{equation}
where
$
A_k
=
\operatorname{su}
\left\{
(\mathbf{h}_k^\top \cdot \mathbf{h}_k)\odot \mathbf{R_x}
\right\}
$,
$\mathbf{h}_k$ is the $k$th
row
vector from $\widehat{\mathbf{T}}_N$,
the function $\operatorname{su}(\cdot)$ returns the sum of the elements of its matrix argument, $\odot$ is the Hadamard matrix product operator~\cite{seber2008matrix},
$\mathbf{R_x}$ is the autocorrelation matrix of the considered first-order Markovian signal,
$B_k = \| \mathbf{g}_k \|^2$
and
$\mathbf{g}_k$ is the $k$th
row vector from~$\widehat{\mathbf{T}}^{-1}_N$, and $\| \cdot \|$ is the Frobenius norm~\cite{seber2008matrix}.

\subsubsection{Transform Efficiency}
Another coding related figure of merit is the transform efficiency, given by~\cite{nikara2001unified}%
\begin{equation*}
\eta({\widehat{\mathbf{T}}_N}) = 100 \frac{ \sum_{i = 1}^{N} \vert r_{i,i}\vert}{ \sum_{i = 1}^{N}  \sum_{j = 1}^{N} \vert r_{i,j} \vert},
\end{equation*}
where $r_{i,j}$ is the $(i,j)$th element from
$\widehat{\mathbf{T}}_N\cdot
\mathbf{R_x}
\cdot \widehat{\mathbf{T}}^\top_N$.

\subsubsection{Mean-Square Error}
The mean-square error (MSE) relative to the KLT is given by~\cite{britanak2010discrete}:
\begin{equation*}
\operatorname{MSE}({\widehat{\mathbf{T}}_N}) = \frac{1}{N}\cdot
\operatorname{tr}
\left\{
(\mathbf{K}^{(\rho)}_N - {\widehat{\mathbf{T}}_N})\cdot \mathbf{R_x} \cdot (\mathbf{K}^{(\rho)}_N - {\widehat{\mathbf{T}}_N})^\top
\right\}
,
\end{equation*}
where
$\operatorname{tr}(\cdot)$ is the trace function~\cite{harville1997trace}.

\subsubsection{Total Error Energy}
The total error energy of an approximation relative to the KLT is computed by~\cite{cintra2011dct}:
\begin{equation*}
\epsilon(\widehat{\mathbf{T}}_N) = \pi \cdot ||\mathbf{K}^{(\rho)}_N - {\widehat{\mathbf{T}}_N}||^2.
\end{equation*}

\subsubsection{Proposed Figures of Merit}\label{ss:unifig}
Because the above discussed figures of merit are defined for a fixed value of $\rho$, we propose the following total metrics which take into account the performance for all values of $0 < \rho < 1$:
\begin{align*}
{\rm Cg}_T(\widehat{\mathbf{T}}_N) &= \int_{0}^{1}|{\rm Cg} ({\mathbf{K}^{(\rho)}_N})-{\rm Cg} ({\widehat{\mathbf{T}}_N})|\operatorname{d}\rho,\\
\eta_T(\widehat{\mathbf{T}}_N) &= \int_{0}^{1}|\eta({\mathbf{K}^{(\rho)}_N})-\eta({\widehat{\mathbf{T}}_N})|\operatorname{d}\rho,\\
{\rm MSE}_T(\widehat{\mathbf{T}}_N) &= \int_{0}^1 {\rm MSE}({\widehat{\mathbf{T}}_N})\operatorname{d}\rho,\\
\epsilon_T(\widehat{\mathbf{T}}_N) &= \int_{0}^ 1 \epsilon({\widehat{\mathbf{T}}_N})\operatorname{d}\rho.
\end{align*}

\subsection{Optimization Problem} \label{S:otimizacao}
In order to identify the overall best-performing approximations, we propose the following optimization problem:
\begin{equation}
\label{eq:optimization}
\widehat{\mathbf{T}}_N^* = \operatornamewithlimits{arg \ min}_{0 < \rho < 1} \operatorname{error} (\widehat{\mathbf{T}}_N^{(\rho)}),
\end{equation}
where $\operatorname{error}(\cdot)$ is one of the proposed measures, ${\rm Cg}_T(\cdot)$, $\eta_T(\cdot)$, ${\rm MSE}_T(\cdot)$, $\epsilon_T(\cdot)$,   presented in the previous subsection. %
Note that, for a fixed transform length $N$,
up to four optimal SKLT can be obtained, each one
optimizing a total metric.
Hereafter we denote the optimal transforms of length $N \in \{4,8,16,32\}$ as $\widehat{\mathbf{T}}_{N,i}$, which are indexed by the subscript $i = 1,2,\ldots,J$, where $J$ is the number of approximate transforms for $N$.

Table~\ref{t:tabela}
summarizes the results for the
optimal SKLT
with
different transform lengths $N$ and for the intervals of $\rho$ that each transform is defined.
DCT, DST, and the SDCT
results are included only for comparison purposes.
It is important to emphasize again that this comparison is not completely fair since the proposed transforms cover different intervals of $\rho$ while the DCT and DST are defined for $\rho$ tending merely to one and zero, respectively.
All metrics are
computed with respect to the exact KLT.
The values in bold are the
best
measurements
for each transform
length $N$.
The transforms $\widehat{\mathbf{T}}_{4,1}$, $\widehat{\mathbf{T}}_{8,2}$, and $\widehat{\mathbf{T}}_{16,3}$ are already known in the literature, and coincide with the SDCT. The remaining transforms are, to the best of our knowledge, new ones.

\begin{table*}[h!]
	\centering
	\caption{Comparison of proposed transforms with KLT}
	\label{t:tabela}
	\begin{tabular}{lcccccc}
		\toprule
		Transform & $\rho$ & ${\rm Cg}_T(\widehat{\mathbf{T}}_N)$ & $\eta_T(\widehat{\mathbf{T}}_N)$ &  ${\rm MSE}_T(\widehat{\mathbf{T}}_N)$   & $\epsilon_T(\widehat{\mathbf{T}}_N)$  \\ \midrule
		\multicolumn{6}{c}{\textbf{$N = 4$}}             \\ \midrule
		$\widehat{\mathbf{T}}_{4,1}$~\cite{haweel2001new} &$(0,1)$  & $\mathbf{0.162}$ & $\mathbf{10.862}$ & $\mathbf{0.039}$ & $\mathbf{0.764}$        \\
		DCT   &    $\rho \rightarrow 1$        & 0.025 & 7.071 &  0.014     &     0.167    \\
			DST   &   $\rho \rightarrow 0$         & 0.317 & 11.859 &   0.016    &   0.167  \\ \midrule
		\multicolumn{6}{c}{\textbf{$N = 8$}}             \\ \midrule
		$\widehat{\mathbf{T}}_{8,1}$ &$(0,0.619]$ & 2.726 & $\mathbf{32.907}$ & $\mathbf{0.110}$ & $\mathbf{3.670}$        \\
		$\widehat{\mathbf{T}}_{8,2}$~\cite{haweel2001new} &$(0.619,1)$         & $\mathbf{2.170}$ & 37.192 & 0.129 & 3.950  \\
		DCT   &    $\rho \rightarrow 1$        & 0.031 & 12.173 &   0.042    &    0.888   \\
		DST   &    $\rho \rightarrow 0$        & 0.362 & 18.940 &   0.032    &    0.778   \\ \midrule
		\multicolumn{6}{c}{\textbf{$N = 16$}}            \\ \midrule
		$\widehat{\mathbf{T}}_{16,1}$ & $(0.540,0.550]$  & 2.825 & 56.924 & 0.157 & $\mathbf{8.499}$        \\
		$\widehat{\mathbf{T}}_{16,2}$ & $(0.550,0.700]$  & 2.748 & 54.668 & $\mathbf{0.153}$ & 8.538        \\
		$\widehat{\mathbf{T}}_{16,3}$~\cite{haweel2001new} & $[0.9,1)$       & $\mathbf{2.227}$ & $\mathbf{48.588}$ & 0.169 & 9.532   \\
		DCT   &    $\rho \rightarrow 1$           & 0.024 & 15.709 &  0.078     &   2.945 \\
				DST   &   $\rho \rightarrow 0$         & 0.307 & 22.772 &   0.044    &    2.283   \\ \midrule
		\multicolumn{6}{c}{\textbf{$N = 32$}}            \\ \midrule
		$\widehat{\mathbf{T}}_{32,1}$ & $(0.139,0.162]$      & $\mathbf{2.352}$ & 66.355 & 0.195 & 21.143  \\
		$\widehat{\mathbf{T}}_{32,2}$ & $(0.487,0.490]$  & 2.442 & 65.047 &  $\mathbf{0.184}$ & 19.806       \\
		$\widehat{\mathbf{T}}_{32,3}$ & $(0.490,0.528]$       & 2.477 & 65.475 & 0.185 & $\mathbf{19.779}$  \\
		$\widehat{\mathbf{T}}_{32,4}$ & $(0.956,0.977]$      & 2.450 & $\mathbf{61.930}$ & 0.244 & 24.288    \\
		SDCT &  $\rho \rightarrow 1$   & 2.491 & 62.550 & 0.256 & 24.856 \\
		DCT   &   $\rho \rightarrow 1$             & 0.015 & 18.005 & 0.116    &    7.891 \\
					DST   &   $\rho \rightarrow 0$         & 0.227 & 24.673 &  0.052   &    5.531   \\ \bottomrule
	\end{tabular}
\end{table*}

\subsection{Fast Algorithms}\label{ss:fast}
The direct implementation of the proposed transforms requires $N(N-1)$ additions and no multiplications. Besides searching for multiplication-free transforms, the development of fast algorithms capable of reducing the arithmetic cost of computing the transforms is important. Using sparse matrix factorization~\cite{blahut2010fast}, such as butterfly-based structures, we can derive the factorization for the optimal proposed transforms.
In order to assess the complexity of the proposed fast algorithms we considered the number of arithmetic operations needed for its implementation.
The arithmetic complexity does not depend on the available architecture or technology, an issue that may occur when considering measures such as computation time~\cite{oppenheim1975digital, blahut2010fast,levitin2008introduction}.
The derived matrix factorization for $N=4$ and $8$ are presented in the following,
and for $N=16$ and $32$,  the respective matrix factorization are detailed in Appendix~\ref{A:Factorization}.

\subsubsection{Matrix Factorization for $N=4$}
For $N=4$, we can factorize $\widehat{\mathbf{T}}_{4,1}$\cite{haweel2001new} as follows:
\begin{equation*}
\widehat{\mathbf{T}}_{4,1} = \frac{1}{2} \cdot \mathbf{P}_4 \cdot\mathbf{A}_{4,2} \cdot \mathbf{A}_{4,1},
\end{equation*}
where
\begin{align*}
\mathbf{P}_4 =
\begin{bmatrix}
1 & & &\\
& & 1 & \\
& 1 & & \\
& & & 1
\end{bmatrix}, \quad
\mathbf{A}_{4,2} =
\begin{bmatrix}
1 & 1 & &\\
1 & - &  & \\
&  & 1 & 1 \\
& &  - & 1
\end{bmatrix},
\end{align*}
and
\begin{align*}
\mathbf{A}_{4,1} =
\begin{bmatrix}
\mathbf{I}_2 & \bar{\mathbf{I}}_2 \\
\bar{\mathbf{I}}_2 \ & -\mathbf{I}_2
\end{bmatrix},
\end{align*}
where $\mathbf{I}_2$ and $\bar{\mathbf{I}}_2$ are, respectively, the identity  and counter-identity matrices of order $2$.
The minus sign $-$ represents $-1$ and blank spaces are zeroes.\\
\subsubsection{Matrix Factorization for $N=8$} \label{A:n8}
For $N=8$, we have:

\begin{align*}
&\widehat{\mathbf{T}}_{8,1} = \frac{1}{\sqrt{8}} \cdot \mathbf{P}_8 \cdot\mathbf{A}_{8,3}^\prime\cdot\mathbf{A}_{8,2} \cdot \mathbf{A}_{8,1}, \\
&\widehat{\mathbf{T}}_{8,2} = \frac{1}{\sqrt{8}} \cdot \mathbf{P}_8 \cdot\mathbf{A}_{8,3}^{\prime\prime}\cdot\mathbf{A}_{8,2} \cdot \mathbf{A}_{8,1}\text{\cite{haweel2001new}},
\end{align*}

where
\begin{align*}
&\mathbf{P}_8 =
\begin{bmatrix}
1 & & & & & & &\\
& & & & 1 & & & \\
& & 1 & & & & & \\
& & & & & 1 & & \\
& 1 & & & & & & \\
& & & & & & 1 & \\
& & & 1 & & & & \\
& & & & & & & 1
\end{bmatrix}
,
\end{align*}
\begin{align*}
\mathbf{A}^{\prime}_{8,3} =
\begin{bmatrix}
1 & 1 & & & & & &\\
1 & -& &  & & & & \\
& & 1 & 1 & & & & \\
& & - & 1 & &  & & \\
& & & & 1 & 1 & & \\
& & & &  & & -  & - \\
& & & & 1 & & 1 & \\
& & & & & & 1 & -
\end{bmatrix},
\end{align*}
\begin{align*}
&\mathbf{A}_{8,2} =
\begin{bmatrix}
\mathbf{A}_{4,1} & \\
& \mathbf{A}_{4,1}
\end{bmatrix}, \quad
\mathbf{A}_{8,1} =
\begin{bmatrix}
\mathbf{I}_4 & \bar{\mathbf{I}}_4 \\
\bar{\mathbf{I}}_4 \ & -\mathbf{I}_4
\end{bmatrix},
\end{align*}
and
\begin{align*}
	\mathbf{A}^{\prime\prime}_{8,3} =
\begin{bmatrix}
1 & 1 & & & & & &\\
1 & -& &  & & & & \\
& & 1 & 1 & & & & \\
& & 1 &-& &  & & \\
& & & & 1 & 1 & & \\
& & & & &- & & - \\
& & & & 1 & & 1 &  \\
& & & & & & 1 & -
\end{bmatrix}.
\end{align*}

Table~\ref{t:fastalg} presents the arithmetic cost of the proposed fast algorithms for the approximate transforms compared with the arithmetic cost of the direct implementation of the exact $N$-point KLT.
Since the proposed transforms are, to the best of our knowledge, the first class of approximations for the KLT following this approach, there is nothing to compare with. However, to show that the performance of the proposed transforms is competitive, we introduced in Table~\ref{t:fastalgapprox} the arithmetic cost of some transforms in general that are already known in the literature.
Thus,
Table~\ref{t:fastalgapprox} presents the arithmetical cost of: the fast algorithms proposed by \cite{loeffler1989practical} and \cite{chen1977fast} for the $8$- and $16$-point  DCT, respectevely; and approximations for the DCT already known in the literature proposed in \cite{haweel2001new, lengwehasatit2004scalable,cintra2011dct, bouguezel2012binary,bayer2012digital,Jridi2015, bayer2012dct,bouguezel2008low,bouguezel2009fast,bouguezel2011low, cintra2011integer,da2016orthogonal,da2017multiplierless}.
We can note that the introduced fast algorithms are multiplierless and offer substantial reductions in the additive complexity when compared with the exact KLT and lower or similar arithmetic cost compared with the DCT approximations.

\begin{table*}[]
	\caption{Computational cost comparison for 4-, 8-, 16-, and 32-point transforms}
	\label{t:fastalg}
	\centering
	\begin{tabular}{lccccc}
		\toprule
		Transform           & Additions & Multiplications & Bit-Shifting
	& Addition reduction $(\%)$ \\
		\midrule
		$\widehat{\mathbf{T}}_{4,1}$~\cite{haweel2001new}   & $4$       & $0$  & $0$
		& $66\%$                 \\
				$\mathbf{K}_4$      & $12$      & $16$        & $0$    & - \\
		\addlinespace[1.5ex]
		\midrule
		\addlinespace[1.5ex]
			$\widehat{\mathbf{T}}_{8,1}$  & $24$      & $0$     & $0$
		& $57\%$                                    \\
		$\widehat{\mathbf{T}}_{8,2}$~\cite{haweel2001new}   & $24$      & $0$         & $0$
	& $57\%$                                   \\
			$\mathbf{K}_8$      & $56$      & $64$    & $0$
			& -                                         \\
		\addlinespace[1.5ex]
		\midrule
		\addlinespace[1.5ex]
				$\widehat{\mathbf{T}}_{16,1}$ & $80$      & $0$    &$0$
			& $67\%$                                    \\
		$\widehat{\mathbf{T}}_{16,2}$ & $75$      & $0$            &$0$
		& $69\%$                                    \\
		$\widehat{\mathbf{T}}_{16,3}$~\cite{haweel2001new}  & $72$      & $0$ &$0$
		 & $70\%$                                    \\
		$\mathbf{K}_{16}$   & $240$     & $256$         &$0$
	& -                                        \\
		\addlinespace[1.5ex]
		\midrule
		\addlinespace[1.5ex]
		$\widehat{\mathbf{T}}_{32,1}$ & $288$     & $0$             & $0$
	&	$71\%$                                   \\
		$\widehat{\mathbf{T}}_{32,2}$ & $288$     & $0$             & $0$
	 &$71\%$                                   \\
		$\widehat{\mathbf{T}}_{32,3}$ & $288$     & $0$             & $0$
	 &$71\%$                                    \\
		$\widehat{\mathbf{T}}_{32,4}$ & $232$     & $0$             &  $0$
	&$77\%$                                    \\
				$\mathbf{K}_{32}$   & $992$     & $1024$          &  $0$                & -                       \\
		\bottomrule
	\end{tabular}
\end{table*}

\begin{table}[]
	\caption{Computational cost comparison for 8-, 16-, and 32-point DCT approximation transforms}
	\label{t:fastalgapprox}
	\centering
	\begin{tabular}{lcccc}
		\toprule
		Transform           & Additions & Multiplications & Bit-Shifting \\
		\midrule                          %
		$\text{DCT}_{8}$~\cite{loeffler1989practical}  & $29$ & $11$  & $0$  \\
				$\text{DST}_{8}$~\cite{106875}  & $29$ & $12$  & $0$  \\
		$\mathbf{T}_{8,\text{LO}}$~\cite{lengwehasatit2004scalable} & $24$ & $0$ & $2$ \\
		$\mathbf{T}_{8,\text{RDCT}}$~\cite{cintra2011dct} & $22$ & $0$ & $0$ \\
		$\mathbf{T}_{8,\text{MRDCT}}$~\cite{bayer2012dct} &$14$& $0$ & $0$ \\
		$\mathbf{T}_{8,\text{BAS-2008a}}$~\cite{bouguezel2008low} &$18$& $0$ & $2$ \\
		$\mathbf{T}_{8,\text{BAS-2008b}}$~\cite{bouguezel2008low} &$21$& $0$ & $2$ \\
		$\mathbf{T}_{8,\text{BAS-2009}}$~\cite{bouguezel2009fast} &$18$& $0$ & $0$ \\
		$\mathbf{T}_{8,\text{BAS-2011}}$~\cite{bouguezel2011low} &$16$& $0$ & $0$ \\
		$\mathbf{T}_{8,\text{BAS-2012}}$~\cite{bouguezel2012binary} &$24$& $0$ & $0$ \\
		$\mathbf{T}_{8,1}^\prime$~\cite{cintra2011integer}   & $18$      & $0$         & $0$      \\
		$\mathbf{T}_{8,4}$~\cite{cintra2011integer}   & $24$      & $0$         & $0$      \\
		$\mathbf{T}_{8,5}$~\cite{cintra2011integer}   & $24$      & $0$         & $4$      \\
		$\mathbf{T}_{8,6}$~\cite{cintra2011integer}   & $24$      & $0$         & $6$      \\
		\addlinespace[1.5ex]
		\midrule
		\addlinespace[1.5ex]
		$\text{DCT}_{16}$~\cite{chen1977fast} & $74$ & $44$ & $0$ \\
				$\text{DST}_{16}$~\cite{106875} & $81$ & $32$ & $0$ \\
		$\mathbf{T}_{16,\text{BAS-2012}}$~\cite{bouguezel2012binary} & $64$ &$0$& $0$  \\
		$\mathbf{T}_{16,\text{BCEM}}$~\cite{bayer2012digital} & $72$ & $0$ &  $0$ \\
		$\mathbf{T}_{16,\text{SBCKMK}}$~\cite{da2016orthogonal} & $60$ & $0$ &  $0$ \\
		$\mathbf{T}_{16,\text{SOBCM}}$~\cite{da2017multiplierless} & $44$ & $0$ &  $0$ \\
		$\mathbf{T}_{16,\text{JAM}}$~\cite{Jridi2015} &$60$& $0$ &$0$ \\
		\addlinespace[1.5ex]
		\midrule
		\addlinespace[1.5ex]
			$\text{DCT}_{32}$~\cite{chen1977fast} & $194$ & $116$ & $0$ \\
				$\text{DST}_{32}$~\cite{106875} & $209$ & $80$ & $0$ \\
		$\mathbf{T}_{32,\text{BAS-2012}}$~\cite{bouguezel2012binary}  & $160$ & $0$ & $0$ \\
		$\mathbf{T}_{32,\text{JAM}}$~\cite{Jridi2015}  & $152$ & $0$ & $0$ \\
		\bottomrule
	\end{tabular}
\end{table}

\begin{figure*}[h]
	\center
	\includegraphics[width=10cm]{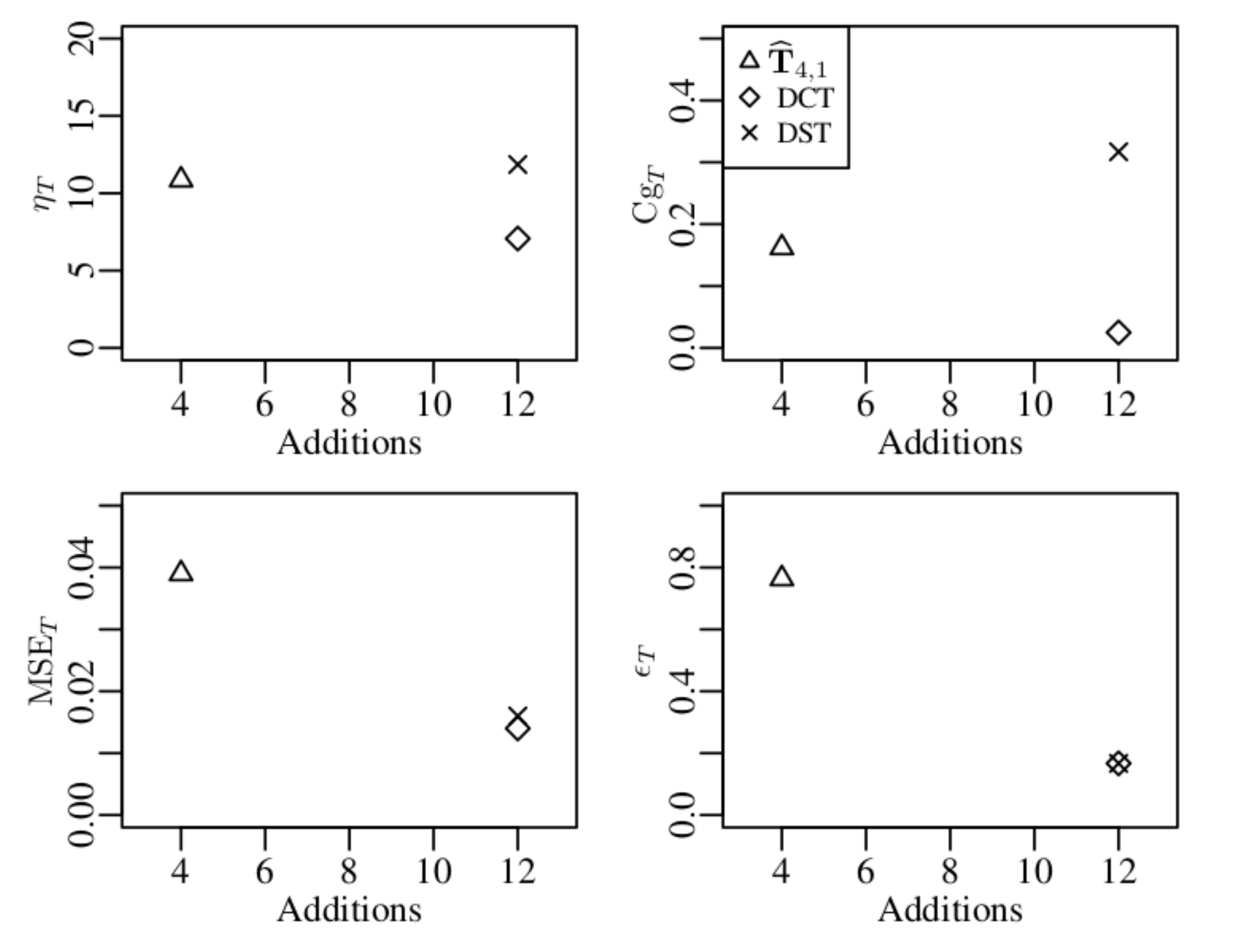}
	\caption{Arithmetic complexity versus the proposed figures of merit for each transform for $N = 4$.}\label{f:compl4}
\end{figure*}

\begin{figure*}[h]
	\center
 \includegraphics[width=10cm]{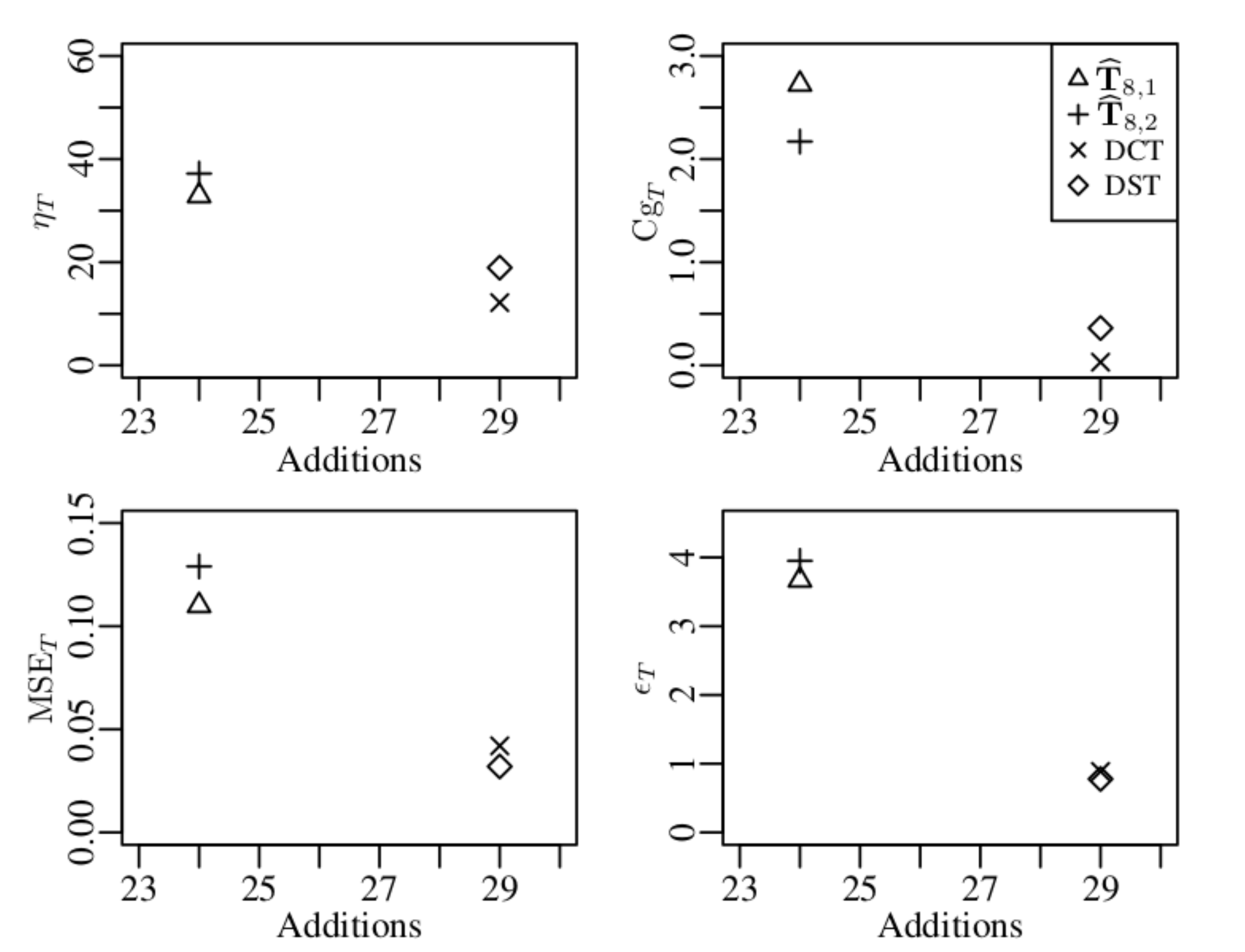}
		\caption{Arithmetic complexity versus the proposed figures of merit for each transform for $N = 8$.}\label{f:compl8}
\end{figure*}

\begin{figure*}[h]
	\center
	\includegraphics[width=10cm]{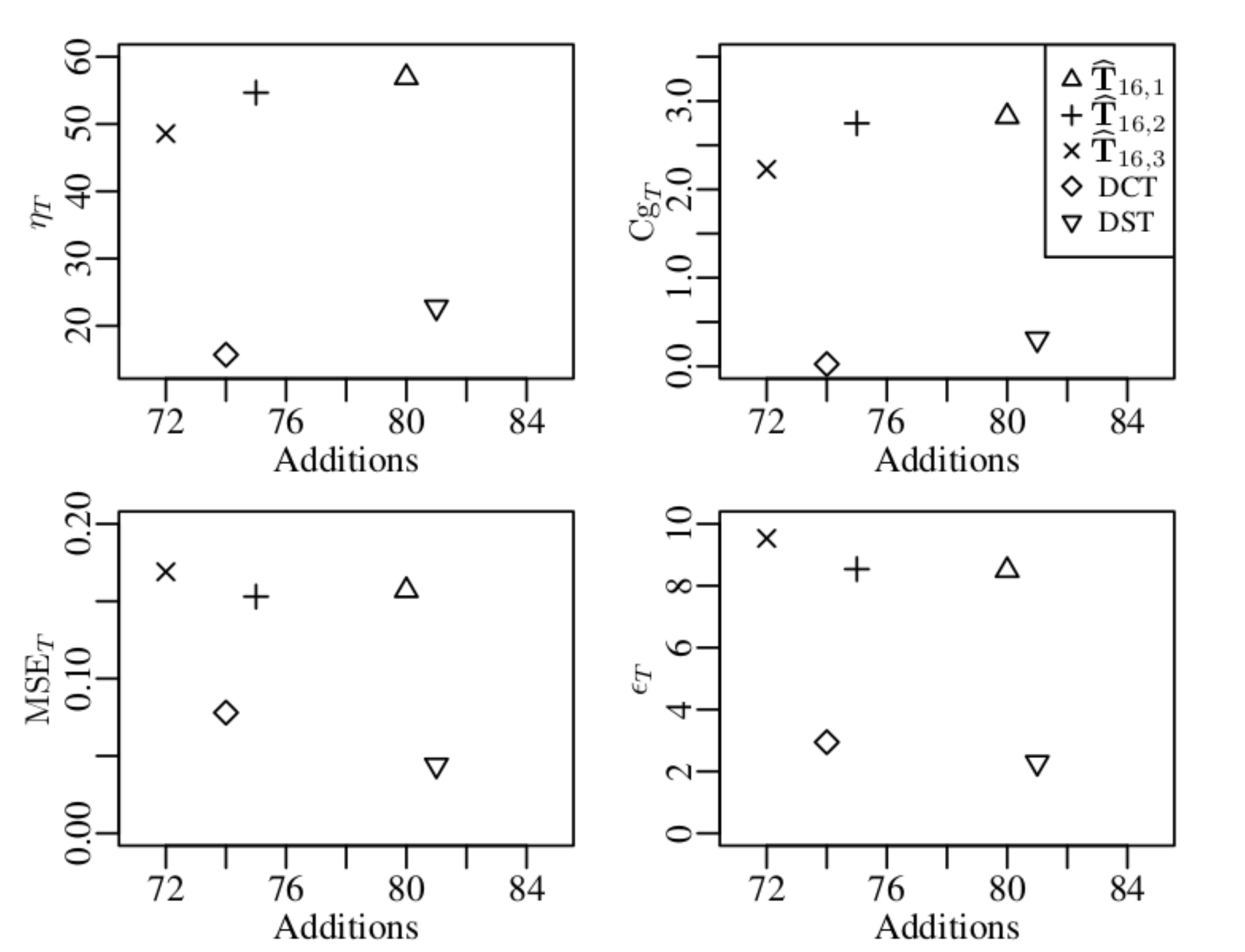}
	\caption{Arithmetic complexity versus the proposed figures of merit for each transform for $N = 16$.}\label{f:compl16}
\end{figure*}

\begin{figure*}[h]
	\center
	\includegraphics[width=10cm]{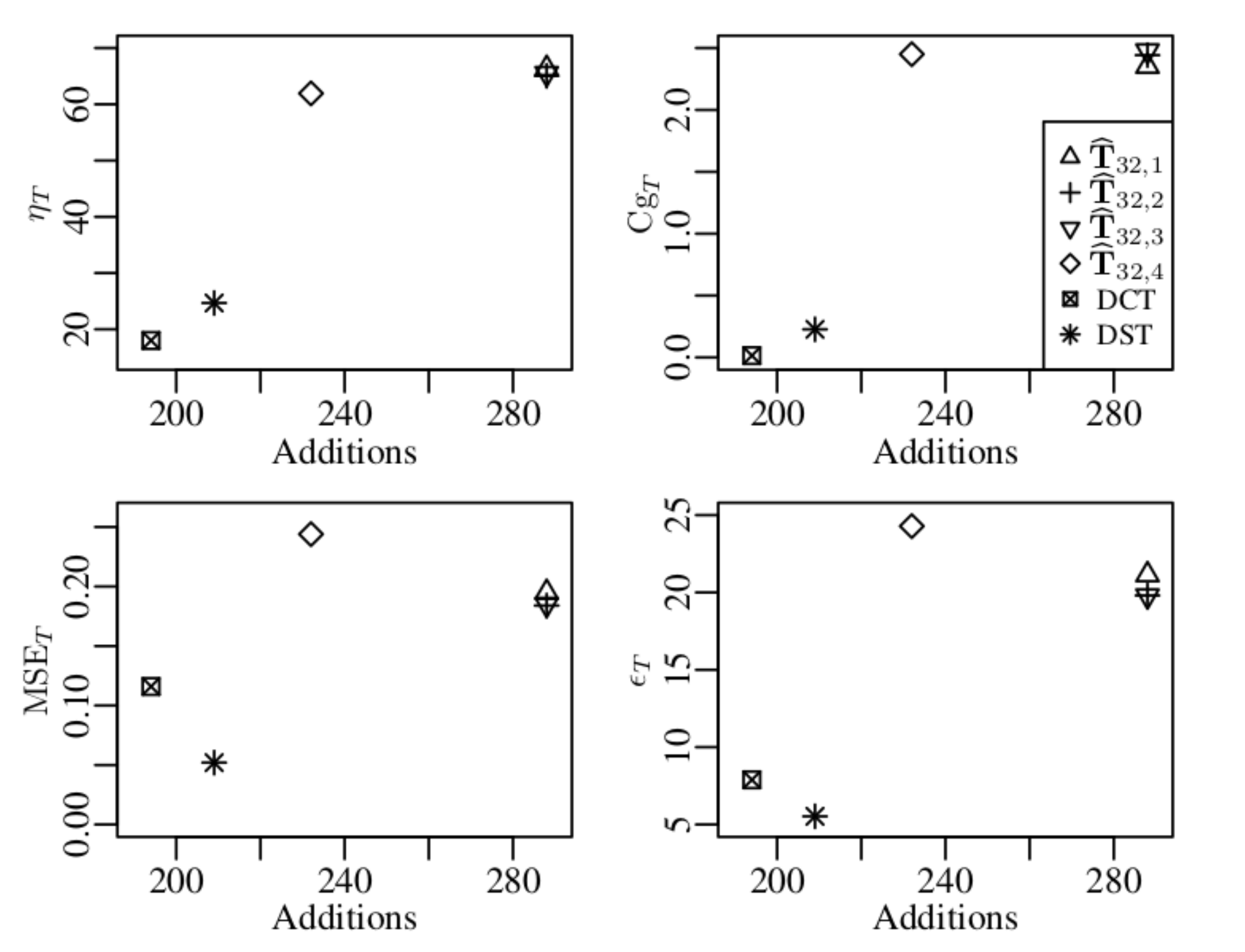}
	\caption{Arithmetic complexity versus the proposed figures of merit for each transform for $N = 32$.}\label{f:compl32}
\end{figure*}

For a better visualization of the results presented in Tables~\ref{t:tabela} and \ref{t:fastalg}, we have combined graphically the number of additions
and the proposed figures of merit of each proposed transform, for $N = 4$, $8$, $16$, and $32$, as presented in Figures~\ref{f:compl4}, \ref{f:compl8}, \ref{f:compl16}, \ref{f:compl32}, respectively.
Note that both DCT and DST also require multiplications when implemented,
differently of the proposed transforms that are multiplierless.

\section{Image and Video Coding} \label{S:applic}
We submitted the proposed transforms to two different contexts that are
classical in the
approximation community area: (i)~still image compression according to a JPEG-like algorithm~\cite{cintra2011dct,bouguezel2008low,cintra2014low}, and (ii)~video encoding as defined in the
HEVC
reference software~\cite{refsoft}.
In this section, we compared the proposed transforms with the exact KLT for $\rho = 0.1$, $0.6$ and $0.9$ ($\mathbf{K}^{(0.1)}$, $\mathbf{K}^{(0.6)}$ and $\mathbf{K}^{(0.9)}$), and with the exact DCT.
The transforms $\mathbf{K}^{(0.1)}$,  $\mathbf{K}^{(0.6)}$, and $\mathbf{K}^{(0.9)}$ were selected because they are suitable
for decorrelation of lowly, moderately, and highly correlated data,
respectively.
Currently literature is restricted to the high correlation scenario,
which is mainly addressed by the DCT and its related approximations.

\subsection{Image Compression}\label{S:compressao}
In this section, we evaluated the performance of the proposed transforms
in
image compression,
similarly to~\cite{cintra2011dct,bouguezel2008low,
	cintra2014low}.
If $\mathbf{A}$ is a two-dimensional (2D) image, then the direct and inverse transformations induced by the SKLT are computed, respectively,
by
\begin{equation}\label{eq:B}
	\mathbf{B} = \widehat{\mathbf{T}}_N^{(\rho)} \cdot \mathbf{A} \cdot ({\widehat{\mathbf{T}}_N^{(\rho)}})^{-1},
\end{equation}
\begin{equation}\label{eq:A}
	\mathbf{A} = ({\widehat{\mathbf{T}}_N^{(\rho)}})^{-1} \cdot \mathbf{B} \cdot  \widehat{\mathbf{T}}_N^{(\rho)}.
\end{equation}
The adopted compression scheme is described as follows~\cite{salomon2004data}:
(i)~the image is divided into disjoint sub-blocks $\mathbf{A}_j$ of size $N \times N$;
(ii)~each sub-block is submitted to a selected transform $\widehat{\mathbf{T}}_N$
according to \eqref{eq:B};
(iii)~using the standard zig-zag sequence~\cite{salomon2004data}, only the initial $r$  coefficients in each sub-block $\mathbf{B}_j$ are retained and the remaining ${N^2-r}$ coefficients are zeroed, resulting in sub-block $\mathbf{\bar{B}}_j$;
(iv)~the two-dimensional inverse transform is applied
according \eqref{eq:A}, and
(v)~the reconstructed sub-blocks $\mathbf{\bar{A}}_j$ are
adequately rearranged.
The
final
reconstructed
image $\mathbf{\bar{A}}$
is
compared with the original image $\mathbf{A}$
for assessing the
performance of $\widehat{\mathbf{T}}_N$.
We adopted $N\in \{4,8,16,32\}$ and used
the peak signal-to-noise ratio (PSNR)~\cite{Huynh-Thu2008Scope},
and the mean structural similarity index (MSSIM)~\cite{wang2004image} as figures of merit for image quality evaluation.
The results were taken individually for
$45$ $512\times512$ $8$-bit greyscale images obtained from \cite{sipi2005usc} and averaged.
For each transform length,
we considered two approaches:  (i)~a qualitative analysis, based on the compressed~\textit{Lena} image with approximately $85 \%$ compression and (ii)~a quantitative one, varying the value of $r$ for compression, considering the average MSSIM and PSNR values of the compressed images.

\begin{figure*}
	\centering
	\includegraphics[width=5cm]{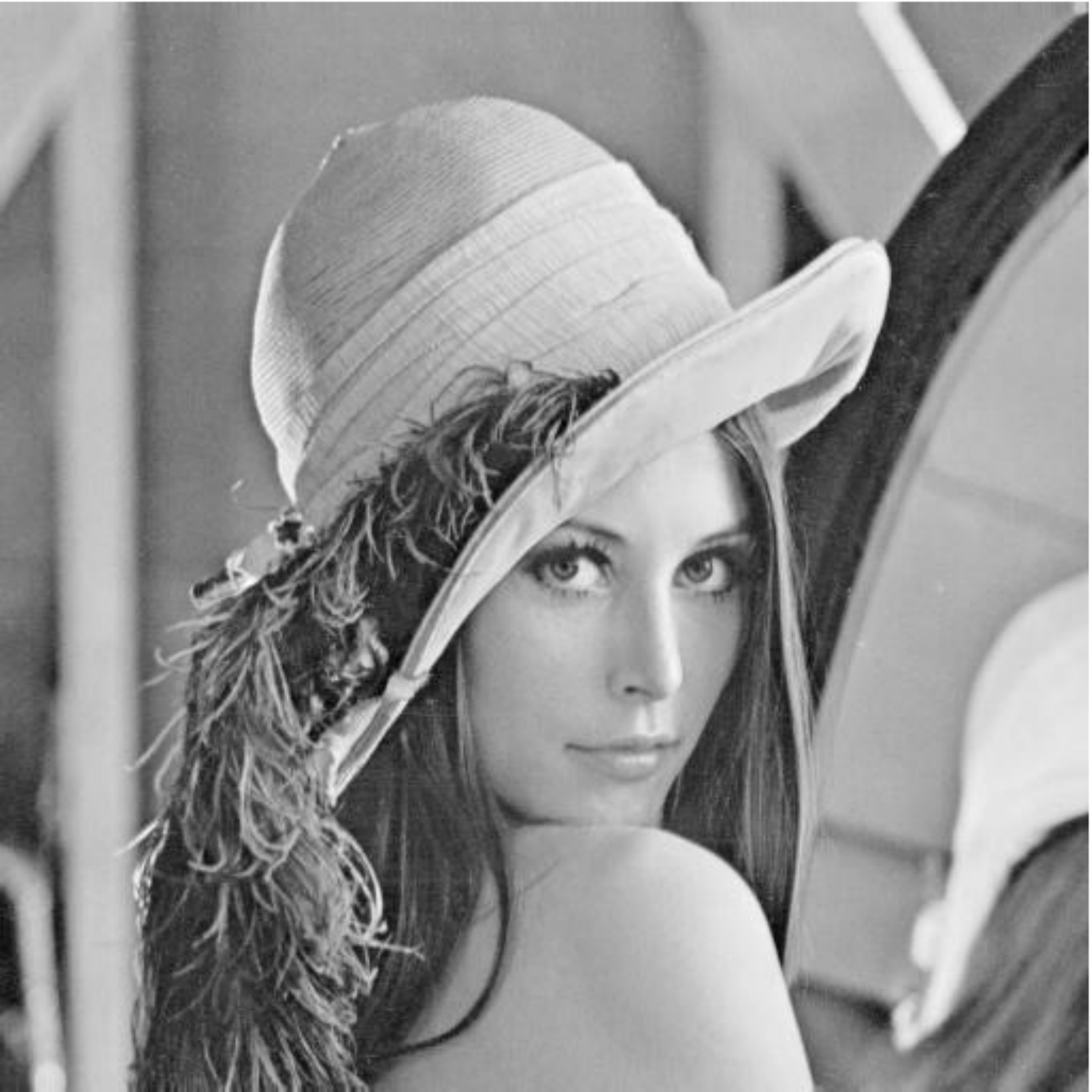}
	\caption{Original \textit{Lena} image.} \label{f:lena-original}
\end{figure*}

\begin{figure*}[h]
	\center
	\subfigure[$\widehat{\mathbf{T}}_{4,1}$~\cite{haweel2001new} , PSNR = $30.758$,  MSSIM = $0.846$]{\label{f:lenaT41}\includegraphics[width=4cm]
		{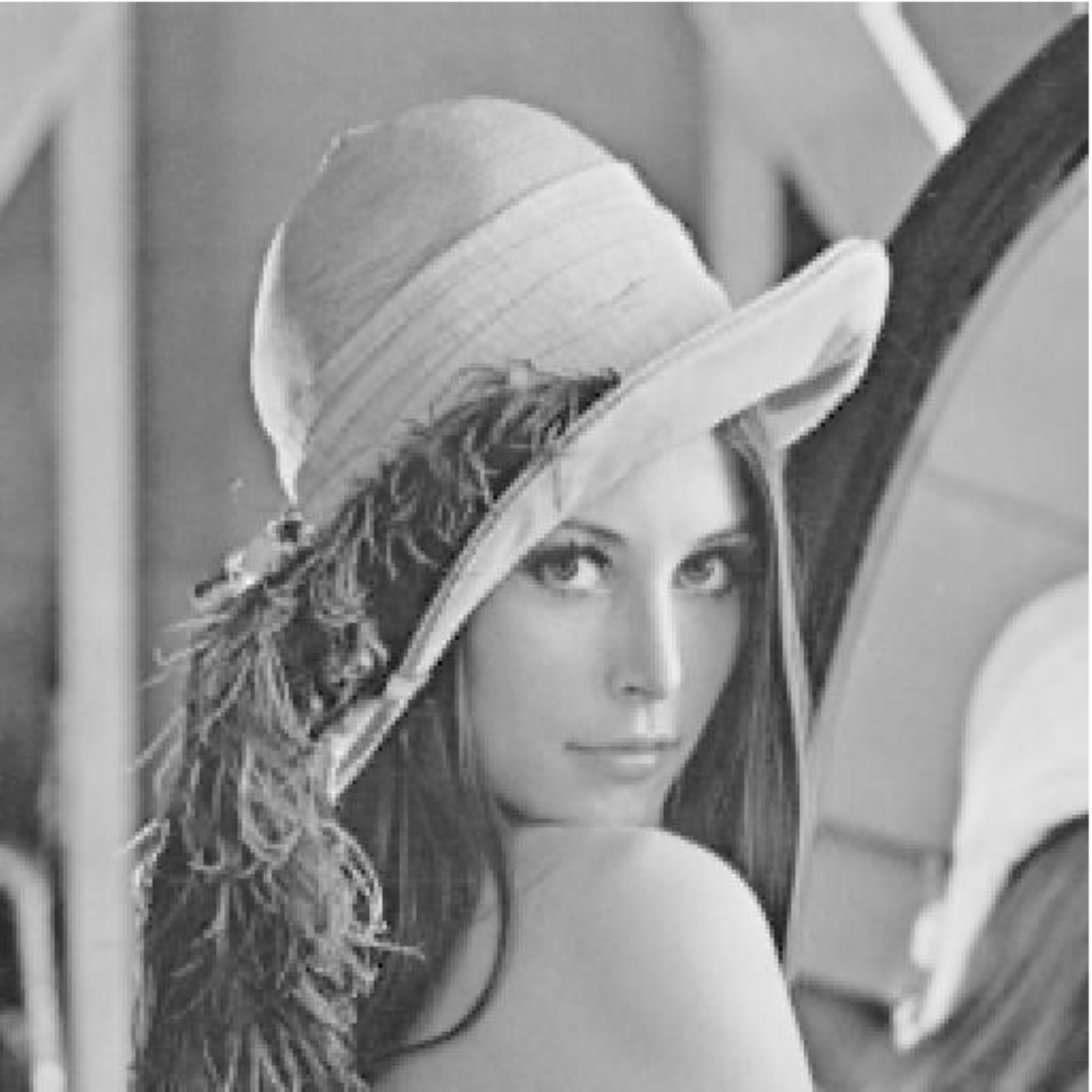}}
		\subfigure[$\text{DCT}_{4} (\rho \rightarrow 1)$, PSNR = $32.001$,  MSSIM = $0.871$]{\label{f:lenadct4}\includegraphics[width=4cm]
		{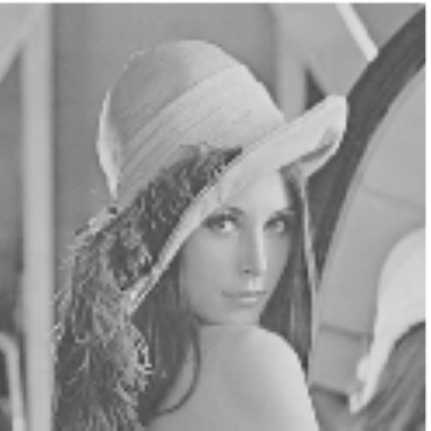}}
		\subfigure[$\text{DST}_{4} (\rho \rightarrow 0)$ , PSNR = $15.491$,  MSSIM = $0.127$]{\label{f:lenadst4}\includegraphics[width=4cm]
		{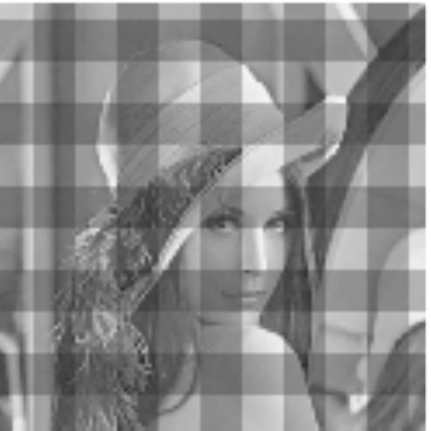}}
	\caption{Compressed \textit{Lena} images for $N = 4$ and $r = 3$.}\label{f:Lenas4}
\end{figure*}

\begin{figure*}[h]
	\center
	\subfigure[$\widehat{\mathbf{T}}_{8,1}$, PSNR = $27.876$, MSSIM = $0.854$]{\label{f:lenaT81}\includegraphics[width=4cm]
				{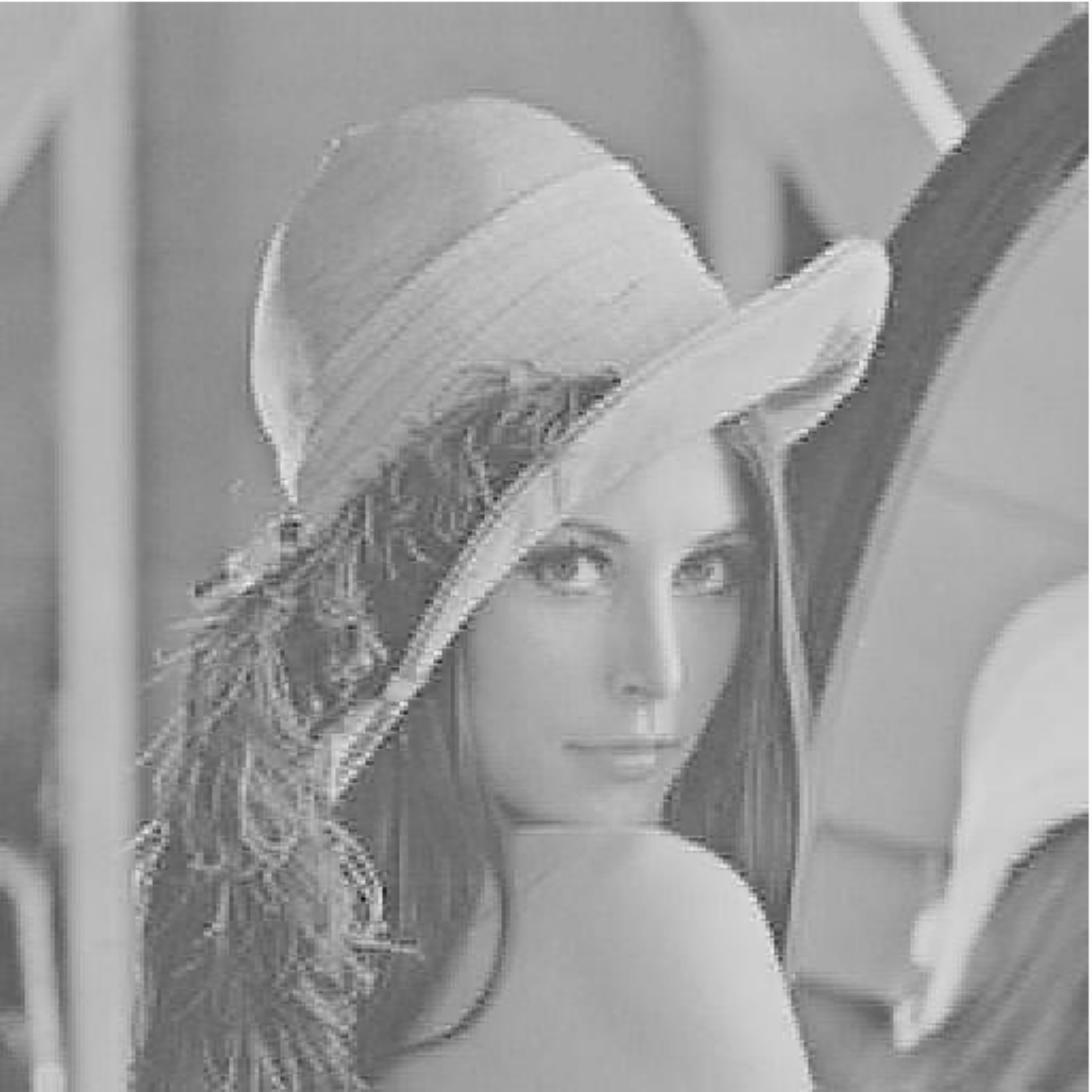}}
			\subfigure[$\widehat{\mathbf{T}}_{8,2}$~\cite{haweel2001new} , PSNR = $27.896$, MSSIM = $0.861$]
			{\label{f:lenaT82}\includegraphics[width=4cm]
				{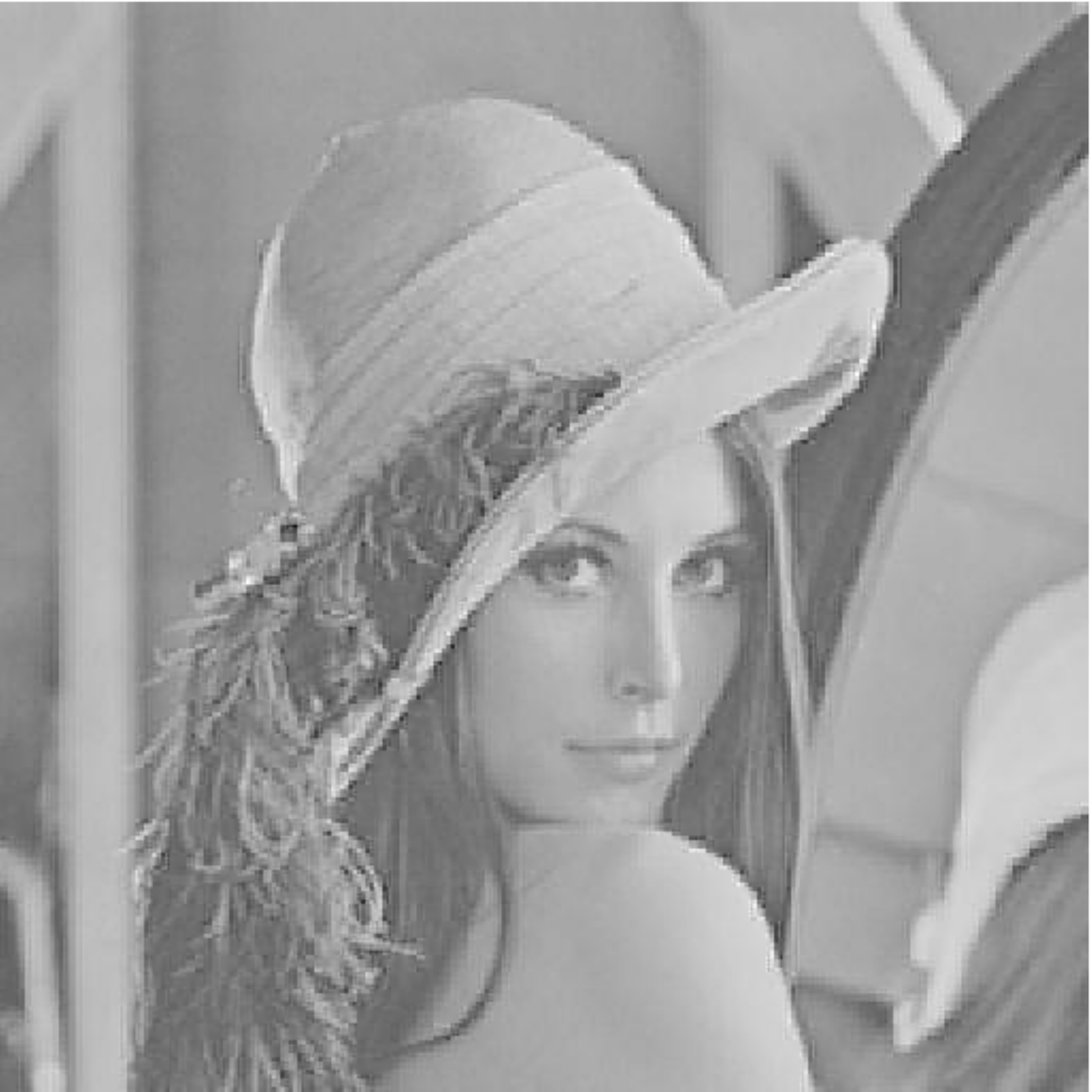}}
	\subfigure[$\text{DCT}_{8} (\rho \rightarrow 1)$, PSNR = $32.081$,  MSSIM = $0.913$]{\label{f:lenadct8}\includegraphics[width=4cm]
		{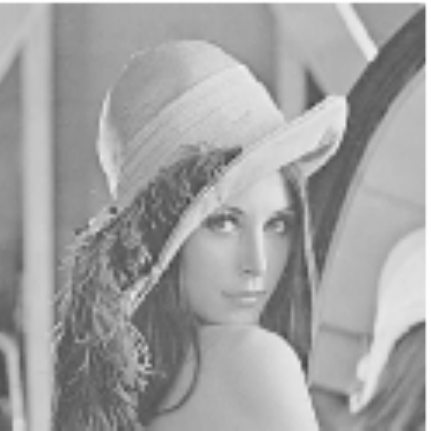}}
	\subfigure[$\text{DST}_{8} (\rho \rightarrow 0)$ , PSNR = $18.297$,  MSSIM = $ 0.272$]{\label{f:lenadst8}\includegraphics[width=4cm]
		{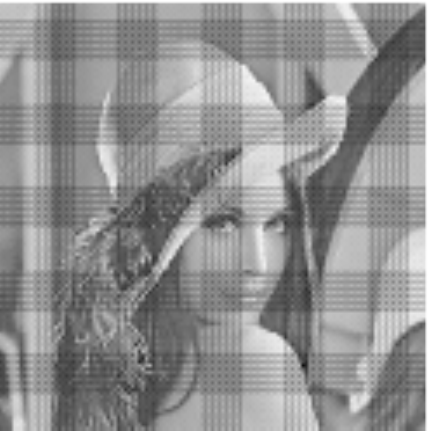}}
	\caption{Compressed \textit{Lena} images for $N = 8$ and $r = 10$.}\label{f:Lenas8}
\end{figure*}

\begin{figure*}[h]
	\center
	\subfigure[$\widehat{\mathbf{T}}_{16,1}$, PSNR = $25.717$, MSSIM = $0.862$]{\label{f:lenaT161}\includegraphics[width=4cm]
				{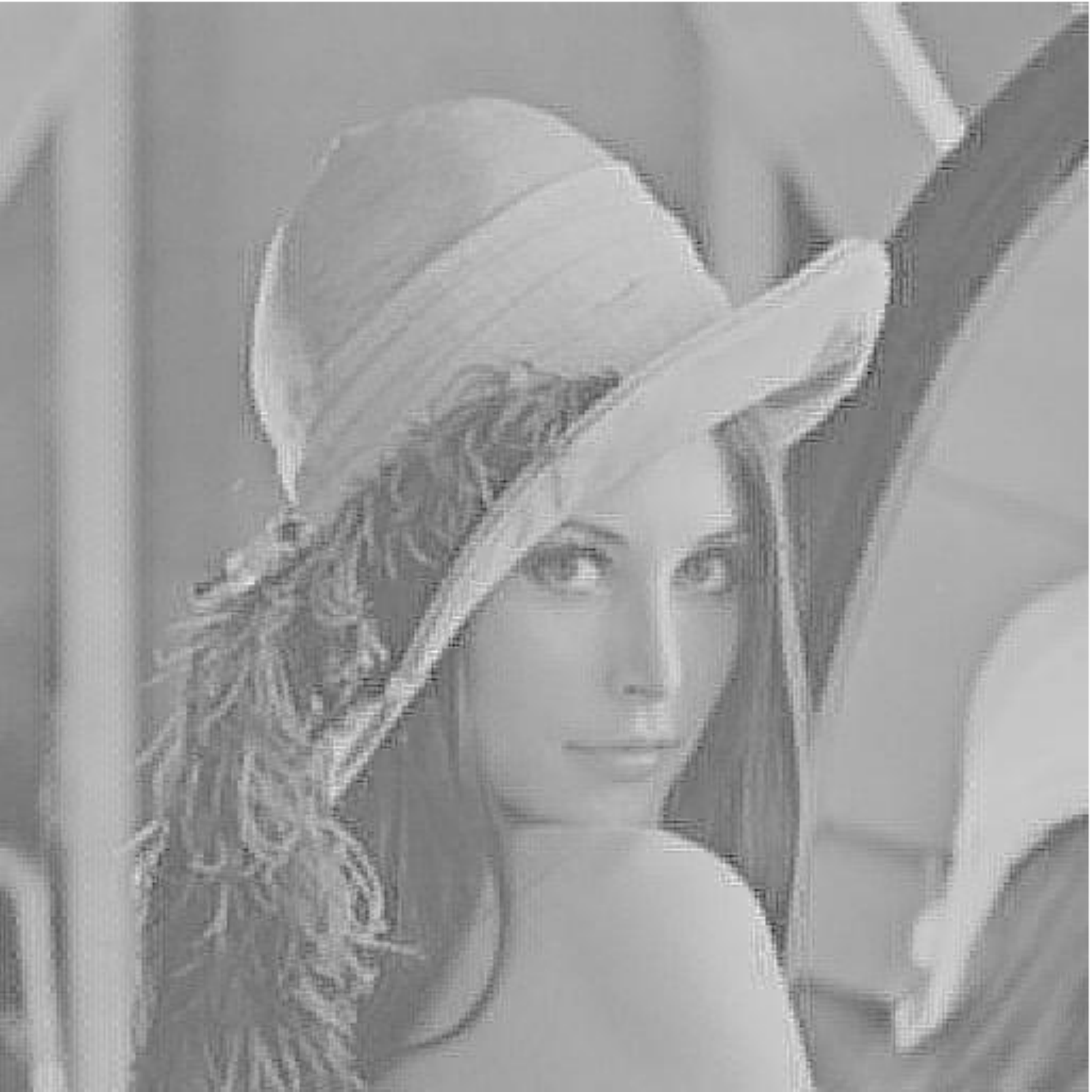}}
			\subfigure[$\widehat{\mathbf{T}}_{16,2}$, PSNR = $25.887$, MSSIM = $0.866$]
			{\label{f:lenaT162}\includegraphics[width=4cm]
				{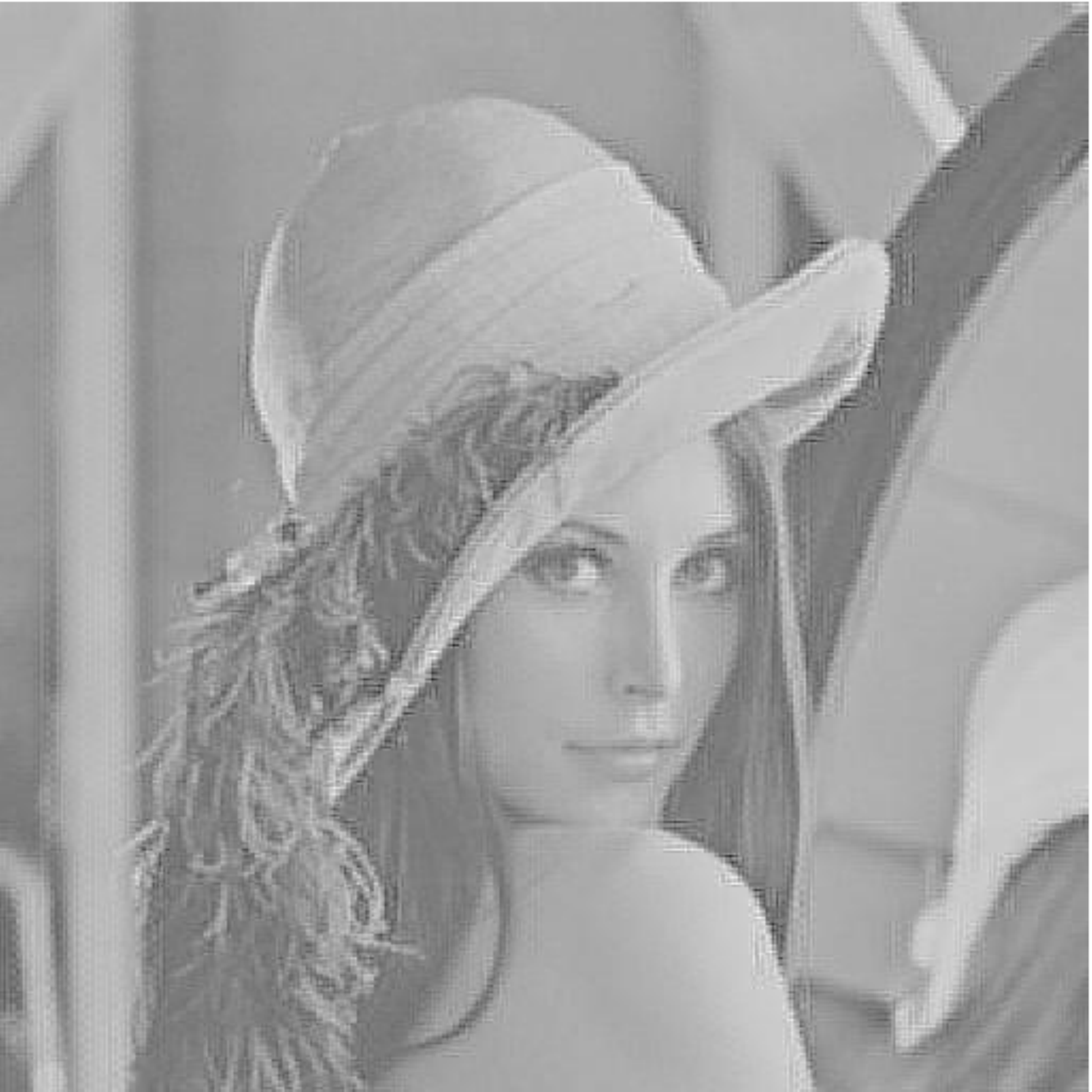}}
			\subfigure[$\widehat{\mathbf{T}}_{16,3}$~\cite{haweel2001new} , PSNR = $27.200$, MSSIM = $0.892$]
			{\label{f:lenaT163}\includegraphics[width=4cm]
				{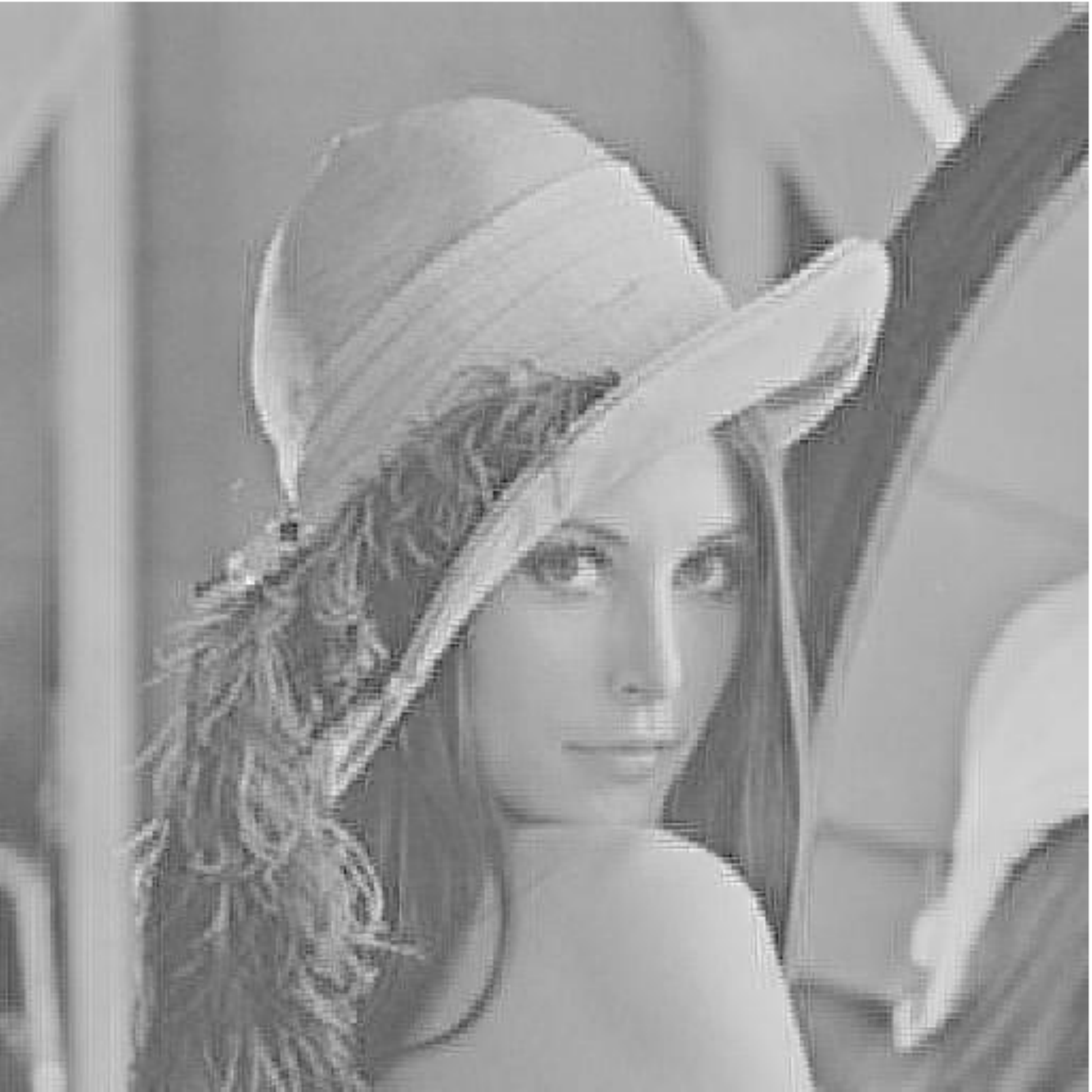}}
	\subfigure[$\text{DCT}_{16} (\rho \rightarrow 1)$, PSNR = $32.494$,  MSSIM = $0.945$]{\label{f:lenadct16}\includegraphics[width=4cm]
		{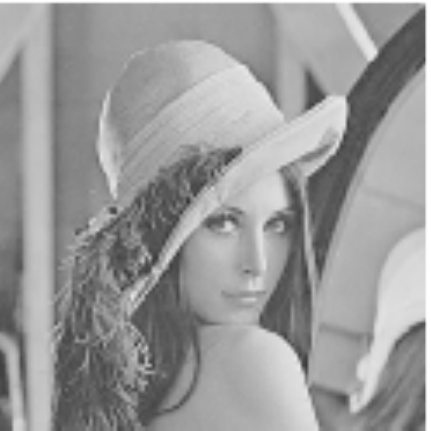}}
	\subfigure[$\text{DST}_{16} (\rho \rightarrow 0)$ , PSNR = $22.639$,  MSSIM = $ 0.532$]{\label{f:lenadst16}\includegraphics[width=4cm]
		{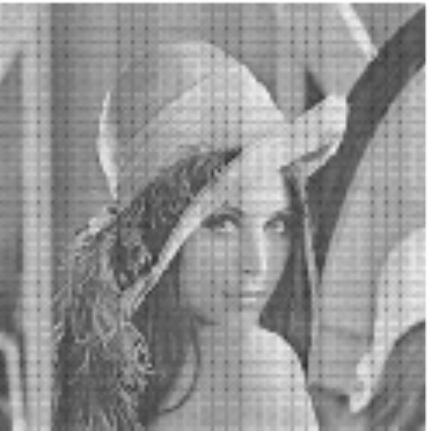}}
	\caption{Compressed \textit{Lena} images for $N = 16$ and $r = 40$.}\label{f:Lenas16}
\end{figure*}

\begin{figure*}[h]
	\center
\subfigure[$\widehat{\mathbf{T}}_{32,1}$, PSNR = $17.519$, MSSIM = $0.473$]{\label{f:lenaT321}\includegraphics[width=4cm]
			{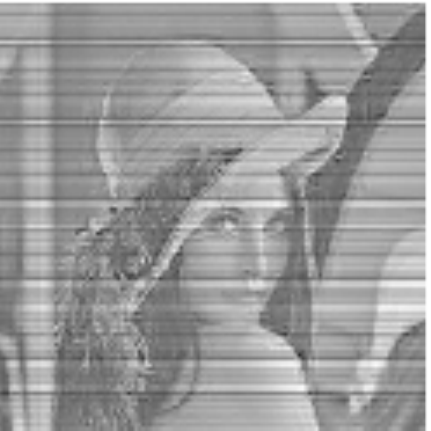}}
		\subfigure[$\widehat{\mathbf{T}}_{32,2}$, PSNR = $17.573$, MSSIM = $0.476$]
		{\label{f:lenaT322}\includegraphics[width=4cm]
			{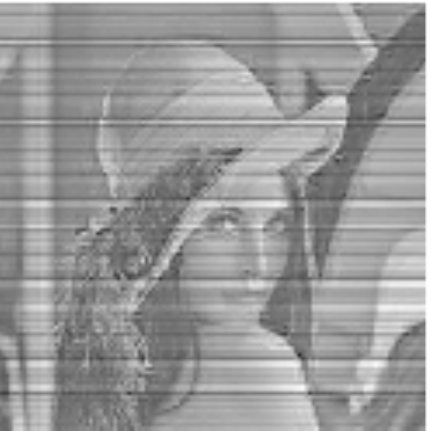}}
		\subfigure[$\widehat{\mathbf{T}}_{32,3}$, PSNR = $20.632$, MSSIM = $0.600$]
		{\label{f:lenaT323}\includegraphics[width=4cm]
			{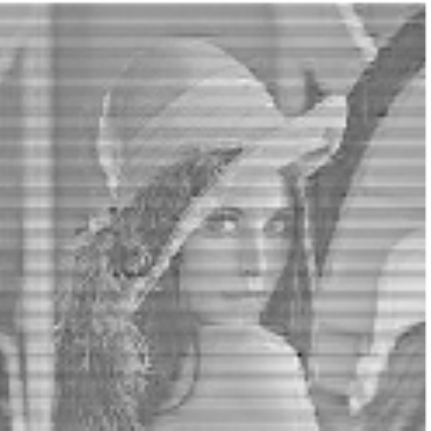}}
		\subfigure[$\widehat{\mathbf{T}}_{32,4}$, PSNR = $25.997$, MSSIM = $0.911$]
		{\label{f:lenaT324}\includegraphics[width=4cm]
			{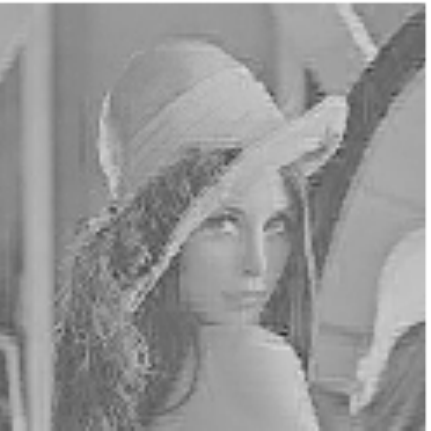}}
	\subfigure[$\text{DCT}_{32} (\rho \rightarrow 1)$, PSNR = $32.996$,  MSSIM = $0.969$]{\label{f:lenadct32}\includegraphics[width=4cm]
		{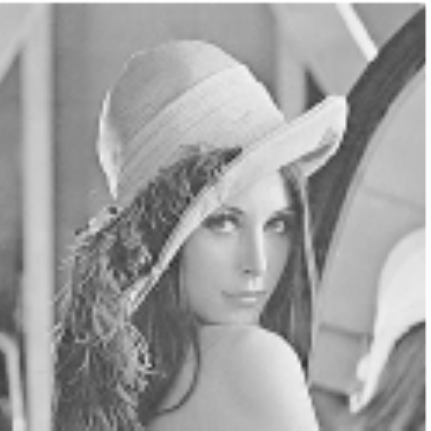}}
	\subfigure[$\text{DST}_{32} (\rho \rightarrow 0)$ , PSNR = $26.089$,  MSSIM = $ 0.781$]{\label{f:lenadst32}\includegraphics[width=4cm]
		{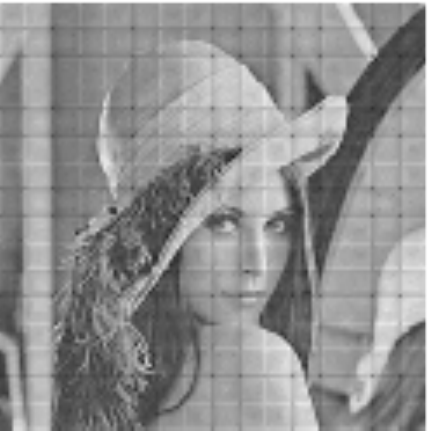}}
	\caption{Compressed \textit{Lena} images for $N = 32$ and $r = 155$.}\label{f:Lenas32}
\end{figure*}

Figure~\ref{f:lena-original}
shows the original \textit{Lena} image.
The reconstructed images using the proposed transforms, DCT, and DST for $N=4$, $8$,
$16$, and $32$ are presented, respectively, in Figures~\ref{f:Lenas4}, \ref{f:Lenas8},
\ref{f:Lenas16}, and \ref{f:Lenas32}.
 The corresponding compression ratio (CR) was CR $=81.25\%$, $84.38\%$, $84.38\%$, $84.86\%$, respectively.
Visually, the reconstructed images after compression exhibit quality comparable with the original image.

\begin{figure*}[h]
	\centering
		\subfigure[Average PSNR ($N=4$)]{\label{f:PSNR4}{\includegraphics[width=7cm]{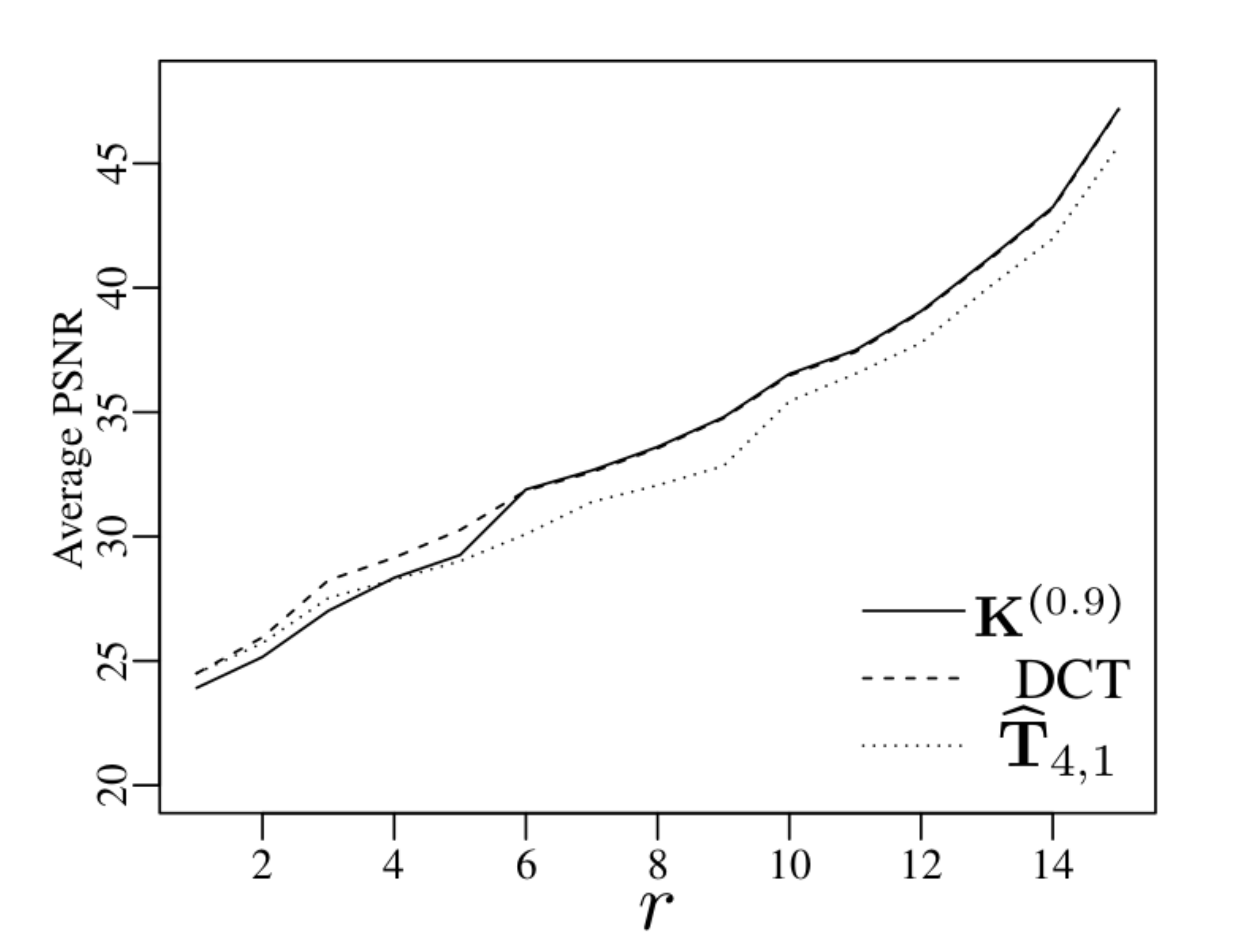}}}
	\subfigure[Average MSSIM ($N=4$)]{\label{f:MSSIM4}\includegraphics[width=7cm]{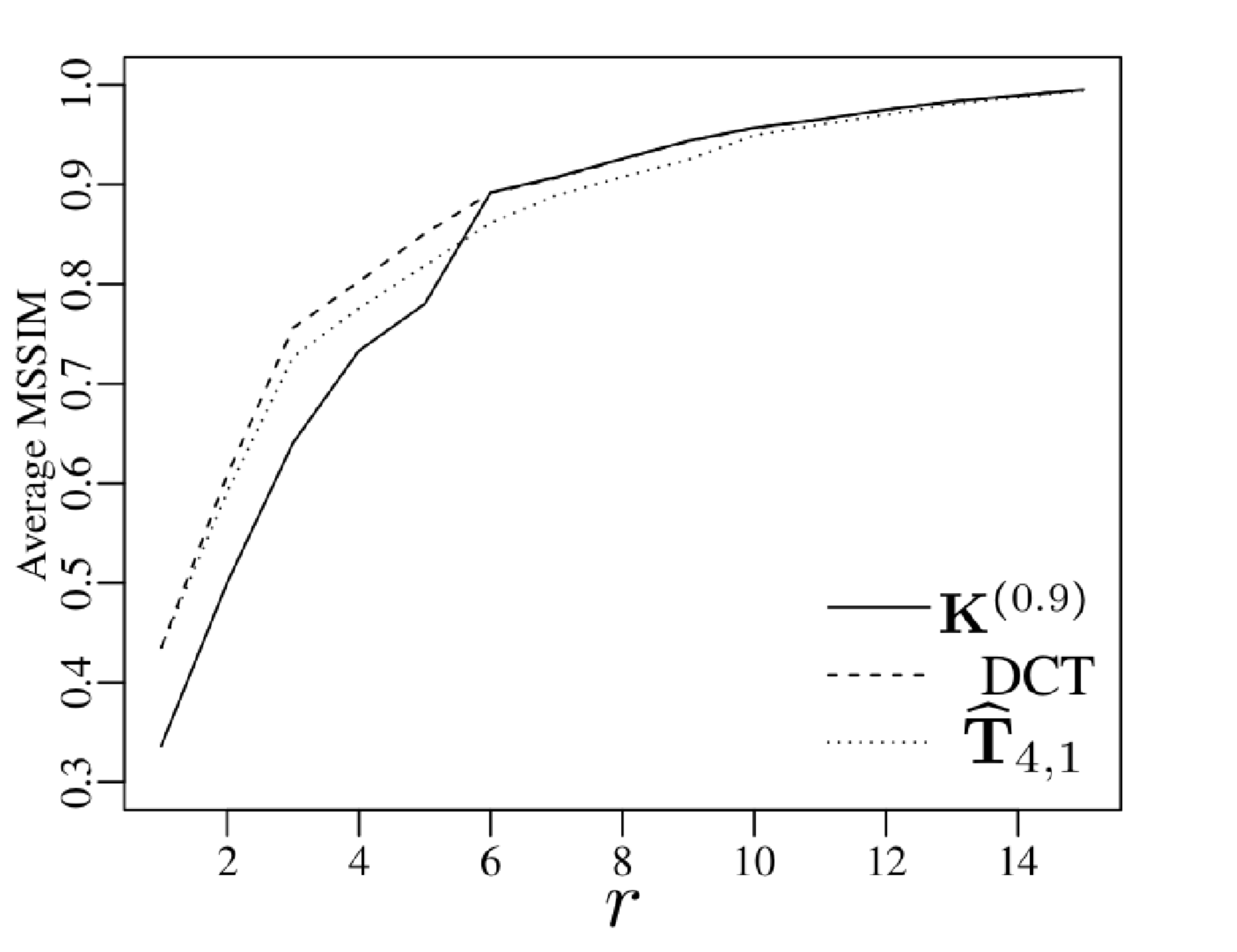}}
	\subfigure[Average PSNR ($N=8$)]{\label{f:PSNR8}{\includegraphics[width=7cm]{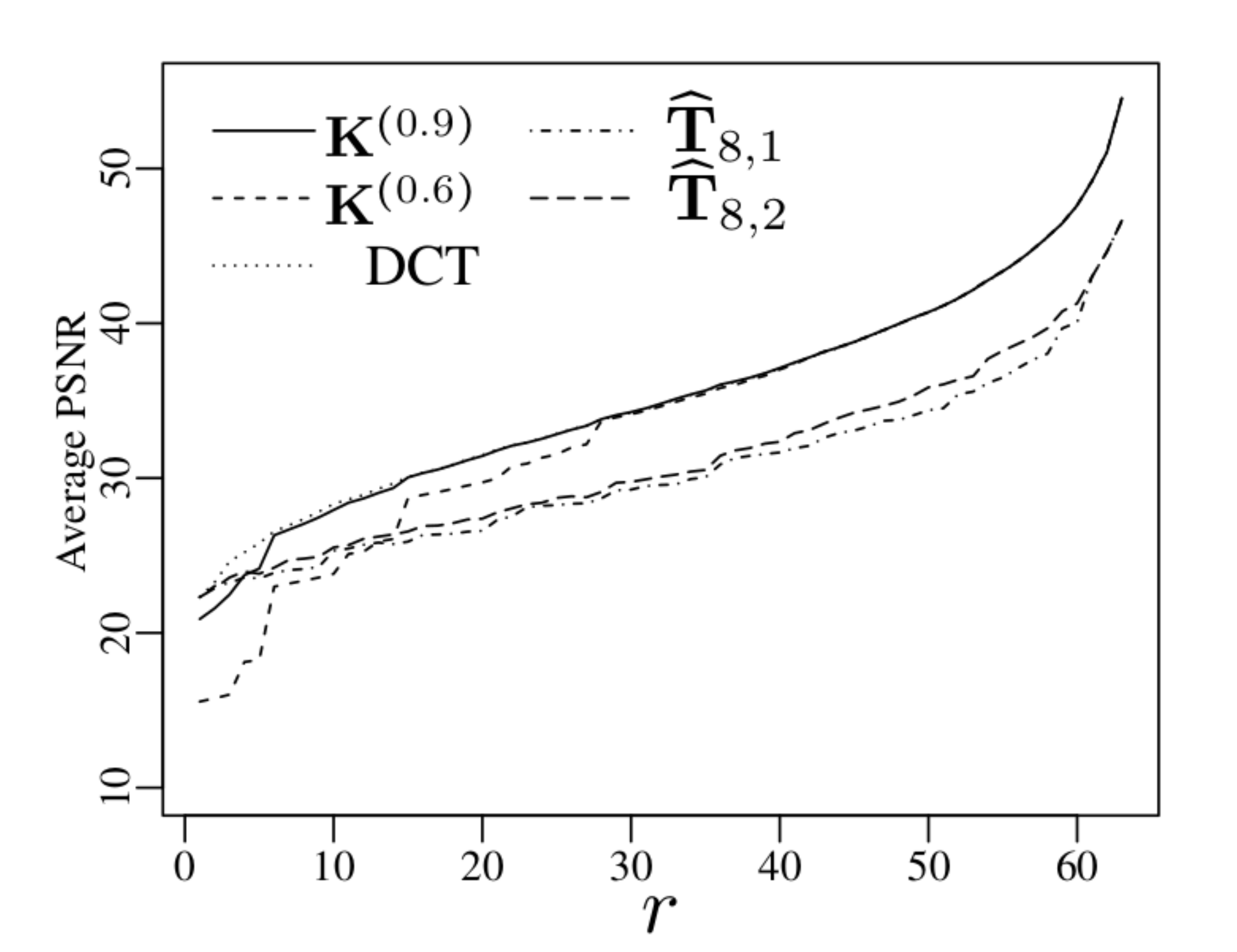}}}
	\subfigure[Average MSSIM ($N=8$)]{\label{f:MSSIM8}\includegraphics[width=7cm]{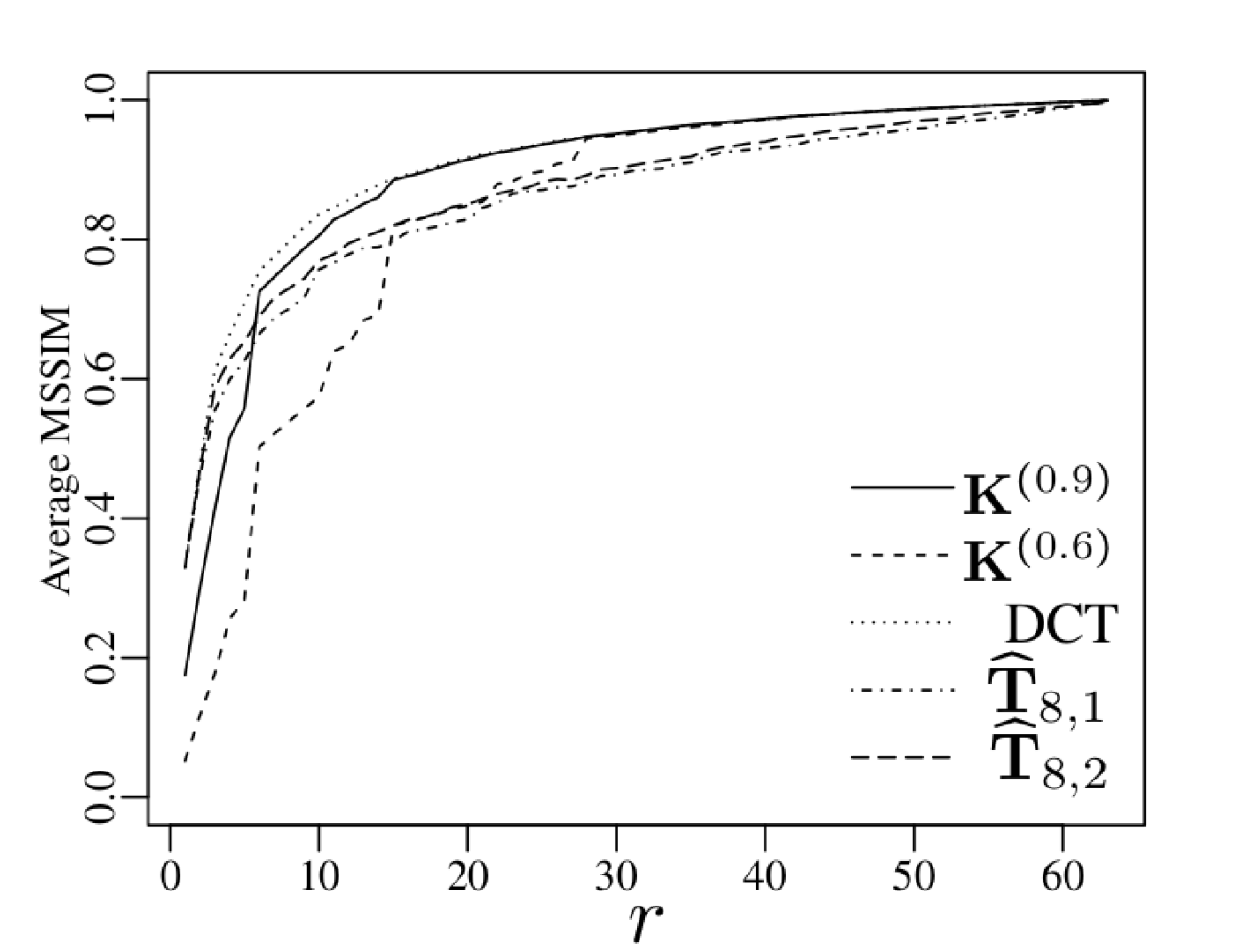}}
	\subfigure[Average PSNR ($N=16$)]{\label{f:PSNR16}{\includegraphics[width=7cm]{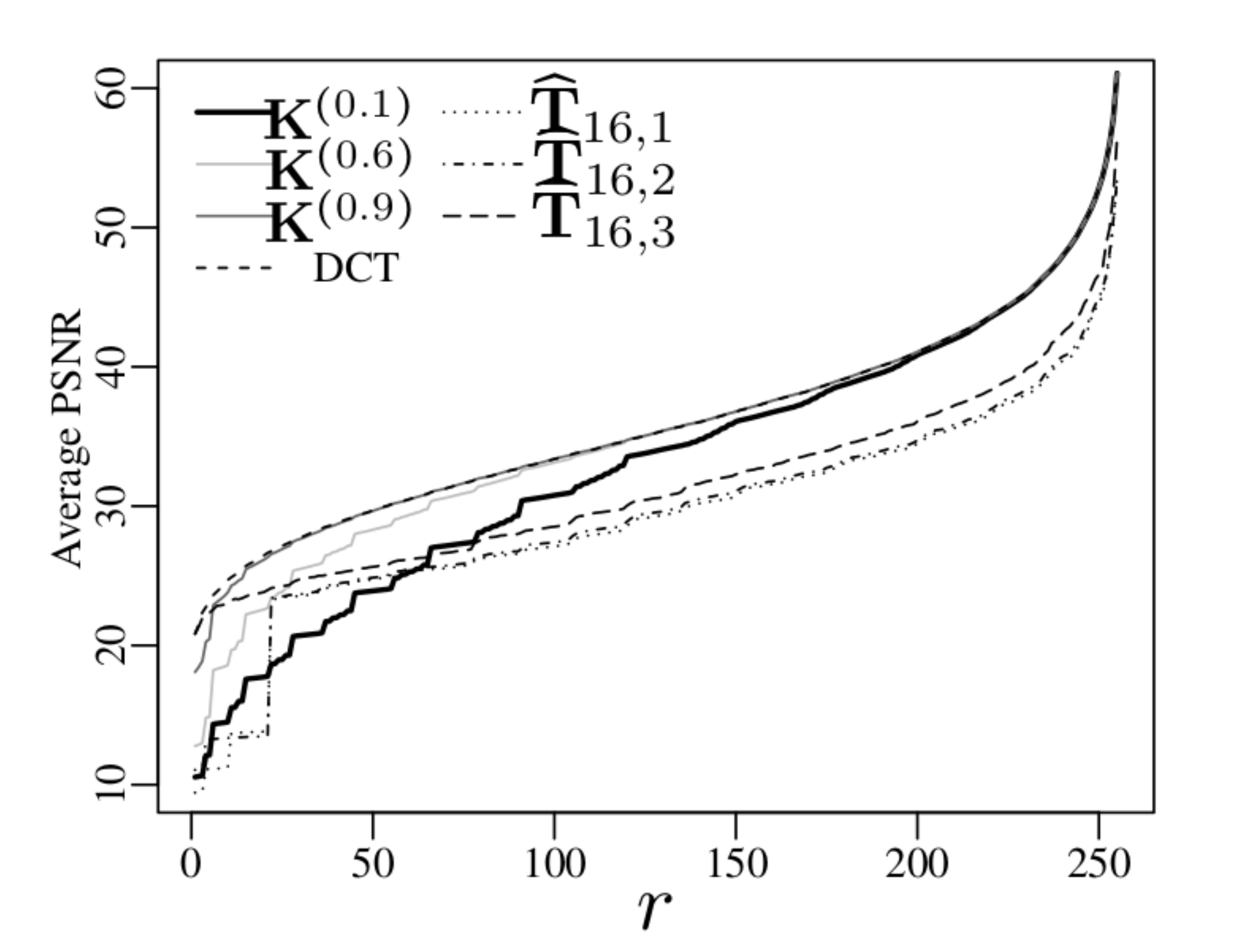}}}
	\subfigure[Average MSSIM ($N=16$)]{\label{f:MSSIM16}\includegraphics[width=7cm]{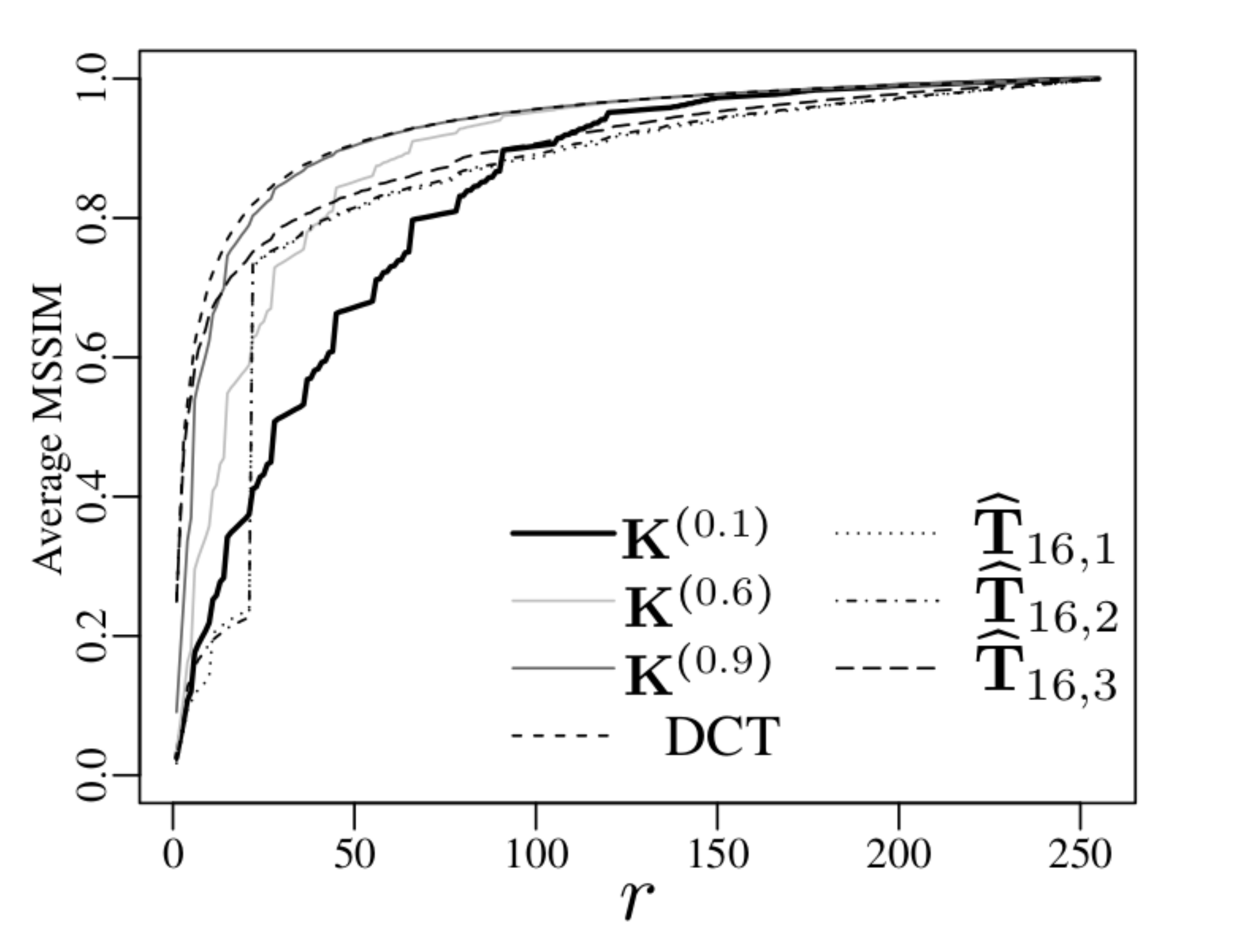}}
	\subfigure[Average PSNR ($N=32$)]{\label{f:PSNR32}{\includegraphics[width=7cm]{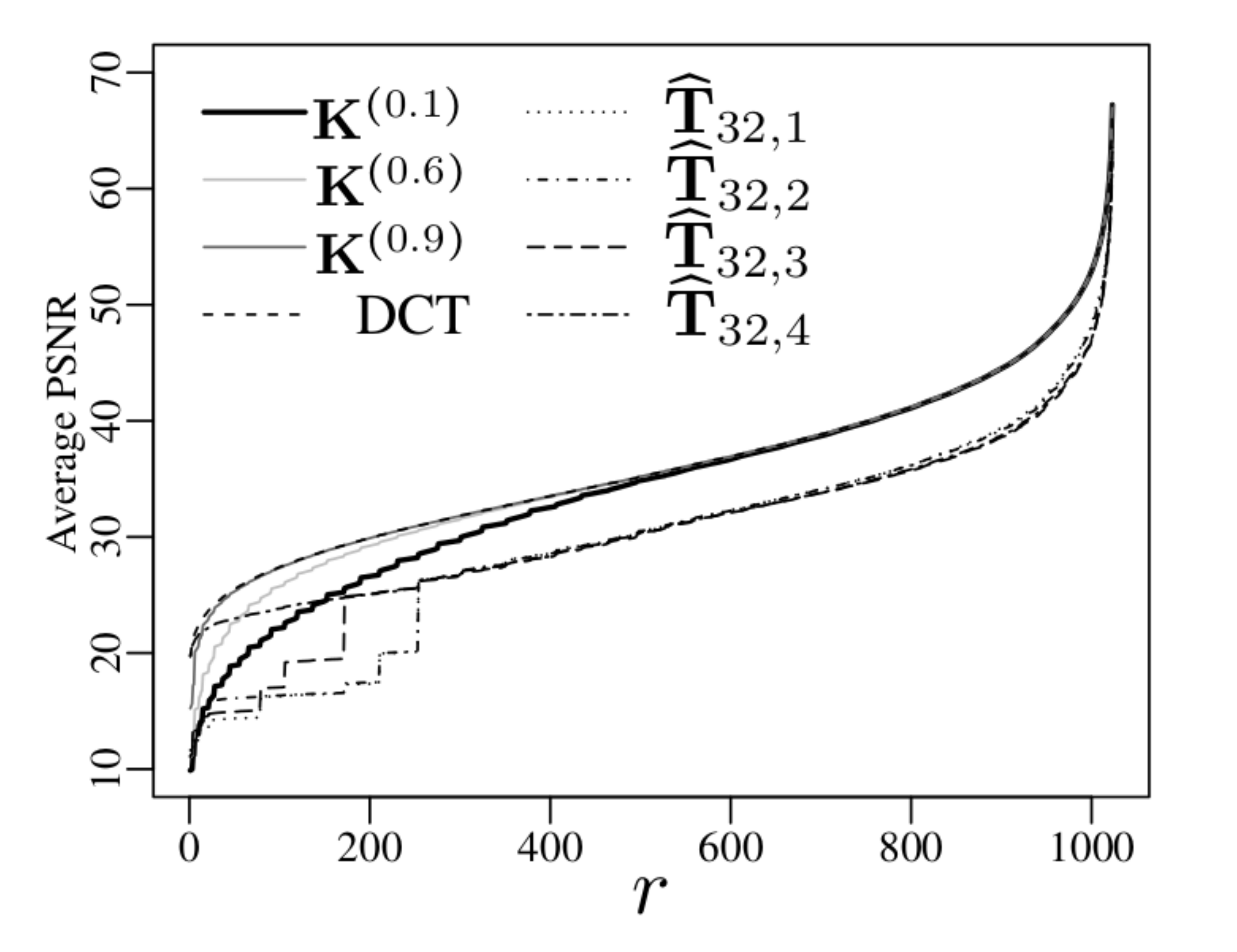}}}
	\subfigure[Average MSSIM ($N=32$)]{\label{f:MSSIM32}\includegraphics[width=7cm]{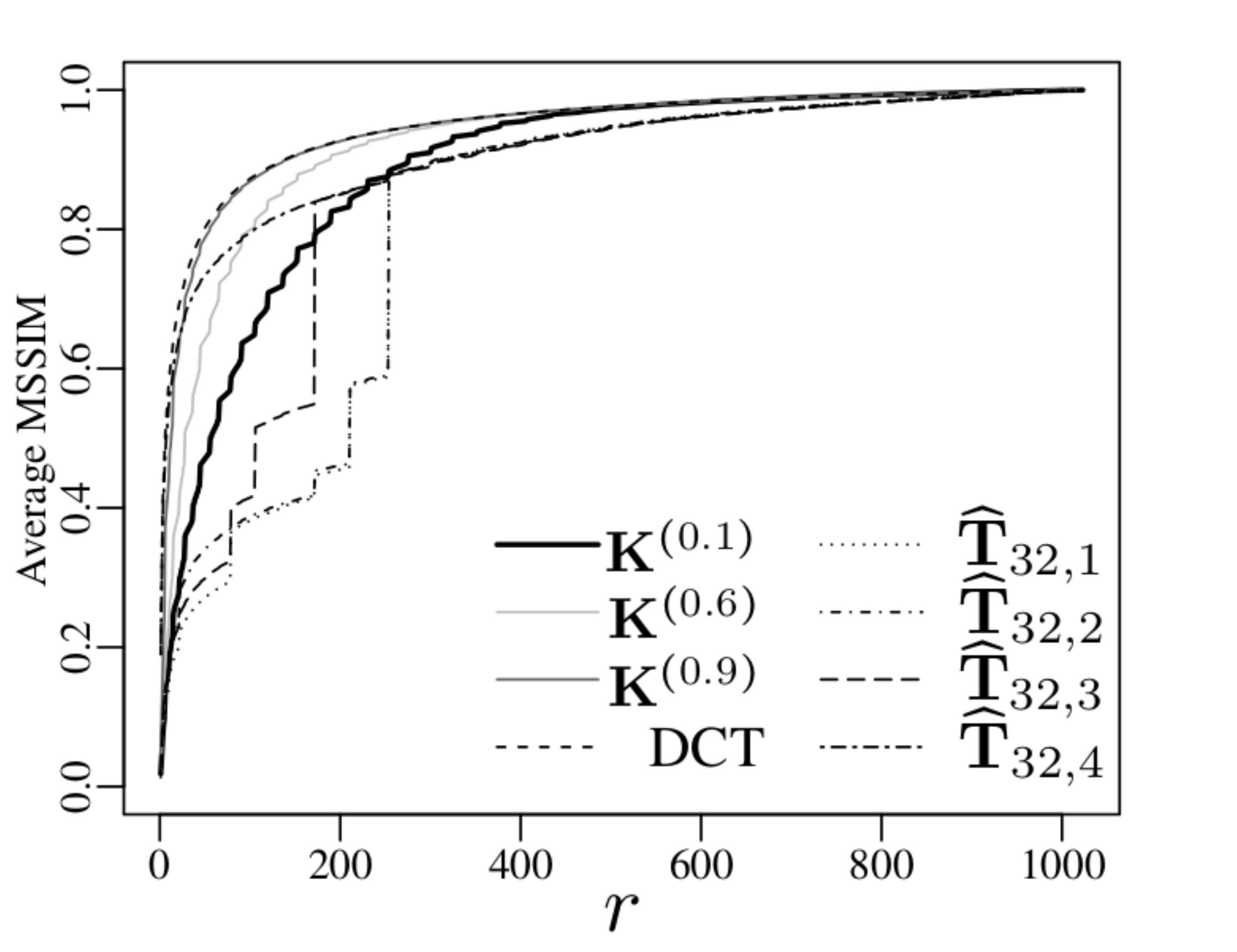}}
	\caption{Image quality measurements for different levels of compression.}
	\label{f:quali8}
\end{figure*}

Figure~\ref{f:quali8} presents
the average
image quality measurements for different values of $N$ considering different levels of compression, comparing with the exact KLT for $\rho = 0.1$, $0.6$, and $0.9$ and the DCT.
Figure~\ref{f:MSSIM4} shows that the average
MSSIM of
$\widehat{\mathbf{T}}_{4,1}$ is very close to the results from the KLT and DCT.
Figure~\ref{f:PSNR4} also shows that the values of the average PSNR of $\widehat{\mathbf{T}}_{4,1}$ are close to the values obtained by KLT and DCT.
We can also notice that the approximation $\widehat{\mathbf{T}}_{4,1}$ presents average MSSIM values better than KLT itself when we retain fewer coefficients, $r$ ranging from zero to six.
For $N = 8$, one can see that
$\widehat{\mathbf{T}}_{8,1}$ and $\widehat{\mathbf{T}}_{8,2}$ behave in a similar way according  to the image quality measures.
For
$N=16$,
$\widehat{\mathbf{T}}_{16,3}$ transform has considerably closer values to the DCT and KLT than the other approximations. This may be related to the fact that $\widehat{\mathbf{T}}_{16,3}$ is the obtained transform with a higher value of $\rho$ ($\rho\geq 0.9$).
One
can
see that, for $r<10$, the approximation $\widehat{\mathbf{T}}_{16,3}$ presents better average MSSIM values than KLT itself.
Figures \ref{f:PSNR32} and \ref{f:MSSIM32} present the image quality
measurements for
$N=32$,
that poses $\widehat{\mathbf{T}}_{32,4}$ as the best-performing approximate KLT
for compression.

\subsection{Video coding}\label{S:HEVC}

In order to demonstrate the suitability of the introduced SKLTs in the video coding context, we embed the proposed approximations into a public available HEVC reference software~\cite{refsoft}.
The HEVC employs an integer DCT (IDCT) of lengths 4, 8, 16, and 32~\cite{Ohm2012}, unlike its predecessors~\cite{le1992mpeg, h261, h263, h2642003}.
According to~\cite{pourazad2012hevc}, the larger transforms generally work better for smooth image regions, whereas the textured areas are better handled by the small sized transforms.

For our experiment, we substituted the original set of
IDCTs
natively defined in the HEVC standard by our proposed KLT approximations.
The original integer
DST-VI~\cite{Harize2017}
of length 4, responsible for residual coding in HEVC, is kept  unchanged in the reference software.
The main reason
is that
the
optimal SKLT for length 4 is unique for all $\rho$ values (confer Table~\ref{t:tabela}), and it approaches the coding capabilities of the DCT~\cite{haweel2001new}.
We separated four suits of approximations, relating to the optimality in Equation \eqref{eq:optimization}:
(i)~$\widehat{\mathbf{T}}_{4,1}$, $\widehat{\mathbf{T}}_{8,1}$, $\widehat{\mathbf{T}}_{16,2}$, and $\widehat{\mathbf{T}}_{32,2}$ (Group I);
(ii)~$\widehat{\mathbf{T}}_{4,1}$, $\widehat{\mathbf{T}}_{8,1}$, $\widehat{\mathbf{T}}_{16,1}$, and $\widehat{\mathbf{T}}_{32,3}$ (Group II);
(iii)~$\widehat{\mathbf{T}}_{4,1}$, $\widehat{\mathbf{T}}_{8,2}$, $\widehat{\mathbf{T}}_{16,3}$, and $\widehat{\mathbf{T}}_{32,1}$ (Group III); and
(iv)~$\widehat{\mathbf{T}}_{4,1}$, $\widehat{\mathbf{T}}_{8,1}$, $\widehat{\mathbf{T}}_{16,3}$, and $\widehat{\mathbf{T}}_{32,4}$ (Group IV).
Namely, Groups I, II, III, and IV are optimal regarding total MSE, total error energy, total unified coding gain, and total transform efficiency, respectively.
Therefore, we substituted the original IDCT by each SKLT Group I--IV in the HEVC reference software.

In our experiments, we encoded the first $100$ frames of one video sequence of each A to F class in accordance with the Common Test Conditions (CTC) document~\cite{ctconditions2013}. The considered 8-bit standard video sequences were:
\texttt{PeopleOnStreet} (2560$\times$1600 at 30~fps),
\texttt{BasketballDrive} (1920$\times$1080 at 50~fps),
\texttt{RaceHorses} (832$\times$480 at 30~fps),
\texttt{BlowingBubbles} (416$\times$240  at 50~fps),
\texttt{KristenAndSara} (1280$\times$720  at 60~fps),
and
\texttt{BasketbalDrillText} (832$\times$480  at 50~fps).
We further considered the \texttt{Foreman} (352$\times$288 at 30~fps)~\cite{xiphVideos}, a standard 8-bit CIF video sequence adopted in related works like~\cite{lengwehasatit2004scalable,da2017multiplierless}.
As done in~\cite{Jridi2015}, we set all the test parameters in accordance with the CTC documentation. Also, we considered the four standard 8-bit coding configurations in the \texttt{Main} profile: \texttt{All Intra} (AI), \texttt{Random Access} (RA) and \texttt{Low-Delay B} and \texttt{P} (LD-B and LD-P).
We selected the frame-wise PSNR for each YUV color channel~\cite{Ohm2012} as figure of merit.
Then, for each video sequence,
we computed the rate distortion (RD) curve considering
quantization parameter (QP) values equal to 22, 27, 32, and 37~\cite{ctconditions2013}.

\begin{figure*}[h]
	\centering
	\subfigure[]{\includegraphics[scale=.55]{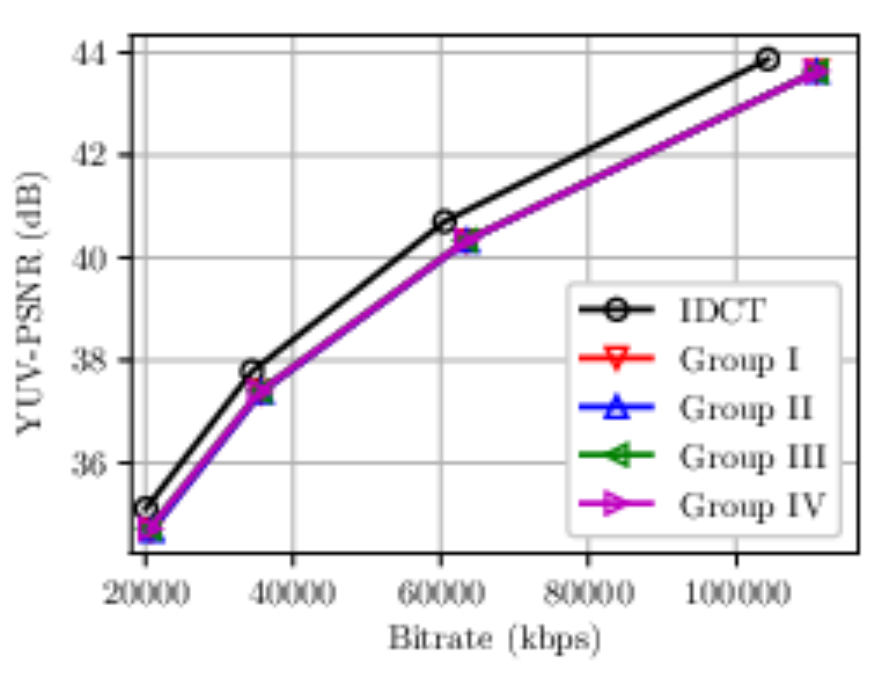}}
	\subfigure[]{\includegraphics[scale=.55]{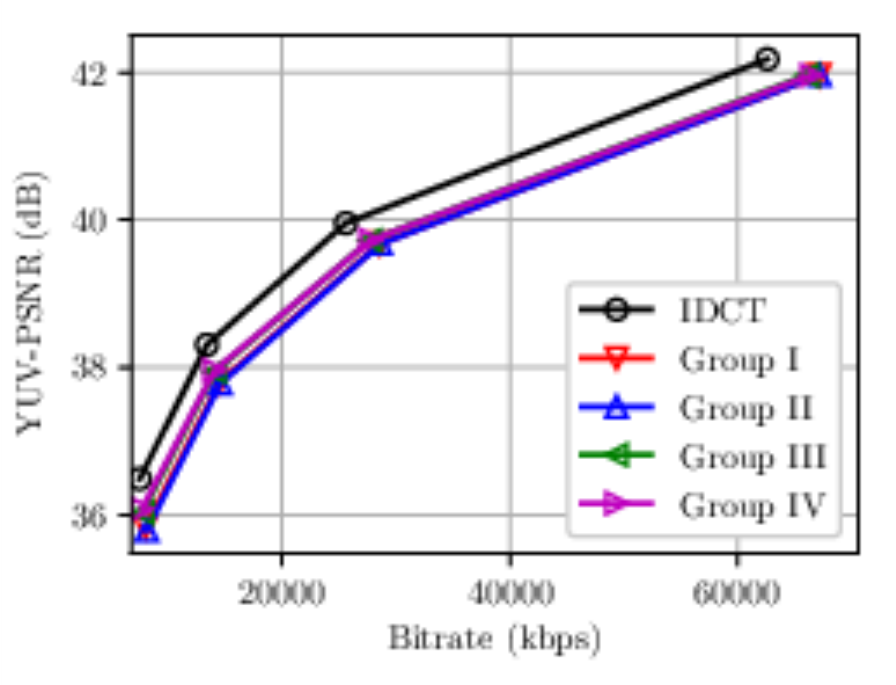}}
	\subfigure[]{\includegraphics[scale=.55]{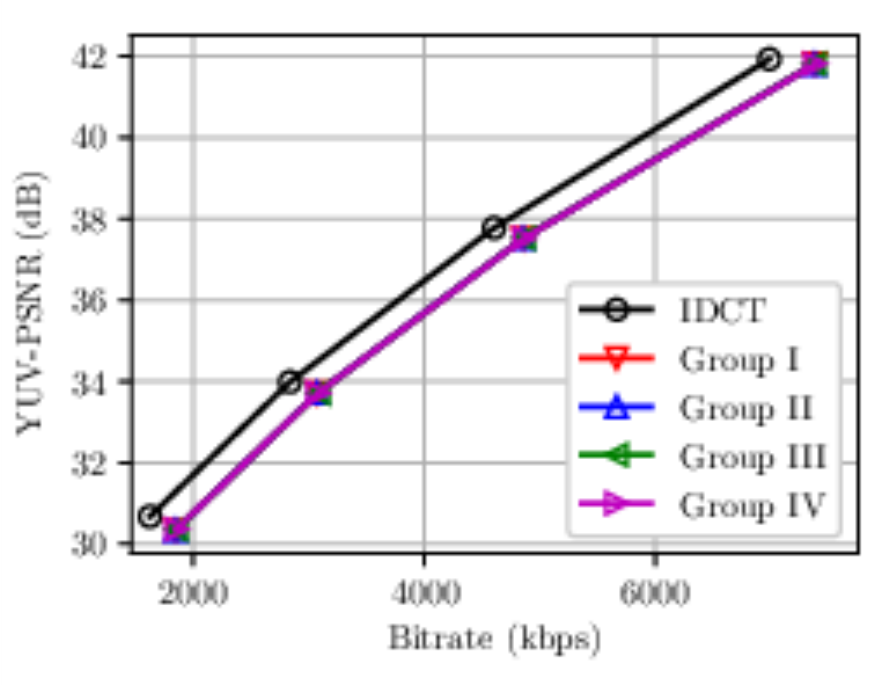}}
	\subfigure[]{\includegraphics[scale=.55]{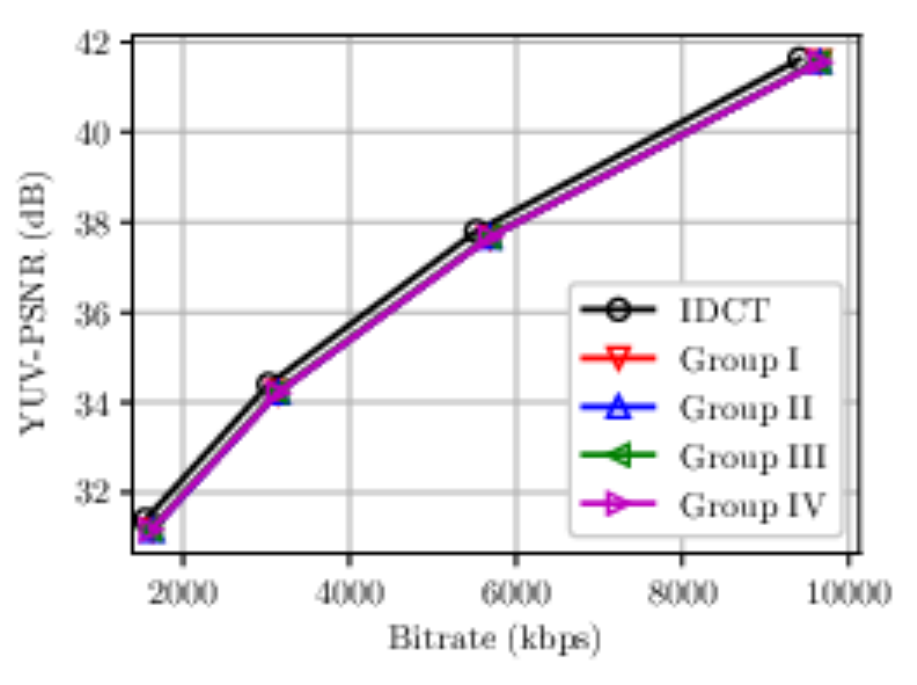}}
	\subfigure[]{\includegraphics[scale=.55]{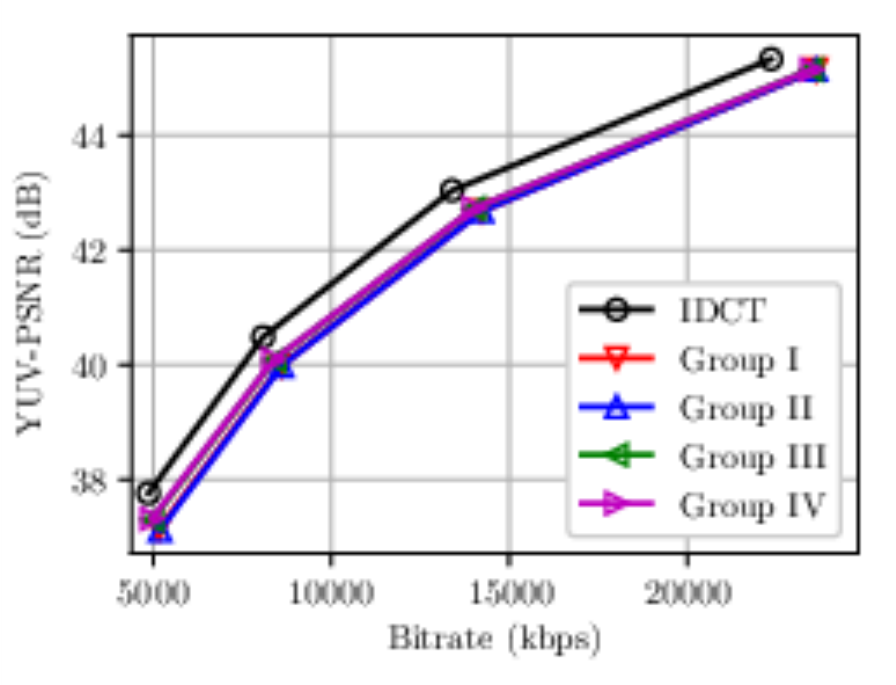}}%
	\subfigure[]{\includegraphics[scale=.55]{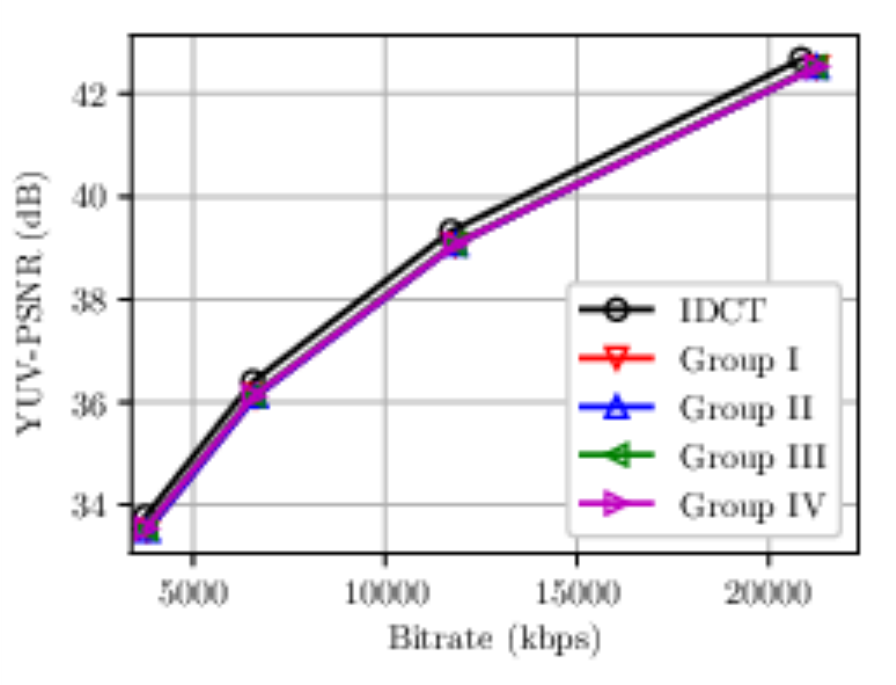}}
	\subfigure[]{\includegraphics[scale=.55]{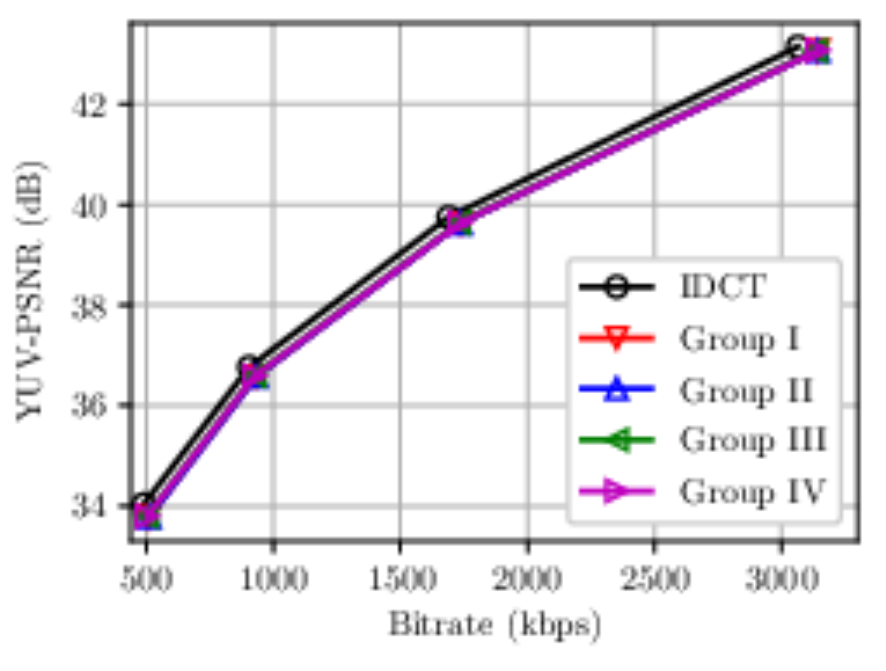}}
	\caption{%
		RD curves of the modified HEVC versions for test sequences in Main profile and AI configuration:
		(a)~\texttt{PeopleOnStreet},
		(b)~\texttt{BasketballDrive},
		(c)~\texttt{RaceHorses},
		(d)~\texttt{BlowingBubbles},
		(e)~\texttt{KristenAndSara},
		(f)~\texttt{BasketbalDrillText}, and (g)~\texttt{Foreman}.
	}
	\label{fig:allbitratesAI}
\end{figure*}

\begin{figure*}[h]
	\centering
	\subfigure[]{\includegraphics[scale=.55]{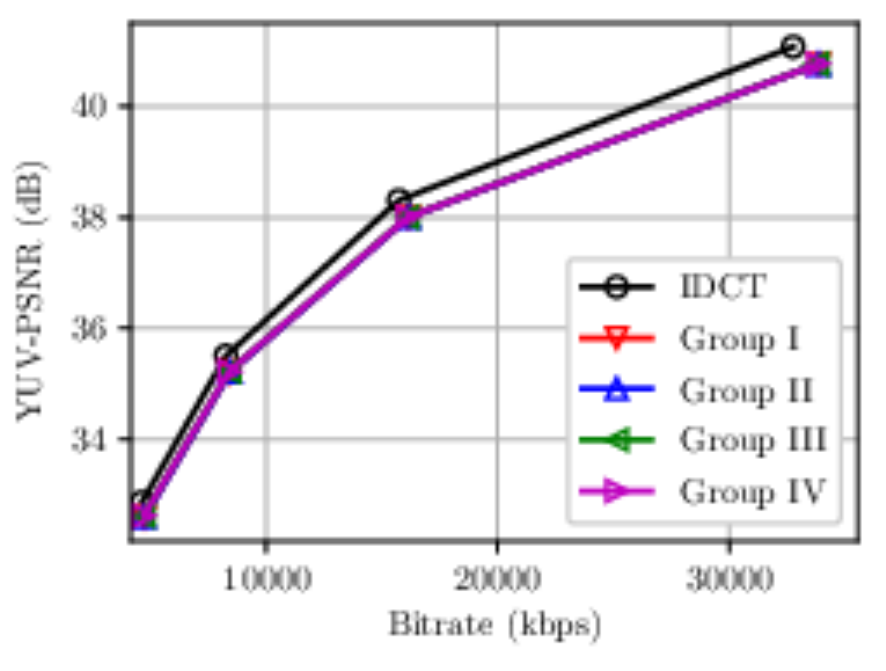}}
	\subfigure[]{\includegraphics[scale=.55]{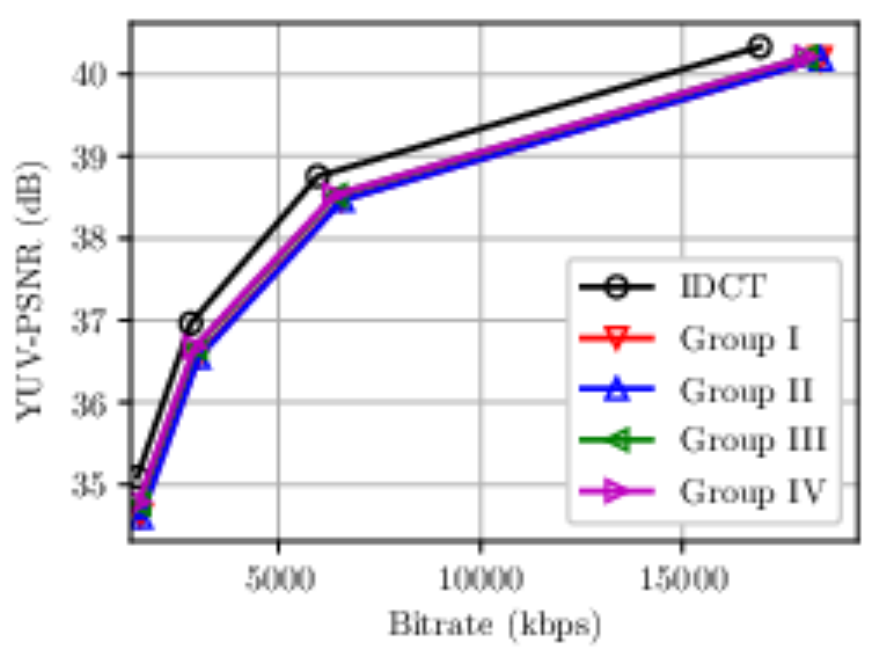}}
	\subfigure[]{\includegraphics[scale=.55]{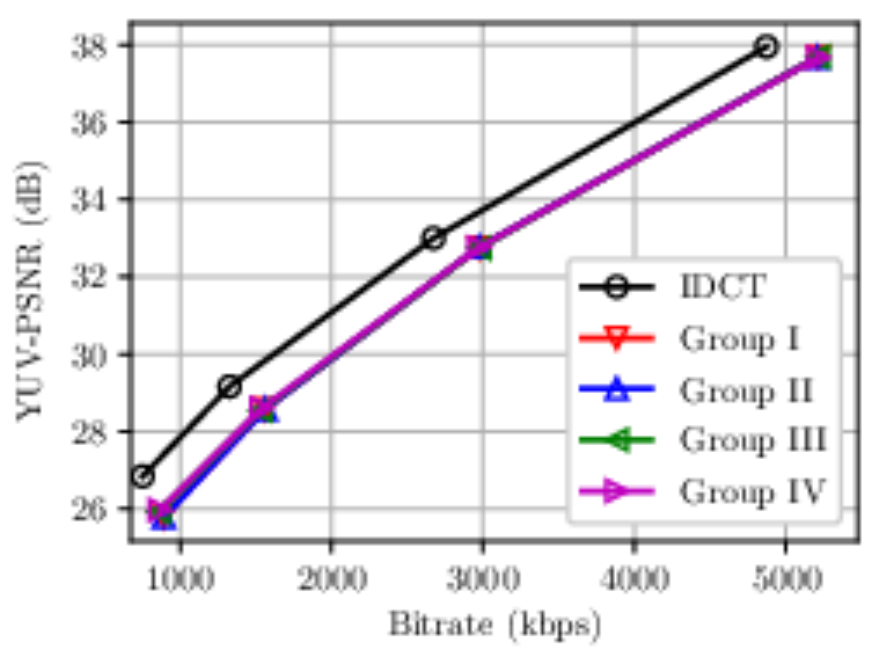}}
	\subfigure[]{\includegraphics[scale=.55]{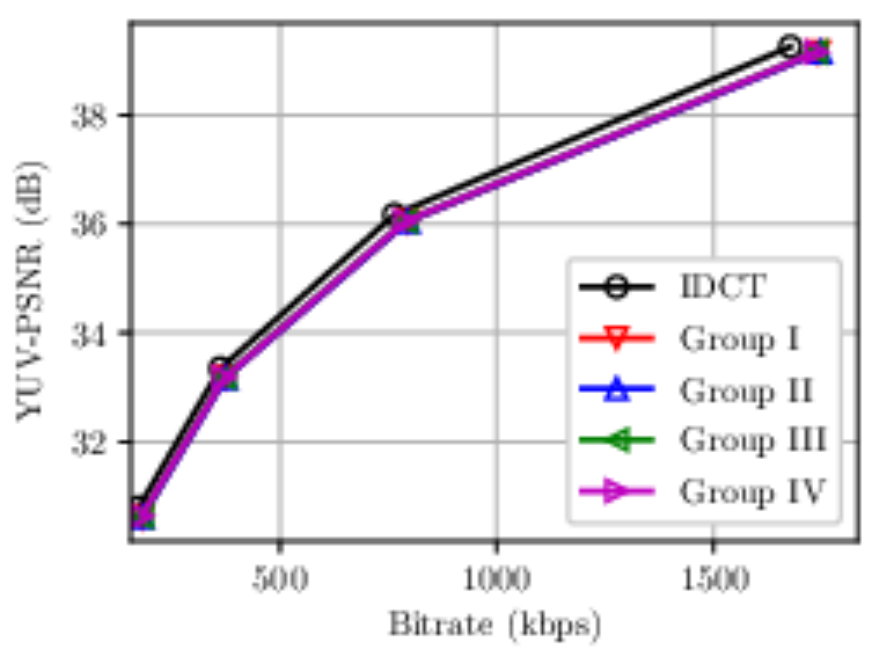}}
	\subfigure[]{\includegraphics[scale=.55]{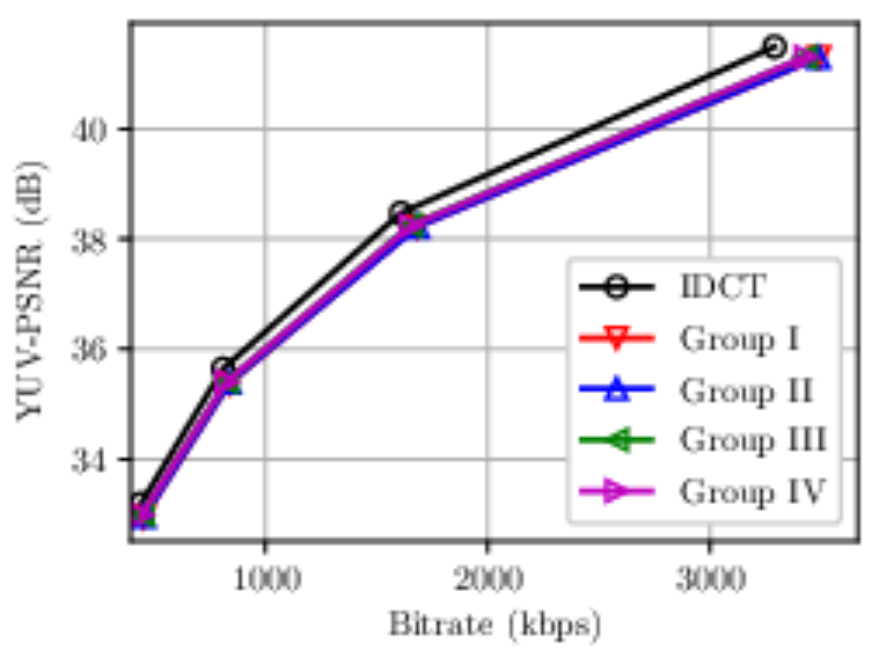}}
	\subfigure[]{\includegraphics[scale=.55]{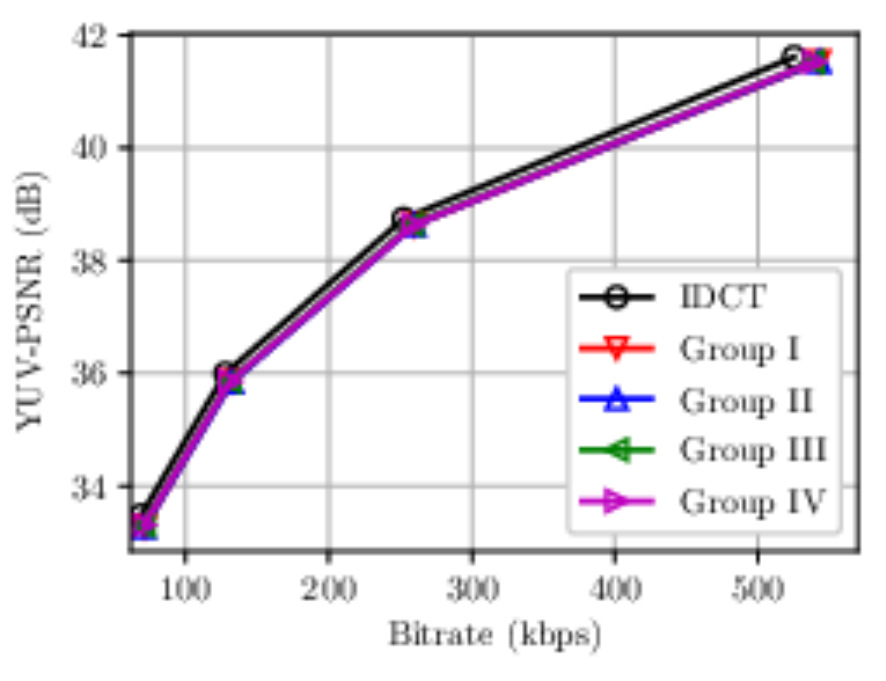}}
	\caption{%
		RD curves of the modified HEVC versions for test sequences in Main profile and RA configuration:
		(a)~\texttt{PeopleOnStreet},
		(b)~\texttt{BasketballDrive},
		(c)~\texttt{RaceHorses},
		(d)~\texttt{BlowingBubbles},
		(e)~\texttt{BasketbalDrillText}, and (f)~\texttt{Foreman}.
	}
	\label{fig:allbitratesRA}
\end{figure*}

\begin{figure*}[h]
	\centering
	\subfigure[]{\includegraphics[scale=.55]{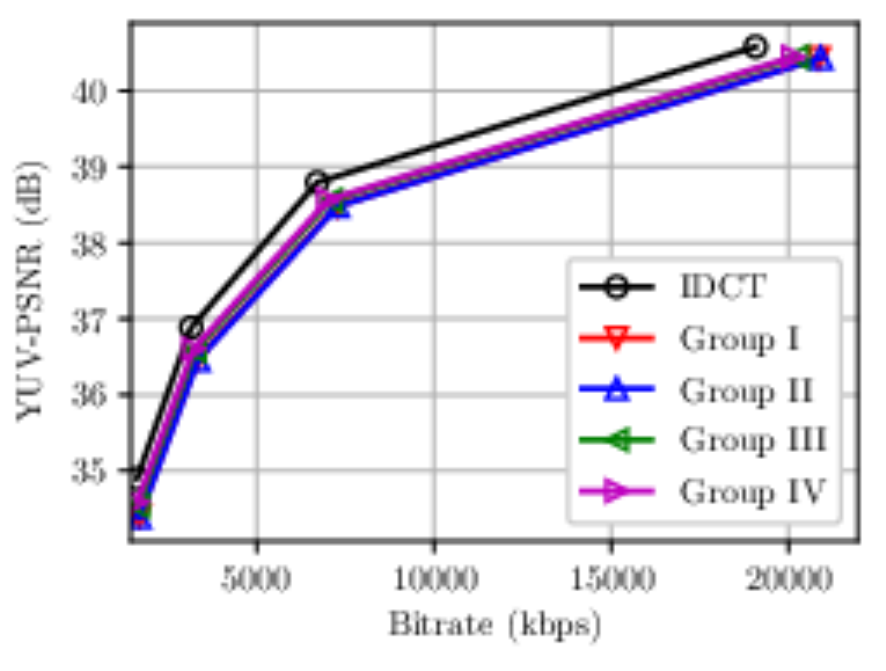}}
	\subfigure[]{\includegraphics[scale=.55]{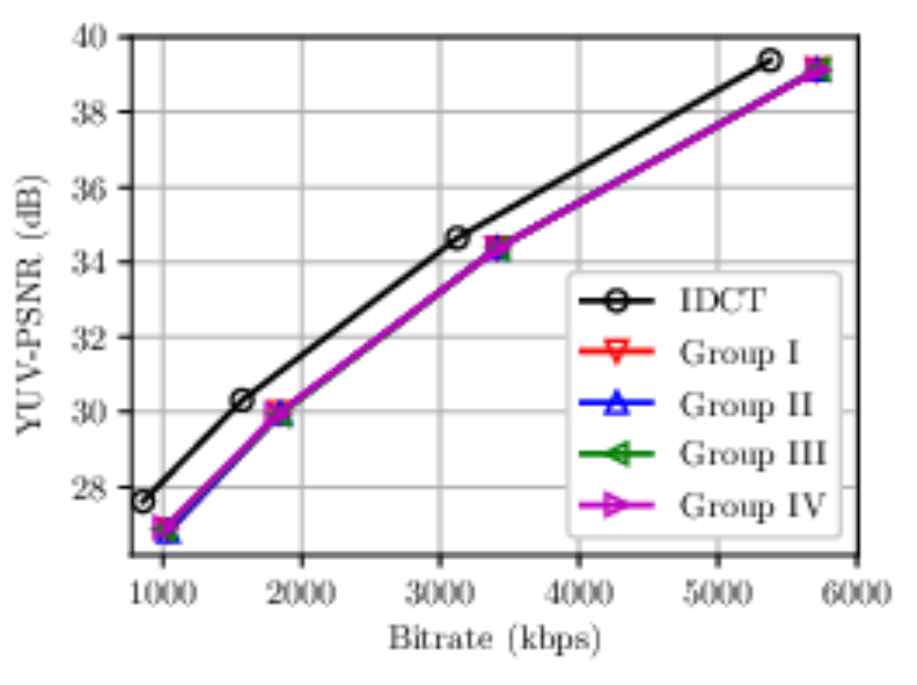}}
	\subfigure[]{\includegraphics[scale=.55]{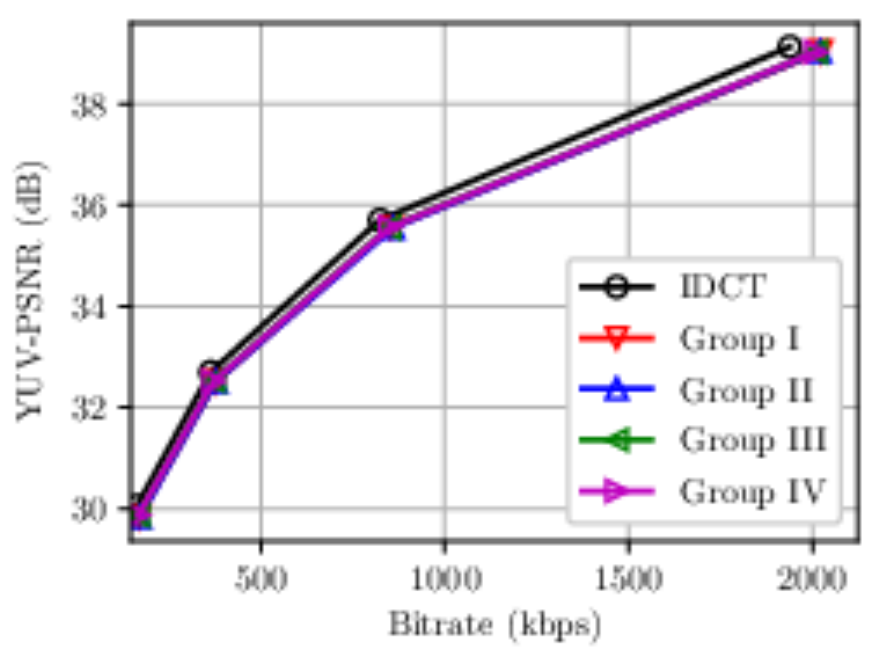}}
	\subfigure[]{\includegraphics[scale=.55]{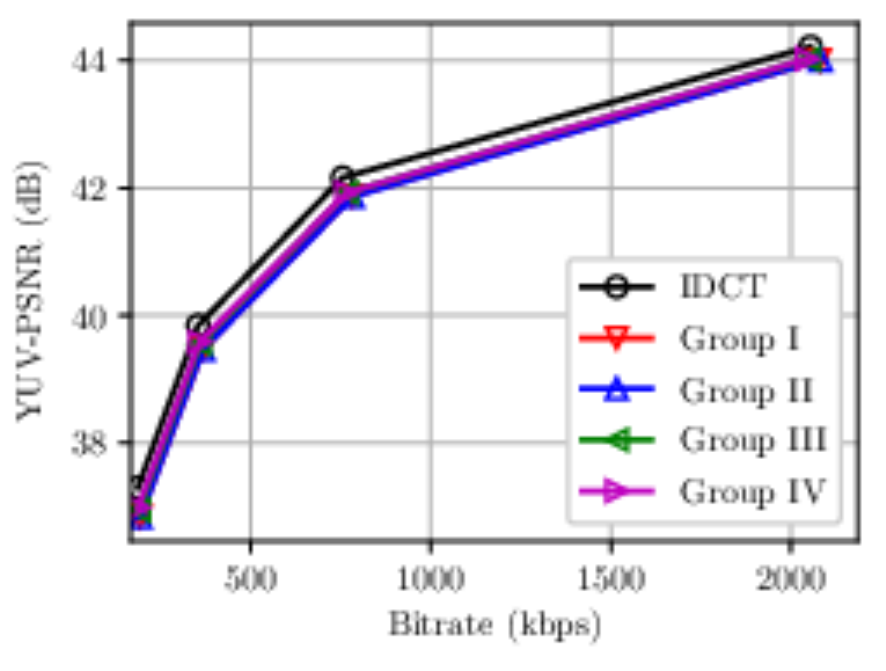}}
	\subfigure[]{\includegraphics[scale=.55]{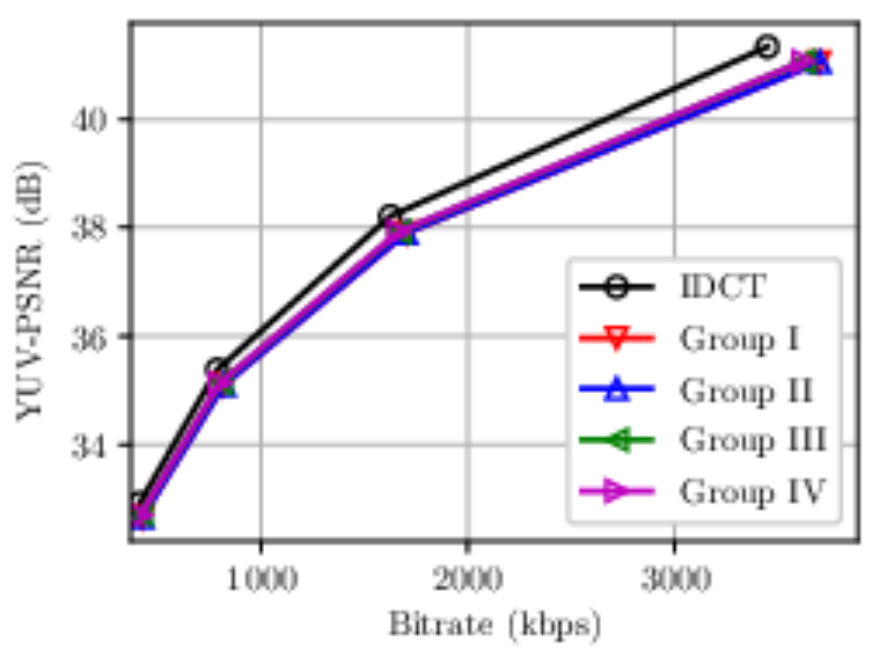}}
	\subfigure[]{\includegraphics[scale=.55]{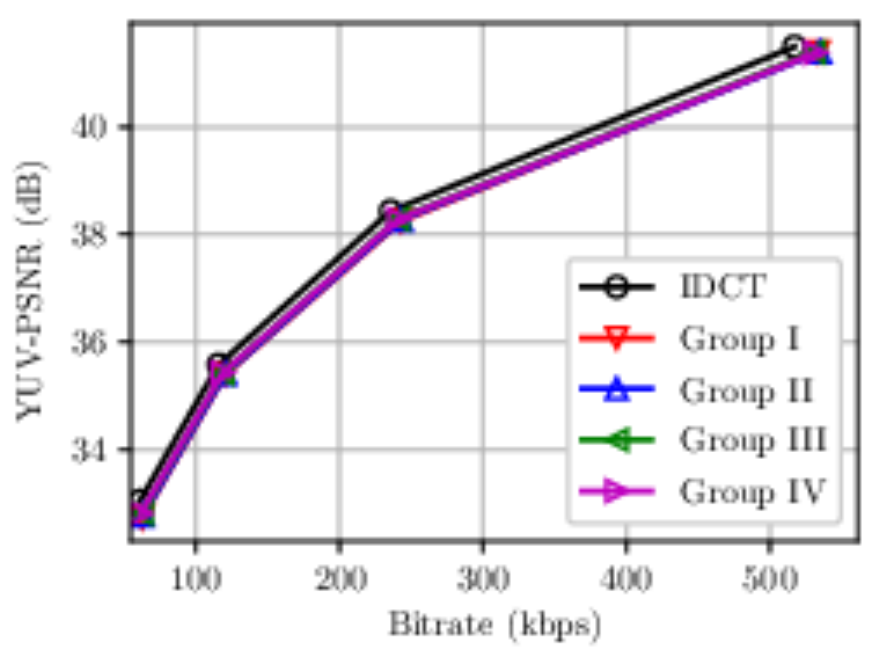}}
	\caption{%
		RD curves of the modified HEVC versions for test sequences in Main profile and LD-B configuration:
		(a)~\texttt{BasketballDrive},
		(b)~\texttt{RaceHorses},
		(c)~\texttt{BlowingBubbles},
		(d)~\texttt{KristenAndSara},
		(e)~\texttt{BasketbalDrillText}, and (f)~\texttt{Foreman}.
	}
	\label{fig:allbitratesLDB}
\end{figure*}

\begin{figure*}[h]
	\centering
	\subfigure[]{\includegraphics[scale=.55]{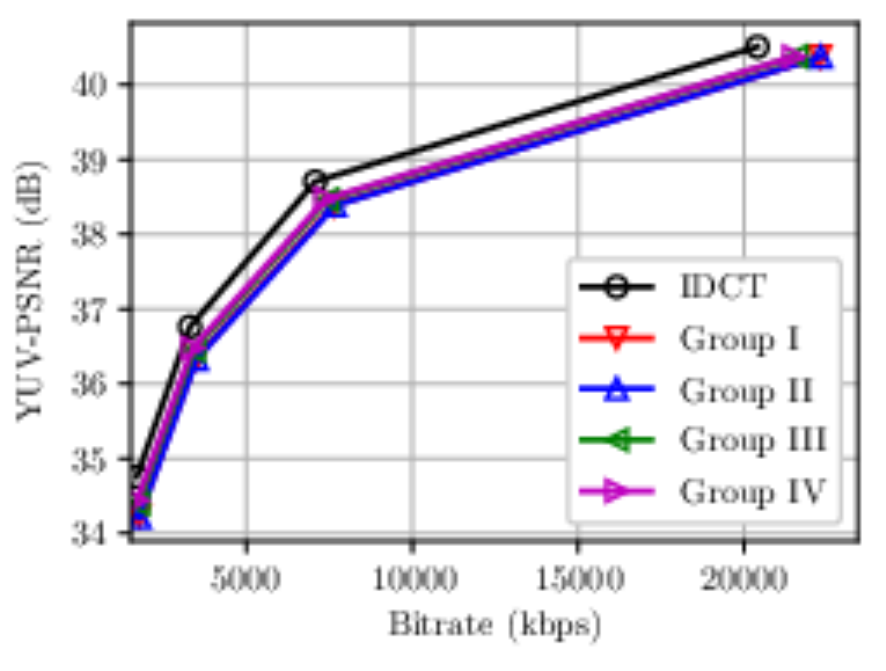}}
	\subfigure[]{\includegraphics[scale=.55]{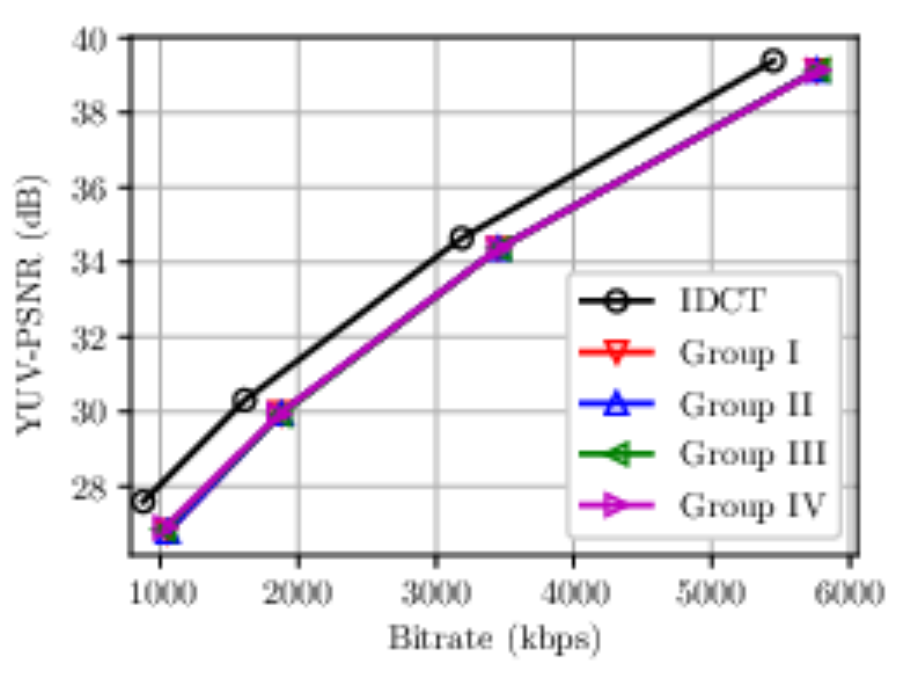}}
	\subfigure[]{\includegraphics[scale=.55]{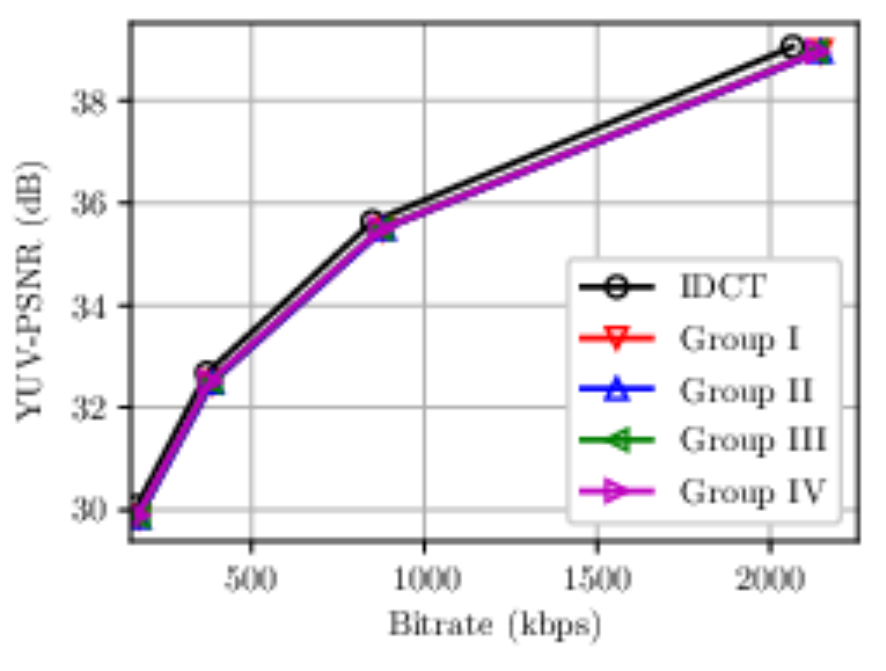}}
	\subfigure[]{\includegraphics[scale=.55]{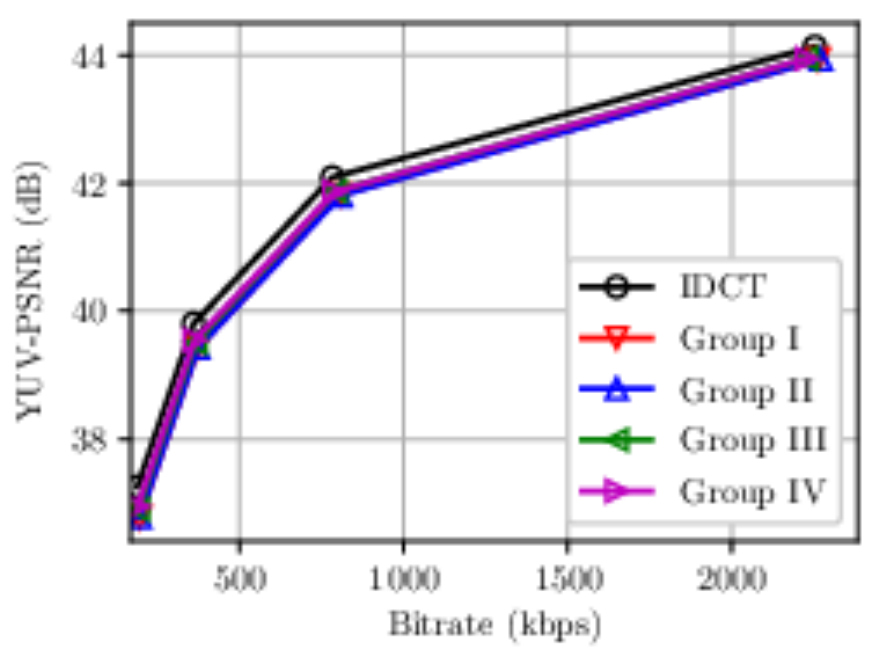}}
	\subfigure[]{\includegraphics[scale=.55]{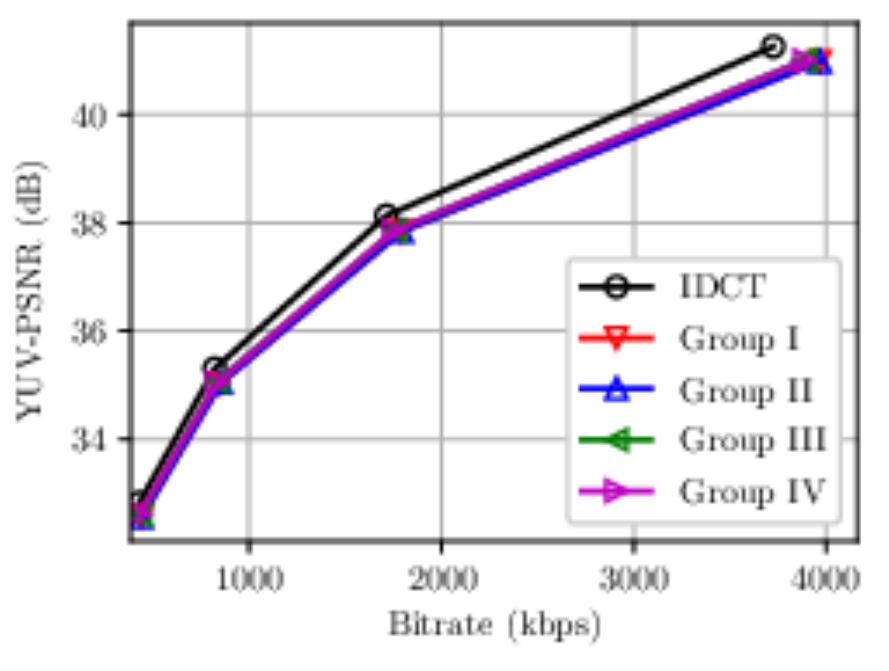}}
	\subfigure[]{\includegraphics[scale=.55]{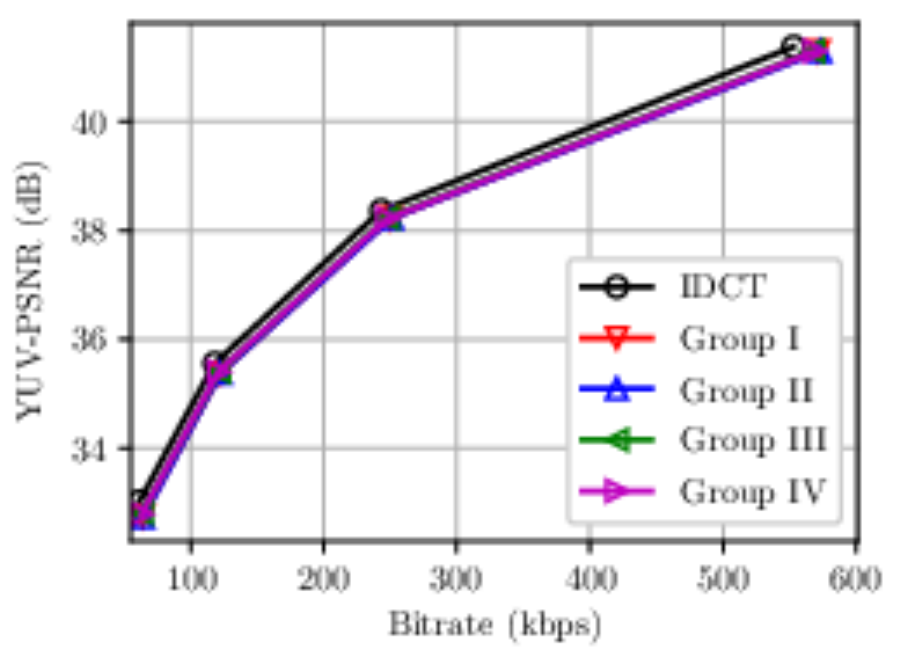}}
	\caption{%
		RD curves of the modified HEVC versions for test sequences in Main profile and LD-P configuration:
		(a)~\texttt{BasketballDrive},
		(b)~\texttt{RaceHorses},
		(c)~\texttt{BlowingBubbles},
		(d)~\texttt{KristenAndSara},
		(e)~\texttt{BasketbalDrillText}, and (f)~\texttt{Foreman}.
	}
	\label{fig:allbitratesLDP}
\end{figure*}

Moreover, we measured the Bj{\o}ntegaard's delta PSNR (BD-PSNR)~\cite{bjontegaard2001, Hanhart2014}
for the modified versions of the HEVC software.
The average results per video for all the transform groups and coding configurations are presented in Table~\ref{tab:bdpsnr}.
One can note from Table~\ref{tab:bdpsnr} that the Group IV performed better than the Group III on average.
The metrics used for selecting both SKLT Groups III and IV maximize the coding efficiency of the transforms.
From the table, Group IV outperformed Groups I and II in most of the cases regardless the configuration mode.
This information can be confirmed in Figures~\ref{fig:allbitratesAI},~\ref{fig:allbitratesRA},~\ref{fig:allbitratesLDB}, and~\ref{fig:allbitratesLDP}, which show the RD curves for the four groups of transforms in AI, RA, LD-B, and LD-P configurations.

\begin{table*}
	\centering
	\caption{Average BD-PSNR of the modified HEVC reference software for tested video sequences}
	\label{tab:bdpsnr}
	\begin{tabular}{cccccc}
		\hline
		\multirow{2}{*}{ Configuration } & \multirow{2}{*}{ Video sequence } & \multicolumn{4}{c}{ Transforms } \\
		\cline{3-6}
		& & Group I & Group II & Group III & Group IV  \\
		\hline
		\multirow{6}{*}{ AI }
		& \texttt{PeopleOnStreet} & $-0.6074$ & $-0.6127$ & $-0.5696$ & $-0.5774$  \\
		& \texttt{BasketballDrive}& $-0.6194$ & $-0.6434$ & $-0.4549$ & $-0.4452$  \\
		& \texttt{RaceHorses} &$-0.8221$  & $-0.8219$ & $-0.8280$ & $-0.8290$ \\
		& \texttt{KristenAndSara} & $-0.3446$ & $-0.3464$ & $-0.3142$ & $-0.3361$  \\
		& \texttt{BlowingBubbles} & $-0.7275$ & $-0.7425$ & $-0.5495$ & $-0.5391$  \\
		& \texttt{BasketballDrillText} & $-0.3163$ & $-0.3217$ & $-0.2680$ & $-0.2844$ \\
		&\texttt{Foreman} & $-0.3030$ & $-0.3075$ & $-0.2636$ & $-0.2794$ \\
		\hline
		\multirow{5}{*}{ RA }
		& \texttt{PeopleOnStreet} & $-0.3995$ & $-0.4030$ & $-0.3748$ & $-0.3797$  \\
		& \texttt{BasketballDrive} & $-0.4855$ & $-0.5054$ & $-0.3629$ & $-0.3423$  \\
		& \texttt{RaceHorses} & $-1.2096$ &  $-1.2150$ & $-1.1687$ & $-1.1435$ \\
		& \texttt{BlowingBubbles} & $-0.2784$ & $-0.2849$ & $-0.2440$ & $-0.2525$  \\
		& \texttt{BasketballDrillText}& $-0.4108$ & $-0.4095$ & $-0.3184$ & $-0.3241$  \\
		& \texttt{Foreman} & $-0.2532$ & $-0.2561$ & $-0.2127$ & $-0.2290$ \\
		\hline
		\multirow{5}{*}{LD-B }
		& \texttt{BasketballDrive}& $-0.4900$ & $-0.5089$ & $-0.3548$ & $-0.3117$  \\
		& \texttt{RaceHorses} & $-1.1279$ & $-1.1342$ & $-1.1324$ & $-1.1028$  \\
		& \texttt{BlowingBubbles} & $-0.2978$ & $-0.3150$ & $-0.2519$ & $-0.2610$  \\
		& \texttt{KristenAndSara} & $-0.4069$ &  $-0.4245$ & $-0.2958$ & $-0.2740$  \\
		& \texttt{BasketballDrillText} & $-0.4609$ & $-0.4629$ & $-0.3616$ & $-0.3528$  \\
		& \texttt{Foreman} & $-0.3147$ & $-0.2988$ & $-0.2421$ & $-0.2436$ \\
		\hline
		\multirow{5}{*}{ LD-P }
		& \texttt{BasketballDrive} & $-0.4932$ & $-0.5139$ & $-0.3542$ & $-0.3153$  \\
		& \texttt{RaceHorses} &$-1.0903$  & $-1.0940$ &$-1.0929$ & $-1.0637$  \\
		& \texttt{BlowingBubbles} &$-0.2905$  & $-0.2958$ &$-0.2438$ & $-0.2404$  \\
		& \texttt{KristenAndSara} &$-0.3889$  & $-0.4062$ &$-0.2791$ & $-0.2544$  \\
		& \texttt{BasketballDrillText} &$-0.4317$  & $-0.4369$ &$-0.3400$ & $-0.3361$ \\
		& \texttt{Foreman} & $-0.2828$ & $-0.2927$ & $-0.2308$ & $-0.2381$ \\
		\hline
	\end{tabular}
\end{table*}

One can notice that the group of transforms that optimize the total transform efficiency metric (Group IV) tended to outperform the other three groups (Groups I, II, and III).
The results in video coding corroborate those of the still-image experiments presented in Section~\ref{S:compressao}.

\begin{figure*}[h]
	\centering
	\subfigure[
	$\mbox{PSNR-Y} = 39.4857$dB,
	$\mbox{PSNR-U} = 43.5883$dB and
	$\mbox{PSNR-V} = 44.4961$dB]{\includegraphics[scale=.15]{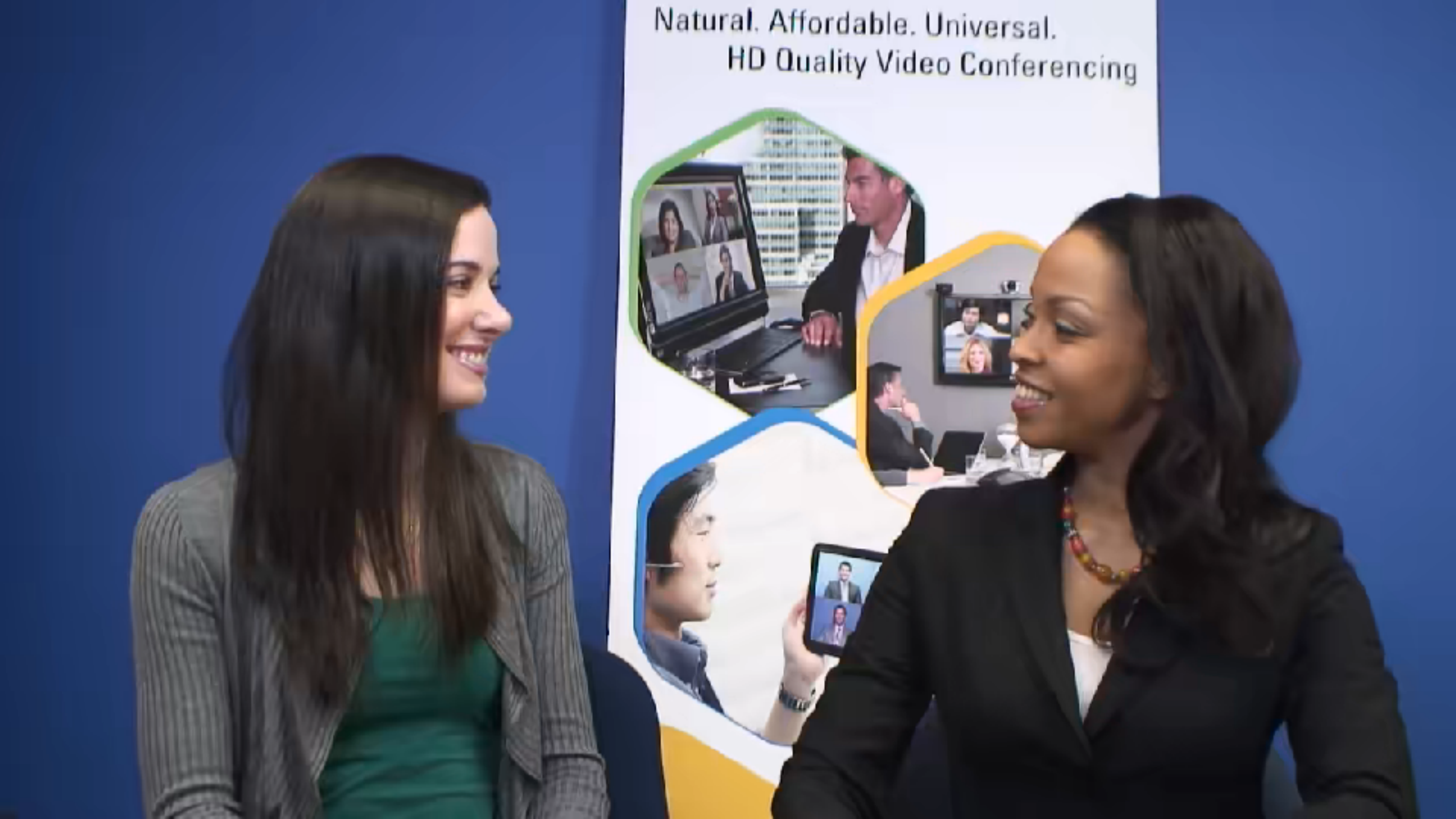}}
	\subfigure[
	$\mbox{PSNR-Y} = 39.0457$dB,
	$\mbox{PSNR-U} = 43.0431$dB and
	$\mbox{PSNR-V} = 43.9074$dB]{\includegraphics[scale=.15]{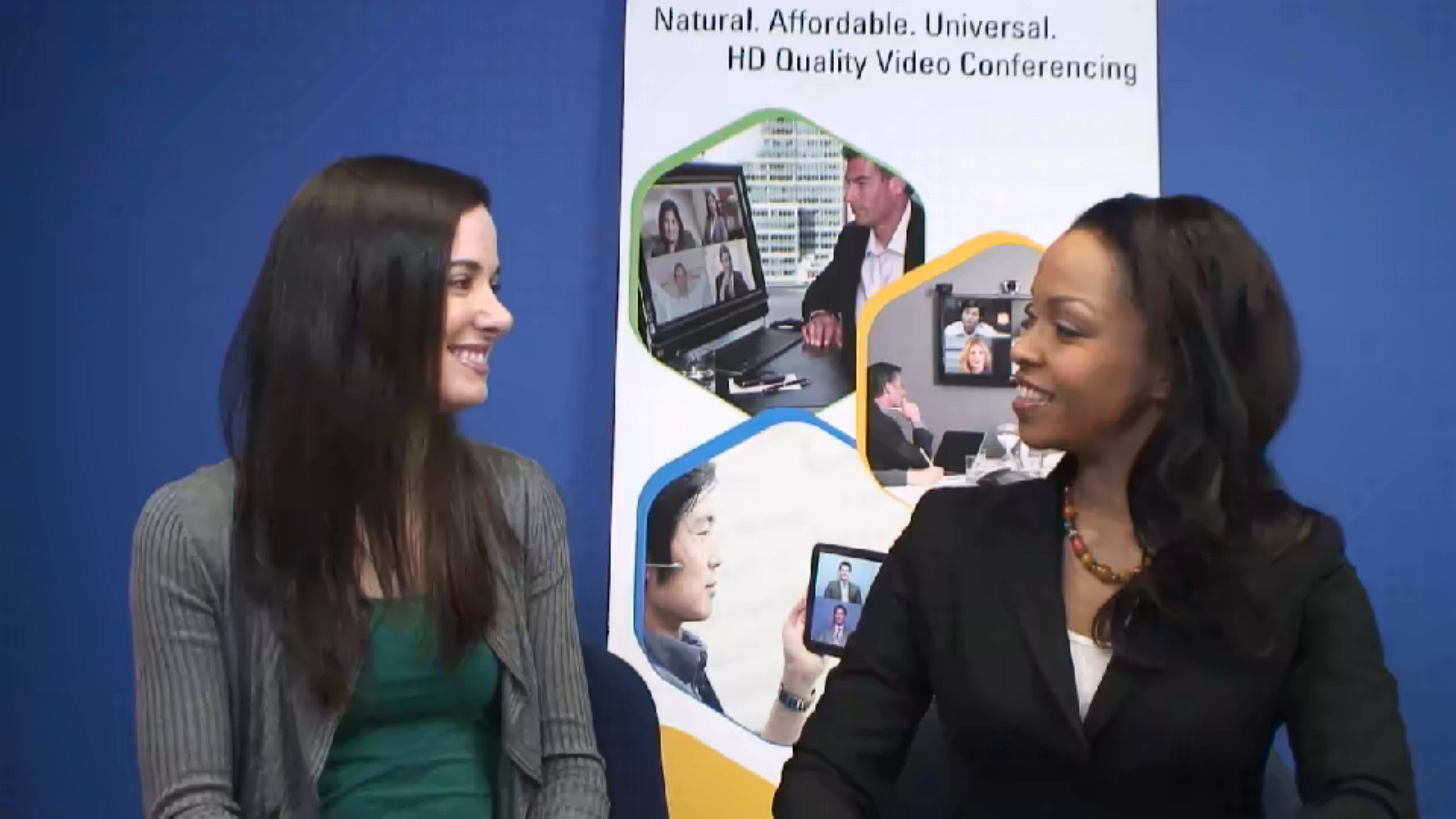}}
	\subfigure[
	$\mbox{PSNR-Y} = 39.0571$dB,
	$\mbox{PSNR-U} = 43.0027$dB and
	$\mbox{PSNR-V} = 43.8623$dB]{\includegraphics[scale=.15]{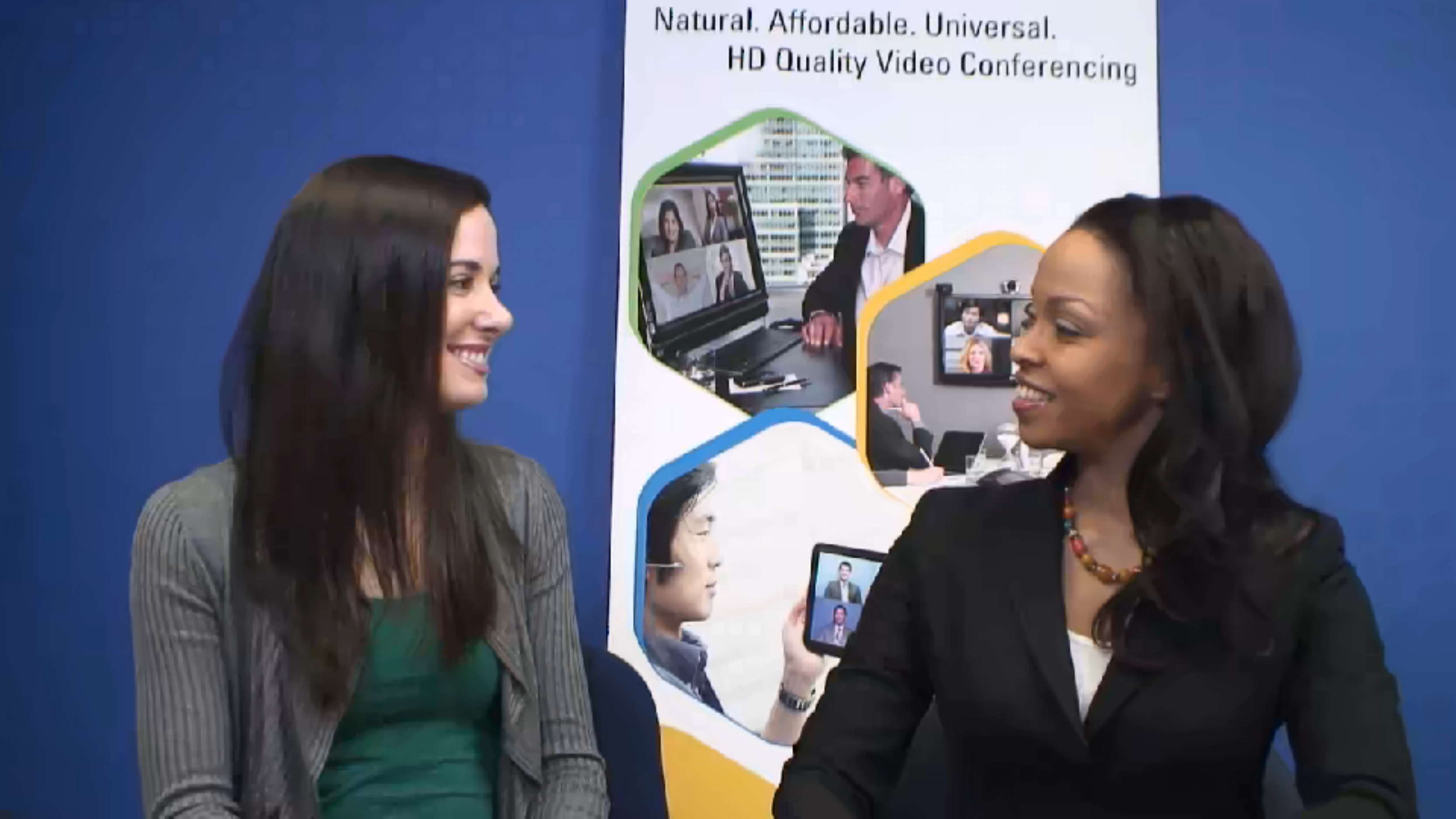}}
	\subfigure[
	$\mbox{PSNR-Y} = 39.1155$dB,
	$\mbox{PSNR-U} = 43.1140$dB and
	$\mbox{PSNR-V} = 44.0624$dB]{\includegraphics[scale=.15]{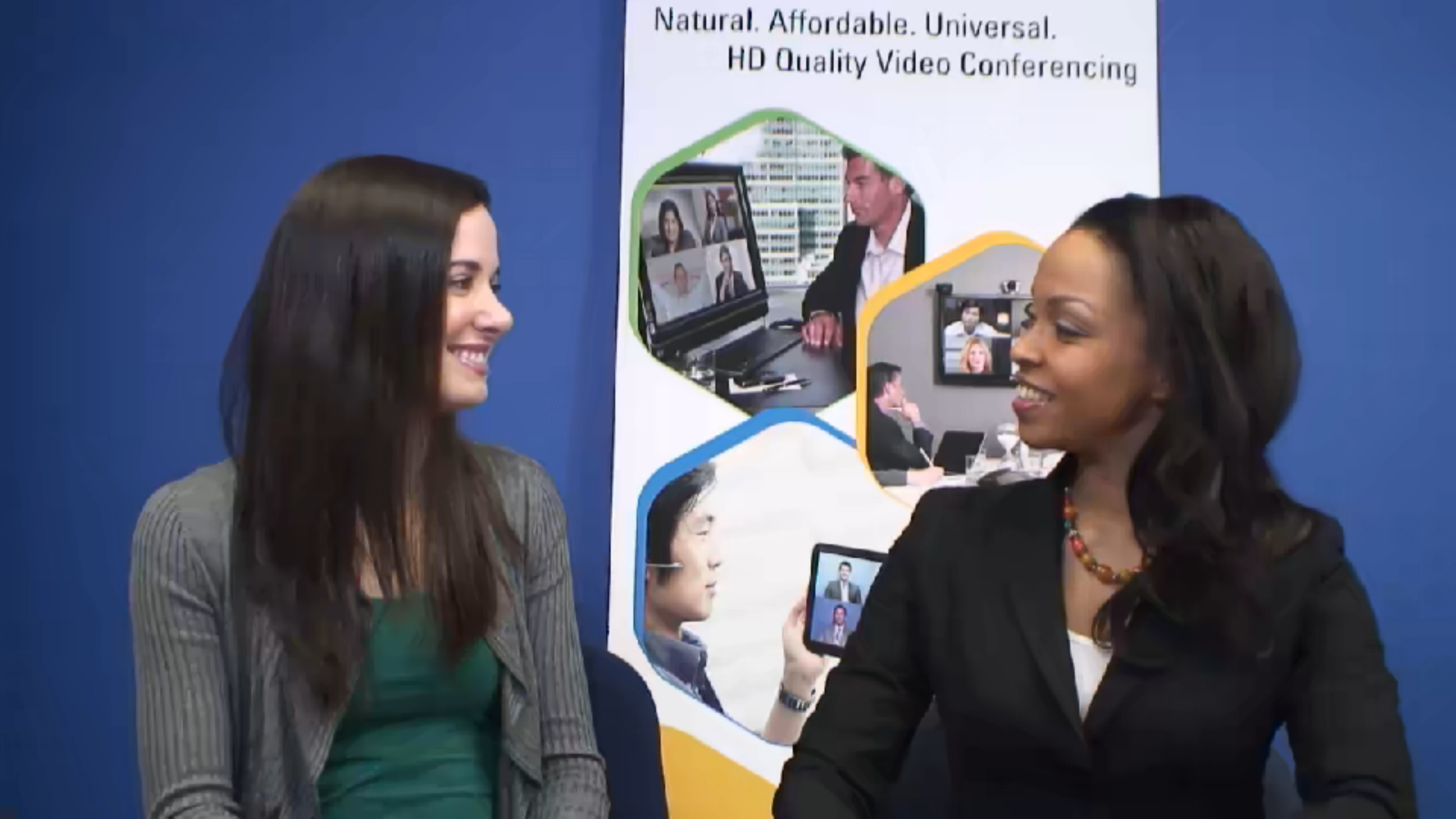}}
	\subfigure[
	$\mbox{PSNR-Y} = 39.1320$dB,
	$\mbox{PSNR-U} = 43.1120$dB and
	$\mbox{PSNR-V} = 44.0912$dB]{\includegraphics[scale=.15]{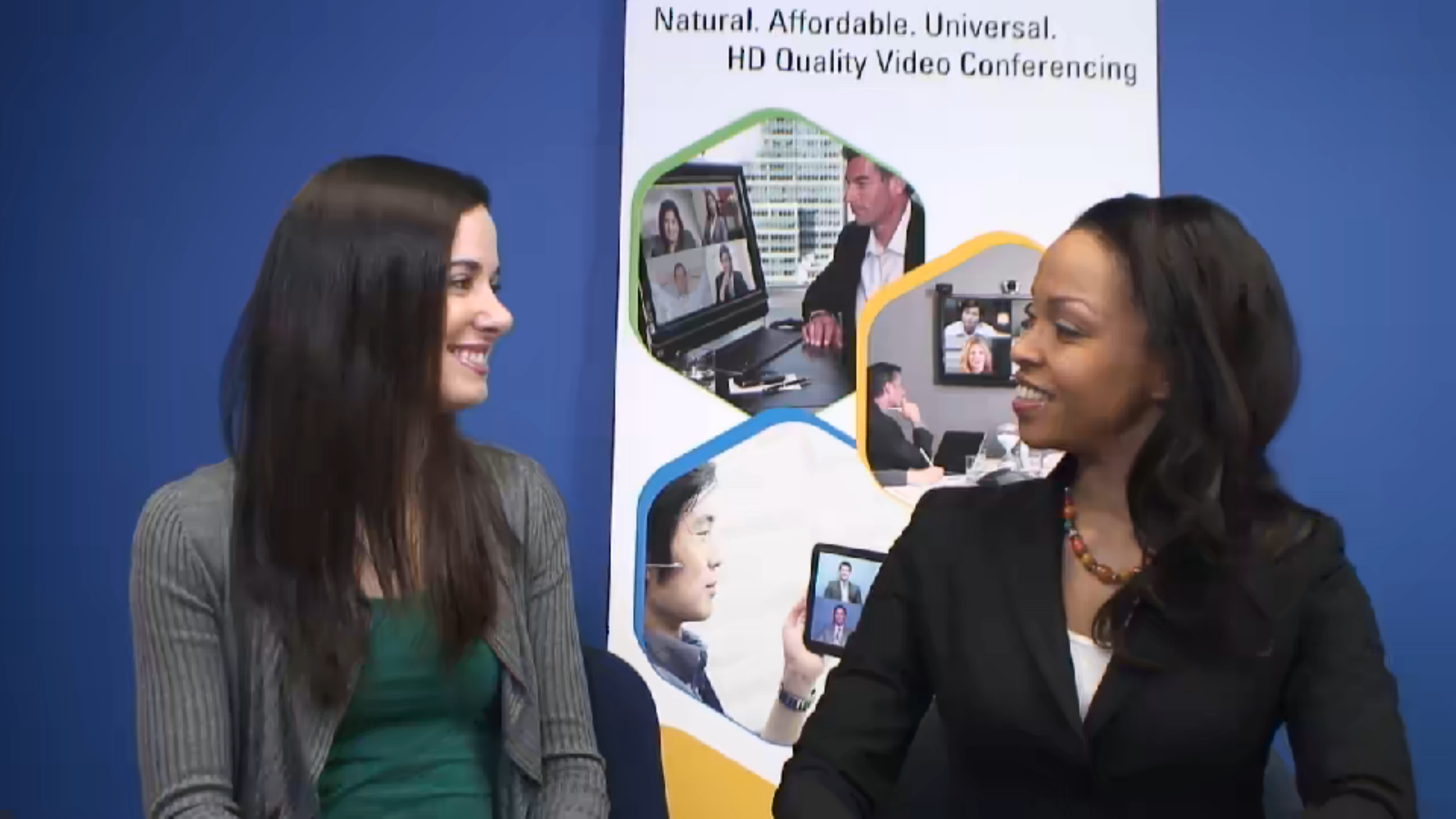}}
	\caption{
		Compression of the tenth frame of \texttt{KristenAndSara} using
		the default  and  modified versions of the HEVC software in configured to the AI mode
		and $\mbox{QP} = 32$. Results for the IDCT are shown in (a), and for the SKLT Groups I to IV in (b) to (e), respectively.
	}
	\label{fig:exemplehevc}
\end{figure*}

As a qualitative example, we present in Figure~\ref{fig:exemplehevc}
the tenth frame of the \texttt{KristenAndSara} video encoded according to~the default HEVC IDCT and the transforms in Groups I--IV in AI configuration.
blue
The presented metrics are a representation from the performance obtained by the proposed transforms on the other video sequences as well.
Here, QP value was set to 32.
Blocking artifacts are not easily perceptible, highlighting the applicability of the proposed SKLT.

\section{Conclusions}\label{S:conclu}

Based on the signum function, we proposed a class of low-complexity KLT approximations, which are suitable for data decorrelation.
These transforms are deterministically defined for processing data at a wide range correlation,
which differs from the fast approximations for the KLT already known in the literature.
Since the proposed transforms are the first approximations for the KLT following this approach, there is nothing to compare with. In principle,
all approximations derived from this method are novelty.
In particular, we explicitly derived new transforms of length  $4$, $8$, $16$, and $32$ and submitted them to a comprehensive assessment in the context of image and video coding. Total
figures of merit
were proposed for the selection of optimal transforms.
The proposed approximations were tailored to decorrelate Markovian first-order data, at a very low arithmetic complexity and multiplierless operation.
Besides the lower complexity, we also derived fast algorithms for the proposed transforms that were capable of reducing, even more, the arithmetic cost of its implementation.
The proposed SKLT showed good
compaction energy properties
at a very low cost.
Still image and video experiments demonstrate the suitability of the proposed approximations for image/video encoding, being capable of generating high quality images according to coding and similarity metrics.
For future works, we wish to consider larger blocklenghts transforms and
other
applications for
fully exploring the potential of the proposed transforms, such as the versatile video coding (VVC) standard~\cite{zeng2021approximated,9498196,ding2020fast}, for example.

\appendix
\section{Matrix Factorization}\label{A:Factorization}
\subsection{ $N=16$} \label{A:n16}
For $N = 16$, we can rewrite $\widehat{\mathbf{T}}_{16,1}$, $\widehat{\mathbf{T}}_{16,2}$, and $\widehat{\mathbf{T}}_{16,3}$~\cite{haweel2001new}  as follows:

\begin{align*}
	&\widehat{\mathbf{T}}_{16,1} = \frac{1}{4} \cdot  \mathbf{P}_{16} \cdot\mathbf{A}_{16,3}^\prime\cdot\mathbf{A}_{16,2} \cdot \mathbf{A}_{16,1}, \\
	&\widehat{\mathbf{T}}_{16,2} =  \frac{1}{4} \cdot  \mathbf{P}_{16} \cdot\mathbf{A}_{16,5}^{\prime}\cdot\mathbf{A}_{16,4}^{\prime} \cdot\mathbf{A}_{16,3}^{\prime\prime}\cdot\mathbf{A}_{16,2} \cdot \mathbf{A}_{16,1}, \\
	&\widehat{\mathbf{T}}_{16,3} =  \frac{1}{4} \cdot  \mathbf{P}_{16} \cdot\mathbf{A}_{16,5}^{\prime\prime}\cdot\mathbf{A}_{16,4}^{\prime\prime} \cdot\mathbf{A}_{16,3}^{\prime\prime\prime}\cdot\mathbf{A}_{16,2} \cdot \mathbf{A}_{16,1},
\end{align*}
where
\begin{align*}
	&\mathbf{P}_{16} = \left[
	\begin{rsmallmatrix}
		1 &  &  &  &  &  &  &  &  &  &  &  &  &  &  &  \\
		&  &  &  &  &  &  &  & 1 &  &  &  &  &  &  &  \\
		&  &  &  & 1 &  &  &  &  &  &  &  &  &  &  &  \\
		&  &  &  &  &  &  &  &  &  &  &  & 1 &  &  &  \\
		& 1 &  &  &  &  &  &  &  &  &  &  &  &  &  &  \\
		&  &  &  &  &  &  &  &  & 1 &  &  &  &  &  &  \\
		&  &  &  &  & 1 &  &  &  &  &  &  &  &  &  &  \\
		&  &  &  &  &  &  &  &  &  &  &  &  & 1 &  &  \\
		&  & 1 &  &  &  &  &  &  &  &  &  &  &  &  &  \\
		&  &  &  &  &  &  &  &  &  & 1 &  &  &  &  &  \\
		&  &  &  &  &  & 1 &  &  &  &  &  &  &  &  &  \\
		&  &  &  &  &  &  &  &  &  &  &  &  &  & 1 &  \\
		&  &  & 1 &  &  &  &  &  &  &  &  &  &  &  &  \\
		&  &  &  &  &  &  &  &  &  &  & 1 &  &  &  &  \\
		&  &  &  &  &  &  & 1 &  &  &  &  &  &  &  &  \\
		&  &  &  &  &  &  &  &  &  &  &  &  &  &  & 1 \\
	\end{rsmallmatrix} \right],
	\end{align*}
	\begin{align*}
	\mathbf{A}_{16,1} =
	\begin{bmatrix}
		\mathbf{I}_8 & \bar{\mathbf{I}}_8 \\
		\bar{\mathbf{I}}_8 \ & -\mathbf{I}_8
	\end{bmatrix}, \quad
	\mathbf{A}_{16,2} =
	\begin{bmatrix}
		\mathbf{A}_{8,1} & \\
		& \mathbf{A}_{8,1}
	\end{bmatrix},
	\end{align*}
	\begin{align*}
	&\mathbf{A}_{16,3}^\prime = \left[
	\begin{smallmatrix}
		1 & 1 & 1 & 1 &  &  & &  &  &  &  &  &  &  &  &  \\
		1 & 1 &  & - &  & 1 &  &  &  &  &  &  &  &  &  &  \\
		1 & - & - & 1 &  &  &  &  &  &  &  &  &  &  &  &  \\
		1 & - & 1 & - &  &  &  &  &  &  &  &  &  &  &  &  \\
		&  & & 1 &  & 1 & 1 & 1 &  &  &  &  &  &  &  &  \\
		& 1 &  &  & - & - &  & 1 &  &  &  & &  &  &  &  \\
		&  &  &  & 1 & 1 & - & 1 &  &  &  &  &  &  &  &  \\
		&  &  &  & - & 1 & - & 1 &  &  &  &  &  &  &  &  \\
		& &  &  &  &  &  &  & 1 & 1 & 1 & 1 &  &  &  &  \\
		&  &  &  &  &  &  &  & 1 & 1 &  & - &  & 1 &  &  \\
		&  &  &  & & &  &  & 1 &  & - & 1 &  &  & 1 & \\
		&  & &  &  &  &  &  & 1 & - & 1 &  & 1 &  &  &  \\
		&  &  &  &  & &  &  &  &  &  & - & & - & - & - \\
		&  &  &  &  &  & & &  & - &  &  & 1 & 1 &  & - \\
		&  &  &  &  &  &  &  &  &  & 1 &  & - &  & 1 & - \\
		&  &  &  &  &  &  &  &  &  &  &  & 1 & - & 1 & - \\
	\end{smallmatrix} \right], \quad
	\end{align*}
	\begin{align*}
	\mathbf{A}_{16,3}^{\prime\prime} = \left[
	\begin{rsmallmatrix}
		1 & & 1 & &  &  & &  &   \\
		& 1 & &  &  &  &  &  &    \\
		1 &  & - &  &  &  &  &  &   \\
		&  &  & 1 &  &  &  &  &  \\
		&  & &  & 1 &  & &  &    \\
		&  &  &  &  & 1 &  & 1 &   \\
		&  &  &  &  &  & 1 &  &    \\
		&  &  &  &  & 1 &  & - &   \\
		& &  &  &  &  &  &  & \mathbf{I}_8  \\
	\end{rsmallmatrix} \right], \quad
	\mathbf{A}_{16,4}^{\prime} = \left[
	\begin{rsmallmatrix}
		1 & &  & &  &  & &  &    \\
		& 1 & &  &  &  &  &  &    \\
		&  & 1 &  &  &  &  &  &    \\
		&  &  & 1 &  &  &  &  &    \\
		&  & &  & 1 &  & &  &    \\
		&  &  &  &  & 1 & 1 &  &   \\
		&  &  &  &  & 1 & - &  &    \\
		& &  &  &  &  &  &  & \mathbf{I}_8  \\
	\end{rsmallmatrix} \right], \\
	\end{align*}
	\begin{align*}
	&\mathbf{A}_{16,5}^{\prime} = \left[
	\begin{rsmallmatrix}
		1 & 1 &  & 1 &  &  &  & &  & &  & &  &  &  &  \\
		& 1 & 1 & - &  & &  &  & & &  &  &  &  &  &  \\
		& - & 1 & 1 & &  &  &  & & &  & &  &  &  & \\
		1 & - &  & - &  &  &  &  & &  &  &  &  &  &  &  \\
		&  &  & 1 &  & 1 &  &  &  &  &  &  &  &  & & \\
		& 1 & &  & - &  &  & - &  &  &  &  &  &  & &  \\
		& &  & & 1 &  & 1 &  &  &  &  &  &  &  &  &  \\
		&  & &  & - &  & 1 &  &  &  &  &  &  &  &  & \\
		&  & &  &  &  &  &  & 1 & 1 & 1 & 1 &  &  &  &  \\
		& &  &  &  &  &  &  & 1 & 1 &  & - &  & 1 &  &  \\
		&  & & &  &  &  &  & 1 &  & - & 1 &  & & 1 & \\
		& &  & &  &  &  &  & 1 & - & 1 &  & 1 &  &  &  \\
		&  & &  &  &  &  & &  & &  & - &  & - & - & - \\
		&  &  &  &  &  &  &  &  & - &  & & 1 & 1 &  & - \\
		&  &  &  & &  &  &  &  &  & 1 &  & - &  & 1 & - \\
		&  &  &  &  &  &  &  &  & &  &  & 1 & - & 1 & - \\
	\end{rsmallmatrix} \right], %
	\end{align*}
	\begin{align*}
	\mathbf{A}_{16,3}^{\prime\prime\prime} =
	\begin{bmatrix}
		\mathbf{A}_{8,2} & \\
		& \mathbf{I}_{8}
	\end{bmatrix}, \quad
	\mathbf{A}_{16,4}^{\prime\prime} = \left[
	\begin{rsmallmatrix}
		1 & 1&  & &  &  & &  &   \\
		1 & - & &  &  &  &  &  &   \\
		&  & 1 & 1  &  &  &  &  &    \\
		&  & 1 & - &  &  &  &  &    \\
		&  & &  & 1 &  & &  &   \\
		&  &  &  &  & 1 & &  &   \\
		&  &  &  &  &  & 1 &  &    \\
		&  &  &  &  &  &  & 1 &   \\
		& &  &  &  &  &  &  & \mathbf{I}_8  \\
	\end{rsmallmatrix} \right], %
	\end{align*}
	\begin{align*}
	&\mathbf{A}_{16,5}^{\prime\prime} = \left[
	\begin{rsmallmatrix}
		1 &  &  &  &  &  &  &  &  &  &  &  &  &  &  &  \\
		&  & 1 &  &  &  &  &  &  &  &  &  &  &  &  &  \\
		& 1 &  &  &  &  &  &  &  &  &  &  &  &  &  &  \\
		&  &  & - &  &  &  &  &  &  &  &  &  &  &  &  \\
		&  &  &  & 1 & 1 &  &  &  &  &  &  &  &  &  &  \\
		&  &  &  &  & - &  & - &  &  &  &  &  &  &  &  \\
		&  &  &  & 1 &  & 1 &  &  &  &  &  &  &  &  &  \\
		&  &  &  &  &  & 1 & - &  &  &  &  &  &  &  &  \\
		&  &  &  &  &  &  &  & 1 & 1 & 1 & 1 &  &  &  &  \\
		&  &  &  &  &  &  &  & 1 & 1 &  & - &  & 1 &  &  \\
		&  &  &  &  &  &  &  & 1 &  & - &  & - &  & 1 &  \\
		&  &  &  &  &  &  &  & 1 & - & 1 &  & 1 &  &  &  \\
		&  &  &  &  &  &  &  &  &  &  & - &  & - & - & - \\
		&  &  &  &  &  &  &  &  & - &  & 1 &  & 1 &  & - \\
		&  &  &  &  &  &  &  &  &  & 1 &  & - &  & 1 & - \\
		&  &  &  &  &  &  &  &  &  &  &  & 1 & - & 1 & - \\
	\end{rsmallmatrix} \right].
\end{align*}
\subsection{ $N=32$}\label{A:n32}
Considering $N = 32$, we have:
\begin{align*}
	&\widehat{\mathbf{T}}_{32,1} =  \frac{1}{\sqrt{32}} \cdot \mathbf{P}_{32} \cdot\mathbf{A}_{32,3}^\prime\cdot\mathbf{A}_{32,2} \cdot \mathbf{A}_{32,1}, \\
	&\widehat{\mathbf{T}}_{32,2} = \frac{1}{\sqrt{32}} \cdot \mathbf{P}_{32} \cdot\mathbf{A}_{32,3}^{\prime\prime}\cdot\mathbf{A}_{32,2} \cdot \mathbf{A}_{32,1}, \\
	&\widehat{\mathbf{T}}_{32,3} = \frac{1}{\sqrt{32}} \cdot \mathbf{P}_{32} \cdot\mathbf{A}_{32,3}^{\prime\prime\prime}\cdot\mathbf{A}_{32,2} \cdot \mathbf{A}_{32,1},\\
	&\widehat{\mathbf{T}}_{32,4} = \frac{1}{\sqrt{32}} \cdot \mathbf{P}_{32} \cdot\mathbf{A}_{32,6}^{\prime} \cdot\mathbf{A}_{32,5}^{\prime} \cdot\mathbf{A}_{32,4}^{\prime} \cdot\mathbf{A}_{32,3}^{\prime\prime\prime\prime}\cdot\mathbf{A}_{32,2} \cdot \mathbf{A}_{32,1},
\end{align*}
where
\begin{align*}
	&\mathbf{P}_{32} = \left[
	\begin{rsmallmatrix}
		1 &  &  &  & &  &  &  &  &  &  &  &  &  &  &  &  &  &  &  &  &  &  &  &  &  &  &  &  &  &  &  \\
		&  &  &  &  &  &  &  &  &  &  &  &  &  &  &  & 1 &  &  &  &  &  &  &  &  &  &  &  &  &  &  &  \\
		&  &  &  &  &  &  &  & 1 &  &  &  &  &  &  &  &  &  &  &  &  &  &  &  &  &  &  &  &  &  &  &  \\
		&  &  &  &  &  &  &  &  &  &  &  &  &  &  &  &  &  &  &  &  &  &  &  & 1 &  &  &  &  &  &  &  \\
		& 1 &  &  &  &  &  &  &  &  &  &  &  &  &  &  &  &  &  &  &  &  &  &  &  &  &  &  &  &  &  &  \\
		&  &  &  &  &  &  &  &  &  &  &  &  &  &  &  &  & 1 &  &  &  &  &  &  &  &  &  &  &  &  &  &  \\
		&  &  &  &  &  &  &  &  & 1 &  &  &  &  &  &  &  &  &  &  &  &  &  &  &  &  &  &  &  &  &  &  \\
		&  &  &  &  &  &  &  &  &  &  &  &  &  &  &  &  &  &  &  &  &  &  &  &  & 1 &  &  &  &  &  &  \\
		&  & 1 &  &  &  &  &  &  &  &  &  &  &  &  &  &  &  &  &  &  &  &  &  &  &  &  &  &  &  &  &  \\
		&  &  &  &  &  &  &  &  &  &  &  &  &  &  &  &  &  & 1 &  &  &  &  &  &  &  &  &  &  &  &  &  \\
		&  &  &  &  &  &  &  &  &  & 1 &  &  &  &  &  &  &  &  &  &  &  &  &  &  &  &  &  &  &  &  &  \\
		&  &  &  &  &  &  &  &  &  &  &  &  &  &  &  &  &  &  &  &  &  &  &  &  &  & 1 &  &  &  &  &  \\
		&  &  & 1 &  &  &  &  &  &  &  &  &  &  &  &  &  &  &  &  &  &  &  &  &  &  &  &  &  &  &  &  \\
		&  &  &  &  &  &  &  &  &  &  &  &  &  &  &  &  &  &  & 1 &  &  &  &  &  &  &  &  &  &  &  &  \\
		&  &  &  &  &  &  &  &  &  &  & 1 &  &  &  &  &  &  &  &  &  &  &  &  &  &  &  &  &  &  &  &  \\
		&  &  &  &  &  &  &  &  &  &  &  &  &  &  &  &  &  &  &  &  &  &  &  &  &  &  & 1 &  &  &  &  \\
		&  &  &  & 1 &  &  &  &  &  &  &  &  &  &  &  &  &  &  &  &  &  &  &  &  &  &  &  &  &  &  &  \\
		&  &  &  &  &  &  &  &  &  &  &  &  &  &  &  &  &  &  &  & 1 &  &  &  &  &  &  &  &  &  &  &  \\
		&  &  &  &  &  &  &  &  &  &  &  & 1 &  &  &  &  &  &  &  &  &  &  &  &  &  &  &  &  &  &  &  \\
		&  &  &  &  &  &  &  &  &  &  &  &  &  &  &  &  &  &  &  &  &  &  &  &  &  &  &  & 1 &  &  &  \\
		&  &  &  &  & 1 &  &  &  &  &  &  &  &  &  &  &  &  &  &  &  &  &  &  &  &  &  &  &  &  &  &  \\
		&  &  &  &  &  &  &  &  &  &  &  &  &  &  &  &  &  &  &  &  & 1 &  &  &  &  &  &  &  &  &  &  \\
		&  &  &  &  &  &  &  &  &  &  &  &  & 1 &  &  &  &  &  &  &  &  &  &  &  &  &  &  &  &  &  &  \\
		&  &  &  &  &  &  &  &  &  &  &  &  &  &  &  &  &  &  &  &  &  &  &  &  &  &  &  &  & 1 &  &  \\
		&  &  &  &  &  & 1 &  &  &  &  &  &  &  &  &  &  &  &  &  &  &  &  &  &  &  &  &  &  &  &  &  \\
		&  &  &  &  &  &  &  &  &  &  &  &  &  &  &  &  &  &  &  &  &  & 1 &  &  &  &  &  &  &  &  &  \\
		&  &  &  &  &  &  &  &  &  &  &  &  &  & 1 &  &  &  &  &  &  &  &  &  &  &  &  &  &  &  &  &  \\
		&  &  &  &  &  &  &  &  &  &  &  &  &  &  &  &  &  &  &  &  &  &  &  &  &  &  &  &  &  & 1 &  \\
		&  &  &  &  &  &  & 1 &  &  &  &  &  &  &  &  &  &  &  &  &  &  &  &  &  &  &  &  &  &  &  &  \\
		&  &  &  &  &  &  &  &  &  &  &  &  &  &  &  &  &  &  &  &  &  &  & 1 &  &  &  &  &  &  &  &  \\
		&  &  &  &  &  &  &  &  &  &  &  &  &  &  & 1 &  &  &  &  &  &  &  &  &  &  &  &  &  &  &  &  \\
		&  &  &  &  &  &  &  &  &  &  &  &  &  &  &  &  &  &  &  &  &  &  &  &  &  &  &  &  &  &  & 1 \\
	\end{rsmallmatrix} \right],
\end{align*}
\begin{align*}
	&\mathbf{A}_{32,1} =
	\begin{bmatrix}
		\mathbf{I}_{16} & \bar{\mathbf{I}}_{16}\\
		\bar{\mathbf{I}}_{16} \ & -\mathbf{I}_{16}
	\end{bmatrix}, \quad
	\mathbf{A}_{32,2} =
	\begin{bmatrix}
		\mathbf{A}_{16,1} & \\
		& \mathbf{A}_{16,1}
	\end{bmatrix},
\end{align*}
	\begin{align*}
	&\mathbf{A}^{\prime}_{32,3} = \left[
	\begin{rsmallmatrix}
		1 & 1 & 1 & 1 & 1 & 1 & 1 & 1 &  &  &  &  &  &  &  &  &  &  &  &  &  &  &  &  &  &  &  &  &  &  &  &  \\
		1 & 1 & 1 &  &  &  & - & - &  &  & 1 & 1 & 1 &  &  &  &  &  &  &  &  &  &  &  &  &  &  &  &  &  &  &  \\
		1 & 1 &  & - & - & - &  & 1 &  & - &  &  &  & 1 &  &  &  &  &  &  &  &  &  &  &  &  &  &  &  &  &  &  \\
		1 &  & - & - &  & 1 &  & - &  & 1 &  & - &  &  & 1 &  &  &  &  &  &  &  &  &  &  &  &  &  &  &  &  &  \\
		1 & - & - & 1 & 1 & - & - & 1 &  &  &  &  &  &  &  &  &  &  &  &  &  &  &  &  &  &  &  &  &  &  &  &  \\
		1 & - &  & 1 & - & - & 1 & - &  &  &  &  &  & - &  &  &  &  &  &  &  &  &  &  &  &  &  &  &  &  &  &  \\
		1 & - & 1 & - & - & 1 & - & 1 &  &  &  &  &  &  &  &  &  &  &  &  &  &  &  &  &  &  &  &  &  &  &  &  \\
		1 & - & 1 & - & 1 & - & 1 & - &  &  &  &  &  &  &  &  &  &  &  &  &  &  &  &  &  &  &  &  &  &  &  &  \\
		&  &  &  &  &  & 1 & 1 &  &  & 1 & 1 & 1 & 1 & 1 & 1 &  &  &  &  &  &  &  &  &  &  &  &  &  &  &  &  \\
		&  & 1 & 1 &  &  &  & - &  & - & - & - &  &  & 1 & 1 &  &  &  &  &  &  &  &  &  &  &  &  &  &  &  &  \\
		&  &  &  &  &  &  &  & 1 & 1 & 1 & - & - & - & 1 & 1 &  &  &  &  &  &  &  &  &  &  &  &  &  &  &  &  \\
		& 1 &  & - &  &  &  &  & - & - & 1 & 1 &  & - &  & 1 &  &  &  &  &  &  &  &  &  &  &  &  &  &  &  &  \\
		&  &  &  & 1 &  &  &  & 1 & 1 & - &  & 1 & - & - & 1 &  &  &  &  &  &  &  &  &  &  &  &  &  &  &  &  \\
		&  &  &  &  &  & 1 &  & - &  & 1 & - & 1 & 1 & - & 1 &  &  &  &  &  &  &  &  &  &  &  &  &  &  &  &  \\
		&  &  &  &  &  &  &  & 1 & - & 1 & 1 & - & 1 & - & 1 &  &  &  &  &  &  &  &  &  &  &  &  &  &  &  &  \\
		&  &  &  &  &  &  &  & - & 1 & - & 1 & - & 1 & - & 1 &  &  &  &  &  &  &  &  &  &  &  &  &  &  &  &  \\
		&  &  &  &  &  &  &  &  &  &  &  &  &  &  &  & 1 & 1 & 1 & 1 & 1 & 1 & 1 & 1 &  &  &  &  &  &  &  &  \\
		&  &  &  &  &  &  &  &  &  &  &  &  &  &  &  & 1 & 1 & 1 & 1 & 1 &  & - & - &  &  & 1 &  &  &  &  &  \\
		&  &  &  &  &  &  &  &  &  &  &  &  &  &  &  & 1 & 1 & 1 & - & - & - &  & 1 &  & - &  &  &  &  &  &  \\
		&  &  &  &  &  &  &  &  &  &  &  &  &  &  &  & 1 & 1 & - & - &  & 1 & 1 & - &  &  &  & - &  &  &  &  \\
		&  &  &  &  &  &  &  &  &  &  &  &  &  &  &  & 1 &  & - &  & 1 &  & - & 1 &  &  & 1 &  & - &  & 1 &  \\
		&  &  &  &  &  &  &  &  &  &  &  &  &  &  &  & 1 &  &  & 1 &  &  & 1 &  & 1 &  & - & 1 &  & - & 1 &  \\
		&  &  &  &  &  &  &  &  &  &  &  &  &  &  &  & 1 & - &  &  & - & 1 & - &  & - &  &  &  & 1 & - &  &  \\
		&  &  &  &  &  &  &  &  &  &  &  &  &  &  &  & 1 & - & 1 & - & 1 &  &  &  & 1 & - & 1 &  &  &  &  &  \\
		&  &  &  &  &  &  &  &  &  &  &  &  &  &  &  &  &  &  &  &  &  &  &  & - & - & - & - & - & - & - & - \\
		&  &  &  &  &  &  &  &  &  &  &  &  &  &  &  &  &  &  &  &  &  &  &  & 1 & 1 & 1 & 1 & - & - & - & - \\
		&  &  &  &  &  &  &  &  &  &  &  &  &  &  &  &  &  & - &  &  & 1 &  &  & - & - &  & 1 & 1 &  & - & - \\
		&  &  &  &  &  &  &  &  &  &  &  &  &  &  &  &  &  &  &  &  &  &  &  & 1 & 1 & - & - & 1 & 1 & - & - \\
		&  &  &  &  &  &  &  &  &  &  &  &  &  &  &  &  & - &  &  & - &  & 1 &  & - &  & 1 &  & - & 1 &  & - \\
		&  &  &  &  &  &  &  &  &  &  &  &  &  &  &  &  &  & 1 &  &  & 1 & - &  & 1 &  &  & 1 & - &  & 1 & - \\
		&  &  &  &  &  &  &  &  &  &  &  &  &  &  &  &  &  &  & - & 1 & - &  &  & - & 1 &  &  &  & - & 1 & - \\
		&  &  &  &  &  &  &  &  &  &  &  &  &  &  &  &  &  &  &  &  &  &  &  & 1 & - & 1 & - & 1 & - & 1 & - \\
	\end{rsmallmatrix} \right],
\end{align*}
\begin{align*}
	&\mathbf{A}^{\prime\prime}_{32,3} = \left[
	\begin{rsmallmatrix}
		1 & 1 & 1 & 1 & 1 & 1 & 1 & 1 &  &  &  &  &  &  &  &  &  &  &  &  &  &  &  &  &  &  &  &  &  &  &  &  \\
		1 & 1 & 1 &  &  &  & - & - &  &  & 1 & 1 & 1 &  &  &  &  &  &  &  &  &  &  &  &  &  &  &  &  &  &  &  \\
		1 & 1 &  & - & - & - &  & 1 &  & - &  &  &  & 1 &  &  &  &  &  &  &  &  &  &  &  &  &  &  &  &  &  &  \\
		1 &  & - & - & 1 & 1 &  & - &  & 1 &  &  &  &  & 1 &  &  &  &  &  &  &  &  &  &  &  &  &  &  &  &  &  \\
		1 & - & - & 1 & 1 & - & - & 1 &  &  &  &  &  &  &  &  &  &  &  &  &  &  &  &  &  &  &  &  &  &  &  &  \\
		1 & - &  & 1 & - & - & 1 & - &  &  &  &  &  & - &  &  &  &  &  &  &  &  &  &  &  &  &  &  &  &  &  &  \\
		1 & - & 1 & - & - & 1 & - & 1 &  &  &  &  &  &  &  &  &  &  &  &  &  &  &  &  &  &  &  &  &  &  &  &  \\
		1 & - & 1 & - & 1 & - & 1 & - &  &  &  &  &  &  &  &  &  &  &  &  &  &  &  &  &  &  &  &  &  &  &  &  \\
		&  &  &  &  &  & 1 & 1 &  &  & 1 & 1 & 1 & 1 & 1 & 1 &  &  &  &  &  &  &  &  &  &  &  &  &  &  &  &  \\
		&  & 1 & 1 &  &  &  & - &  & - & - & - &  &  & 1 & 1 &  &  &  &  &  &  &  &  &  &  &  &  &  &  &  &  \\
		&  &  &  &  &  &  &  & 1 & 1 & 1 & - & - & - & 1 & 1 &  &  &  &  &  &  &  &  &  &  &  &  &  &  &  &  \\
		& 1 &  & - &  &  &  &  & - & - & 1 & 1 &  & - &  & 1 &  &  &  &  &  &  &  &  &  &  &  &  &  &  &  &  \\
		&  &  &  & 1 &  &  &  & 1 & 1 & - &  & 1 & - & - & 1 &  &  &  &  &  &  &  &  &  &  &  &  &  &  &  &  \\
		&  &  &  &  &  & 1 &  & - &  & 1 & - & 1 & 1 & - & 1 &  &  &  &  &  &  &  &  &  &  &  &  &  &  &  &  \\
		&  &  &  &  &  &  &  & 1 & - & 1 & 1 & - & 1 & - & 1 &  &  &  &  &  &  &  &  &  &  &  &  &  &  &  &  \\
		&  &  &  &  &  &  &  & - & 1 & - & 1 & - & 1 & - & 1 &  &  &  &  &  &  &  &  &  &  &  &  &  &  &  &  \\
		&  &  &  &  &  &  &  &  &  &  &  &  &  &  &  & 1 & 1 & 1 & 1 & 1 & 1 & 1 & 1 &  &  &  &  &  &  &  &  \\
		&  &  &  &  &  &  &  &  &  &  &  &  &  &  &  & 1 & 1 & 1 & 1 & 1 &  & - & - &  &  & 1 &  &  &  &  &  \\
		&  &  &  &  &  &  &  &  &  &  &  &  &  &  &  & 1 & 1 & 1 & - & - & - &  & 1 &  & - &  &  &  &  &  &  \\
		&  &  &  &  &  &  &  &  &  &  &  &  &  &  &  & 1 & 1 & - & - &  & 1 & 1 & - &  &  &  & - &  &  &  &  \\
		&  &  &  &  &  &  &  &  &  &  &  &  &  &  &  & 1 &  & - &  & 1 &  & - & 1 &  &  & 1 &  & - &  & 1 &  \\
		&  &  &  &  &  &  &  &  &  &  &  &  &  &  &  & 1 &  &  & 1 &  &  & 1 &  & 1 &  & - & 1 &  & - & 1 &  \\
		&  &  &  &  &  &  &  &  &  &  &  &  &  &  &  & 1 & - &  &  & - & 1 & - &  & - &  &  &  & 1 & - &  &  \\
		&  &  &  &  &  &  &  &  &  &  &  &  &  &  &  & 1 & - & 1 & - & 1 &  &  &  & 1 & - & 1 &  &  &  &  &  \\
		&  &  &  &  &  &  &  &  &  &  &  &  &  &  &  &  &  &  &  &  &  &  &  & - & - & - & - & - & - & - & - \\
		&  &  &  &  &  &  &  &  &  &  &  &  &  &  &  &  &  &  &  &  &  &  &  & 1 & 1 & 1 & 1 & - & - & - & - \\
		&  &  &  &  &  &  &  &  &  &  &  &  &  &  &  &  &  & - &  &  & 1 &  &  & - & - &  & 1 & 1 &  & - & - \\
		&  &  &  &  &  &  &  &  &  &  &  &  &  &  &  &  & - &  &  &  &  &  &  & 1 & 1 & - & - & 1 & 1 &  & - \\
		&  &  &  &  &  &  &  &  &  &  &  &  &  &  &  &  & - &  &  & - &  & 1 &  & - &  & 1 &  & - & 1 &  & - \\
		&  &  &  &  &  &  &  &  &  &  &  &  &  &  &  &  &  & 1 &  &  & 1 & - &  & 1 &  &  & 1 & - &  & 1 & - \\
		&  &  &  &  &  &  &  &  &  &  &  &  &  &  &  &  &  &  & - & 1 & - &  &  & - & 1 &  &  &  & - & 1 & - \\
		&  &  &  &  &  &  &  &  &  &  &  &  &  &  &  &  &  &  &  &  &  &  &  & 1 & - & 1 & - & 1 & - & 1 & - \\
	\end{rsmallmatrix} \right],
\end{align*}
	\begin{align*}
	&\mathbf{A}^{\prime\prime\prime}_{32,3} = \left[
	\begin{rsmallmatrix}
		1 & 1 & 1 & 1 & 1 & 1 & 1 & 1 &  &  &  &  &  &  &  &  &  &  &  &  &  &  &  &  &  &  &  &  &  &  &  &  \\
		1 & 1 & 1 &  &  &  & - & - &  &  & 1 & 1 & 1 &  &  &  &  &  &  &  &  &  &  &  &  &  &  &  &  &  &  &  \\
		1 & 1 &  & - & - & - &  & 1 &  & - &  &  &  & 1 &  &  &  &  &  &  &  &  &  &  &  &  &  &  &  &  &  &  \\
		1 &  & - & - & 1 & 1 & 1 & - &  &  &  &  &  &  & 1 &  &  &  &  &  &  &  &  &  &  &  &  &  &  &  &  &  \\
		1 & - & - & 1 & 1 & - & - & 1 &  &  &  &  &  &  &  &  &  &  &  &  &  &  &  &  &  &  &  &  &  &  &  &  \\
		1 & - & 1 & 1 & - & - & 1 & - &  &  &  &  &  &  &  &  &  &  &  &  &  &  &  &  &  &  &  &  &  &  &  &  \\
		1 & - & 1 & - & - & 1 & - & 1 &  &  &  &  &  &  &  &  &  &  &  &  &  &  &  &  &  &  &  &  &  &  &  &  \\
		1 & - & 1 & - & 1 & - & 1 & - &  &  &  &  &  &  &  &  &  &  &  &  &  &  &  &  &  &  &  &  &  &  &  &  \\
		&  &  &  &  &  & 1 & 1 &  &  & 1 & 1 & 1 & 1 & 1 & 1 &  &  &  &  &  &  &  &  &  &  &  &  &  &  &  &  \\
		&  & 1 & 1 &  &  &  & - &  & - & - & - &  &  & 1 & 1 &  &  &  &  &  &  &  &  &  &  &  &  &  &  &  &  \\
		&  &  &  &  &  &  &  & 1 & 1 & 1 & - & - & - & 1 & 1 &  &  &  &  &  &  &  &  &  &  &  &  &  &  &  &  \\
		&  &  & - &  &  &  &  & - & - & 1 & 1 &  & - & - & 1 &  &  &  &  &  &  &  &  &  &  &  &  &  &  &  &  \\
		&  &  &  & 1 &  &  &  & 1 & 1 & - &  & 1 & - & - & 1 &  &  &  &  &  &  &  &  &  &  &  &  &  &  &  &  \\
		&  &  &  &  &  &  &  & - & 1 & 1 & - & 1 & 1 & - & 1 &  &  &  &  &  &  &  &  &  &  &  &  &  &  &  &  \\
		&  &  &  &  &  &  &  & 1 & - & 1 & 1 & - & 1 & - & 1 &  &  &  &  &  &  &  &  &  &  &  &  &  &  &  &  \\
		&  &  &  &  &  &  &  & - & 1 & - & 1 & - & 1 & - & 1 &  &  &  &  &  &  &  &  &  &  &  &  &  &  &  &  \\
		&  &  &  &  &  &  &  &  &  &  &  &  &  &  &  & 1 & 1 & 1 & 1 & 1 & 1 & 1 & 1 &  &  &  &  &  &  &  &  \\
		&  &  &  &  &  &  &  &  &  &  &  &  &  &  &  & 1 & 1 & 1 & 1 &  &  & - & - &  &  & 1 & 1 &  &  &  &  \\
		&  &  &  &  &  &  &  &  &  &  &  &  &  &  &  & 1 & 1 &  & - & - & - &  & 1 &  & - &  &  &  & 1 &  &  \\
		&  &  &  &  &  &  &  &  &  &  &  &  &  &  &  & 1 & 1 & - & - &  & 1 &  & - &  & 1 &  & - &  &  &  &  \\
		&  &  &  &  &  &  &  &  &  &  &  &  &  &  &  & 1 &  & - &  & 1 &  & - & 1 &  &  & 1 &  & - &  & 1 &  \\
		&  &  &  &  &  &  &  &  &  &  &  &  &  &  &  & 1 &  &  & 1 &  &  & 1 &  & 1 &  & - & 1 &  & - & 1 &  \\
		&  &  &  &  &  &  &  &  &  &  &  &  &  &  &  & 1 & - &  &  & - & 1 & - &  & - &  &  &  & 1 & - &  &  \\
		&  &  &  &  &  &  &  &  &  &  &  &  &  &  &  & 1 & - & 1 & - & 1 &  &  &  & 1 & - & 1 &  &  &  &  &  \\
		&  &  &  &  &  &  &  &  &  &  &  &  &  &  &  &  &  &  &  &  &  &  & - &  & - & - & - & - & - & - & - \\
		&  &  &  &  &  &  &  &  &  &  &  &  &  &  &  &  &  &  & - &  &  &  & 1 &  & 1 & 1 & 1 &  & - & - & - \\
		&  &  &  &  &  &  &  &  &  &  &  &  &  &  &  &  &  & - &  &  & 1 &  &  & - & - &  & 1 & 1 &  & - & - \\
		&  &  &  &  &  &  &  &  &  &  &  &  &  &  &  &  & - &  & 1 &  & - &  &  & 1 & 1 &  & - &  & 1 &  & - \\
		&  &  &  &  &  &  &  &  &  &  &  &  &  &  &  &  & - &  &  & - &  & 1 &  & - &  & 1 &  & - & 1 &  & - \\
		&  &  &  &  &  &  &  &  &  &  &  &  &  &  &  &  &  & 1 &  &  & 1 & - &  & 1 &  &  & 1 & - &  & 1 & - \\
		&  &  &  &  &  &  &  &  &  &  &  &  &  &  &  &  &  &  & - & 1 & - &  &  & - & 1 &  &  &  & - & 1 & - \\
		&  &  &  &  &  &  &  &  &  &  &  &  &  &  &  &  &  &  &  &  &  &  &  & 1 & - & 1 & - & 1 & - & 1 & - \\
	\end{rsmallmatrix} \right],
\end{align*}
	\begin{align*}
	&\mathbf{A}_{32,3}^{\prime\prime\prime\prime} =
	\begin{bmatrix}
		\mathbf{A}_{16,2} & \\
		& \mathbf{I}_{16}
	\end{bmatrix}, \quad
	\mathbf{A}_{32,4}^{\prime} =
	\begin{bmatrix}
		\mathbf{A}_{16,3}^{\prime\prime\prime} & \\
		& \mathbf{I}_{16}
	\end{bmatrix}, \quad
	\mathbf{A}_{32,5}^{\prime} =
	\begin{bmatrix}
		\mathbf{A}_{16,4}^{\prime\prime} & \\
		& \mathbf{I}_{16}
	\end{bmatrix},
\end{align*}
and
\begin{align*}
	\mathbf{A}^{\prime}_{32,6} = \left[
	\begin{rsmallmatrix}
		1 &  &  &  &  &  &  &  &  &  &  &  &  &  &  &  &  &  &  &  &  &  &  &  &  &  &  &  &  &  &  &  \\
		&  &  &  & 1 & 1 &  &  &  &  &  &  &  &  &  &  &  &  &  &  &  &  &  &  &  &  &  &  &  &  &  &  \\
		&  & 1 &  &  &  &  &  &  &  &  &  &  &  &  &  &  &  &  &  &  &  &  &  &  &  &  &  &  &  &  &  \\
		&  &  &  &  & - &  & - &  &  &  &  &  &  &  &  &  &  &  &  &  &  &  &  &  &  &  &  &  &  &  &  \\
		& 1 &  &  &  &  &  &  &  &  &  &  &  &  &  &  &  &  &  &  &  &  &  &  &  &  &  &  &  &  &  &  \\
		&  &  &  & 1 &  & 1 &  &  &  &  &  &  &  &  &  &  &  &  &  &  &  &  &  &  &  &  &  &  &  &  &  \\
		&  &  & - &  &  &  &  &  &  &  &  &  &  &  &  &  &  &  &  &  &  &  &  &  &  &  &  &  &  &  &  \\
		&  &  &  &  &  & 1 & - &  &  &  &  &  &  &  &  &  &  &  &  &  &  &  &  &  &  &  &  &  &  &  &  \\
		&  &  &  &  &  &  &  & 1 & 1 & 1 & 1 &  &  &  &  &  &  &  &  &  &  &  &  &  &  &  &  &  &  &  &  \\
		&  &  &  &  &  &  &  &  &  &  & - &  & - & - & - &  &  &  &  &  &  &  &  &  &  &  &  &  &  &  &  \\
		&  &  &  &  &  &  &  & 1 & 1 &  & - &  & 1 &  &  &  &  &  &  &  &  &  &  &  &  &  &  &  &  &  &  \\
		&  &  &  &  &  &  &  &  & - &  & 1 &  & 1 &  & - &  &  &  &  &  &  &  &  &  &  &  &  &  &  &  &  \\
		&  &  &  &  &  &  &  & 1 &  & - &  & - &  & 1 &  &  &  &  &  &  &  &  &  &  &  &  &  &  &  &  &  \\
		&  &  &  &  &  &  &  &  &  & 1 &  & - &  & 1 & - &  &  &  &  &  &  &  &  &  &  &  &  &  &  &  &  \\
		&  &  &  &  &  &  &  & 1 & - & 1 &  & 1 &  &  &  &  &  &  &  &  &  &  &  &  &  &  &  &  &  &  &  \\
		&  &  &  &  &  &  &  &  &  &  &  & 1 & - & 1 & - &  &  &  &  &  &  &  &  &  &  &  &  &  &  &  &  \\
		&  &  &  &  &  &  &  &  &  &  &  &  &  &  &  & 1 & 1 & 1 & 1 & 1 & 1 & 1 & 1 &  &  &  &  &  &  &  &  \\
		&  &  &  &  &  &  &  &  &  &  &  &  &  &  &  & 1 & 1 & 1 &  &  &  & - & - &  &  & 1 & 1 & 1 &  &  &  \\
		&  &  &  &  &  &  &  &  &  &  &  &  &  &  &  & 1 & 1 &  &  & - &  &  & 1 &  & - & - &  & 1 & 1 &  &  \\
		&  &  &  &  &  &  &  &  &  &  &  &  &  &  &  & 1 &  & - & - &  & 1 &  & - &  & 1 &  & - &  &  & 1 &  \\
		&  &  &  &  &  &  &  &  &  &  &  &  &  &  &  & 1 &  & - &  & 1 &  & - &  & - &  & 1 &  & - &  & 1 &  \\
		&  &  &  &  &  &  &  &  &  &  &  &  &  &  &  & 1 &  &  & 1 &  &  & 1 &  & 1 &  & - & 1 &  & - & 1 &  \\
		&  &  &  &  &  &  &  &  &  &  &  &  &  &  &  & 1 & - &  &  &  & 1 & - &  & - &  &  & - & 1 & - &  &  \\
		&  &  &  &  &  &  &  &  &  &  &  &  &  &  &  & 1 & - & 1 & - & 1 &  &  &  & 1 & - & 1 &  &  &  &  &  \\
		&  &  &  &  &  &  &  &  &  &  &  &  &  &  &  &  &  &  &  &  &  & - & - &  &  & - & - & - & - & - & - \\
		&  &  &  &  &  &  &  &  &  &  &  &  &  &  &  &  &  & - & - & - &  &  & 1 &  & 1 & 1 &  &  &  & - & - \\
		&  &  &  &  &  &  &  &  &  &  &  &  &  &  &  &  & - & - &  & 1 & 1 &  & - &  & - &  &  & 1 &  &  & - \\
		&  &  &  &  &  &  &  &  &  &  &  &  &  &  &  &  & - &  & 1 &  & - &  & 1 &  & 1 &  & - &  & 1 &  & - \\
		&  &  &  &  &  &  &  &  &  &  &  &  &  &  &  &  & - &  &  & - &  & 1 &  & - &  & 1 &  & - & 1 &  & - \\
		&  &  &  &  &  &  &  &  &  &  &  &  &  &  &  &  &  & 1 & - &  & 1 & - &  & 1 &  &  & 1 &  &  & 1 & - \\
		&  &  &  &  &  &  &  &  &  &  &  &  &  &  &  &  &  &  & - & 1 & - &  &  & - & 1 &  &  &  & - & 1 & - \\
		&  &  &  &  &  &  &  &  &  &  &  &  &  &  &  &  &  &  &  &  &  &  &  & 1 & - & 1 & - & 1 & - & 1 & - \\
	\end{rsmallmatrix} \right].
\end{align*}

\section*{Acknowledgments}
We gratefully acknowledge partial financial support from  \textit{Coordena\c c\~ao de Aperfei\c coamento de Pessoal de N\'ivel Superior (CAPES)}, \textit{Conselho Nacional de Desenvolvimento Cient\'ifico e Tecnol\'ogico (CNPq)} and \textit{Funda\c c\~ao de Amparo a Ci\^encia e Tecnologia de Pernambuco (FACEPE)}, Brazil.

\onecolumn

{\small
\singlespacing
\bibliographystyle{ieeetr}
\bibliography{references-sklt.bib}
}

\end{document}